\documentclass[a4paper,12pt]{article}
\usepackage{amsmath,amssymb,amsthm,mathrsfs,graphicx,float,colordvi,url,wrapfig}
\usepackage{here,color,ulem,enumitem,bm,tikz-cd,blkarray,comment}
\usepackage[all]{xy}  
   
\makeatletter

\newcommand{\xleftrightarrow}[2][]{\ext@arrow 3359\leftrightarrowfill@{#1}{#2}} 
\newcommand{\xdashrightarrow}[2][]{\ext@arrow 0359\rightarrowfill@@{#1}{#2}}
\newcommand{\xdashleftarrow}[2][]{\ext@arrow 3095\leftarrowfill@@{#1}{#2}}  
\newcommand{\xdashleftrightarrow}[2][]{\ext@arrow 3359\leftrightarrowfill@@{#1}{#2}} 
\def\rightarrowfill@@{\arrowfill@@\relax\relbar\rightarrow}
\def\leftarrowfill@@{\arrowfill@@\leftarrow\relbar\relax}
\def\leftrightarrowfill@@{\arrowfill@@\leftarrow\relbar\rightarrow}
\def\arrowfill@@#1#2#3#4{
$\m@th\thickmuskip0mu\medmuskip\thickmuskip\thinmuskip\thickmuskip
\relax#4#1 
\xleaders\hbox{$#4#2$}\hfill
#3$
}     
\makeatother 

\begin{document}
\renewcommand{\thefootnote}{\fnsymbol{footnote}} 
\begin{titlepage}

\vspace*{5mm}

\begin{center}

{\large 
\textbf{Canonical Quantization of the U(1) Gauge Field}
\\[1.0mm]
\textbf{in the Rindler Coordinates}\\[1.0mm]}

\vspace*{7.0mm}

\normalsize
{\large Shingo Takeuchi
}
\vspace*{4.0mm} 

\textit{
\small Faculty of Environmental and Natural Sciences, Duy Tan University, Da Nang, Vietnam}\\
\vspace*{5mm}  
\end{center}

\begin{abstract}
This paper describes the canonical quantization of the U(1) gauge field 
across all four regions in the Rindler coordinates in the Lorentz-covariant gauge.  
Concretely, in the four regions (future, past, left and right Rindler-wedges) in the Rindler coordinates, 
the gauge-fixed Lagrangian in the Lorentz-covariant gauge is obtained, 
which is composed of the U(1) gauge field, the $B$-field and ghost fields.
Since the U(1) gauge and $B$-fields are decoupled from the ghost fields by the property of the U(1) gauge theory,
the U(1) gauge field and the $B$-field are examined in this study.

Then, by solving the equations of motion obtained from the gauge-fixed Lagrangian, 
the solutions of each mode of the U(1) gauge field and  the $B$-field can be obtained. 
Following this, with the Klein-Gordon inner-product defined in the Rindler coordinates,  
the normalization constants of each of those mode-solutions are determined.

Subsequently, formulating the canonical commutation relations of the U(1) gauge field and its canonical conjugate momentum, 
the equal-time commutation relations of the coefficient of each mode-solution 
in each direction of the U(1) gauge field 
in each region of the Rindler coordinates are obtained.
From these, it can be seen that those coefficients have physical meaning as creation/annihilation operators. 
The polarization vectors in each region of the Rindler coordinates are also given in this study.
\end{abstract}
\end{titlepage}

\newpage  
  
\allowdisplaybreaks 
\setcounter{footnote}{0}

\section{Introduction}  
\label{sec:int}   

From the analysis of the uniformly accelerated motion, 
the Unruh temperature is derived as 
$T_U = \hbar a/2\pi c k_{\rm B} \approx 4 \times 10^{-23} \, a/$(cm/$s^2$)[K]~\cite{Fulling:1972md,Davies:1974th,Unruh:1976db}.
Since the Unruh temperature which can currently be produced is less than $3$[K] CMB, its detection remains extremely challenging. 
While detecting the Unruh temperature is a technologically important issue.
It is also important  as experimental confirmation of Hawking radiation, 
and in terms of the problem of how an event which is observed in the accelerated systems can be observed in an inertial system. 

Recent studies in the experimental field have touched upon the following topics:~BEC~\cite{Retzker}, 
neutrino oscillation~\cite{Dvornikov:2015eqa,Blasone:2018czm,Blasone:2020vtm}, 
anti-Unruh effect~\cite{Chen:2021evr,Garay:2016cpf,Liu:2016ihf,Brenna:2015fga},
cold atoms~\cite{Kosior:2018vgx,Rodriguez-Laguna:2016kri}, 
Berry phases~\cite{Quach:2021vzo,Martin-Martinez:2010gnz}, 
Casimir effect~\cite{Marino:2014rfa}, 
classical analog~\cite{Leonhardt:2017lwm} 
and others~\cite{Lynch:2019hmk,Lochan:2019osm}. 

In addition, the effects of the thermal excitation of the Unruh effect on the phenomena currently reproducible in the laboratory 
have been analyzed as follows:
analysis of critical temperatures of the phase transitions~\cite{
Ohsaku:2004rv,Ebert:2006bh,Castorina:2007eb,Takeuchi:2015nga};
analysis of the thermal radiation from a particle in constant accelerated motion~\cite{
Schutzhold:2006gj,Schutzhold:2009scb}\cite{Iso:2010yq,Oshita:2015xaa,Iso:2016lua};
analysis of the Schwinger effect in a constant accelerated systems~\cite{
Parentani:1996gd,Kim:2016dmm,Kaushal:2022las};
and analysis of quantum corrections in the energy-momentum tensors of gases in constant accelerated systems~\cite{
Prokhorov:2019cik,Prokhorov:2019hif,Prokhorov:2019yft,Prokhorov:2019sss}. 
Additionally, the Hagedorn transition in string theory due to the Unruh effect has also been analyzed~\cite{Parentani:1989gq}.
\newline

Theoretically, the coordinates of a constant accelerated motion are generalized to the Rindler coordinates, 
which is equivalent to the geometry in the neighborhood of the event-horizon of the Schwarzschild black hole, 
and in which the Killing horizons exist. 
Therefore,  
the thermal radiation and the entropy akin to the Hawking radiation and the Bekenstein-Hawking entropy, respectively, can be considered in the Rindler coordinates. 
In addition, since the Rindler coordinates is the issue of the coordinates, it can be considered at any points in the spacetime based on the general coordinate transformation invariance.
Exploiting these properties of the Rindler coordinates, it can be shown that Einstein's equation can be derived as a state equation at any point in spacetime~\cite{Jacobson:1995ab}.  

Furthermore, since spacetime is causally separated by the Killing horizon in the Rindler coordinates, quantum entanglement arises between those regions in the Rindler coordinates. 
Analysis of the entanglement entropy (EE) between the separated regions in the Rindler coordinates has been performed in many studies, 
and as a result, the area-proportional low is obtained, 
based on which the suggestion that the Bekenstein-Hawking entropy would be a kind of EE between the inside and outside of the black hole can be obtained~\cite{Kabat:1994vj,Dowker:1994fi,Iellici:1996gv}.

In addition, the Rindler coordinates are extended to the AdS space:
the Unruh temperature in the AdS space has been given in \cite{Deser:1997ri}, 
and the AdS/CFT given with the AdS space patched by the Rindler coordinates has been studied in \cite{Parikh:2012kg,Fareghbal:2014oba,Sugishita:2022ldv}.

In terms of studies on the properties of the Rindler coordinates, the following studies seem particularly intriguing: 
the upper bound for acceleration~\cite{Rovelli:2013osa} 
and another on inversion between the bosonic and fermionic statistics occurring in the odd dimensional Rindler coordinates~\cite{Terashima:1999xp}.
\newline

A review of the literature shows that until now, 
the canonical quantization (CQ) of the gauge fields in the Rindler coordinates has not been properly performed yet,
while that of scalar and spinor fields has been done~\cite{Higuchi:2017gcd}\cite{Soffel:1980kx,Ueda:2021nln}. 
Studies treating the U(1) gauge field in the Rindler coordinates can be found as cited in 
\cite{Higuchi:1992td,Moretti:1996zt,Lenz:2008vw,Zhitnitsky:2010ji,Soldati:2015xma,Blommaert:2018rsf}.
However, in these studies, the normalization constants (NC) of the mode-solutions contain some speculation and have not been obtained exclusively from the analysis, 
or the discussion proceeds without the NC. 
Furthermore, it appears that the solutions for the gauge field in all of the directions in spacetime have not been explicitly obtained.

The NC have not been determined until now likely because the integrals in Appendix.\ref{r3g67kb} in this study could not be calculated.
Since those integrals appear in the Klein-Gordon (KG) inner-products between the mode-solutions of the U(1) gauge field in the Rindler coordinates, 
what those integrals cannot be calculated leads to the following four situations; 
{\bf 1)} their NC cannot be determined, 
{\bf 2)} accordingly, 
the equal-time canonical commutation relations (CCR) (referred to as CCR in what follows, omitting the term ``equal-time'') 
cannot be robustly formulated (it is possible to formally write down the CCR, if the form of the KG inner-product is known\footnote{
If the KG inner-product (\ref{stwgi}) is known, 
from the demand that the commutation relations of the creation/annihilation operators 
can be finally obtained like (\ref{roere}), it can be seen that the form of the CCR is given as (\ref{reawv}).};
however, since the mode-solutions of the field are not completely determined due to the lack of the NC,
the concrete CCR, the CCR for each mode-solution, cannot be completely known),
{\bf 3)} analysis performed using the KG inner-products cannot be done without ambiguities or speculation, 
and {\bf 4)} the creation/annihilation operators cannot be taken out.
All of these would finally lead to the first situation mentioned.

In these situations, we will calculate the integrals as demonstrated in Appendix.\ref{r3g67kb}. 
In addition, in this study, constituting the Lagrangian in the Lorentz-covariant gauge in the Rindler coordinates, 
we solve the equations of motion obtained from that Lagrangian for the mode-solutions of the U(1) gauge field by a very explicit manner. 
Therefore, we are able to give the mode-solutions of the U(1) gauge field across all directions with the properly determined normalization constants (NC) 
in each of the four regions of the Rindler coordinates: future, past, left, and right Rindler-wedges (see Fig.\ref{wsdd57}).
Here, in the process to solve the equations of motion to obtain the mode-solutions, 
we put an ansatz; in this sense, the mode-solutions we obtain are solutions but not general solutions.

However, in this study, all directions of the gauge field will be solved by a very explicit manner, 
and no reference has been found in which all directions of the gauge field are solved by such an explicit manner.
Therefore, it is believed that there is usefulness in the mode-solutions obtained  in this study.
The details are described in Sec.\,\ref{nnklf}.

Next, by formulating the CCR of the U(1) gauge field in each of the four regions in the Rindler coordinates, 
we examine the U(1) gauge field in these CCR using the mode-expanded form based on the normalized mode-solutions we have obtained. 
Subsequently, employing the Klein-Gordon (KG) inner-product, 
we derive the commutation relations of the creation/annihilation operators for the coefficients of the mode-solutions of the U(1) gauge field.
\newline

The following section discusses the organization of this paper.
In Sec.\,\ref{vtiss}, the Rindler coordinates used in this study are reviewed. 
In Sec.\,\ref{vesbv}, the Lagrangian of the U(1) gauge theory in the Lorentz-covariant gauge in the Rindler coordinates is obtained.
In Sec.\,\ref{f2v4t}, the mode-solutions of the U(1) gauge field and the $B$-field in each region 
in the Rindler coordinates and their normalization constants are obtained. 
The mode-solutions obtained here are not general solutions as mentioned above, 
and this is also commented at the end of Sec.\,\ref{nnklf}.

In Sec.\,\ref{bywbd}, the canonical quantization of the U(1) gauge field is computed for each region in the Rindler coordinates, 
through which the commutation relations which the coefficients of each mode-solution satisfy are finally obtained. 
From this, it can be confirmed that 
the coefficients of the each mode-solution have roles as creation/annihilation operators in each region in the Rindler coordinates.

In Sec.\,\ref{yervd}, we address the creation/annihilation operators of the U(1) gauge field, 
providing the polarization vector for $(S,L,\pm)$-directions. 
From the expression of this vector, it becomes apparent that 
the region where the polarization vector can be defined is constrained in each region, 
as the norms of the 1-particle states in the $S$- and $L$-directions can potentially be negative. 
The origins of these restrictions and their implications are discussed.

In Sec.\,\ref{Summary}, 
summarizing this study, the future directions of the study are discussed.
In Appendix.\ref{buobhs}, 
the analysis from (\ref{sknuek}) of the Lagrangian to the Lagrangian in the Lorentz-covariant gauge (\ref{etsiph}) is noted.  
In Appendix.\ref{r3g67kb}, 
the integral formulas, 
which are essential for the calculations of the KG inner-products to obtain the NC in Sec.\,\ref{brpnevt}, are given.
\newline

Since there are some overlaps between this study and the study \cite{Takeuchi:2023nxi},
we mention the difference in these. 
\begin{itemize}
\item
While only the RRW is addressed in \cite{Takeuchi:2023nxi},
all four regions are addressed in this study.
\item
On the other hand, the discussion on the difference with preceding studies is given in Sec.\,3.4 in \cite{Takeuchi:2023nxi}.
\item 
In addition, in Sec.\,4 of \cite{Takeuchi:2023nxi},
the Minkowski ground state is given 
by a quantum entangled state of the left and right Rindler-wedges 
excited by the creation operators of the U(1) gauge field 
defined on the Rindler coordinates. 
Then, obtaining the density matrix from that by integrating out its left sector, 
it is shown that the U(1) gauge field in the constant accelerated system will feel the Unruh temperature.
\end{itemize}

\section{The Rindler coordinates used in this study}
\label{vtiss}

In this section, we review the Rindler coordinates used in this study.

\subsection{$ds^2$ in the each region in the Rindler coordinates}
\label{vtiss5}

\begin{figure}[H]  
\vspace{0mm} 
\begin{center}
\includegraphics[clip,width=4.5cm,angle=0]{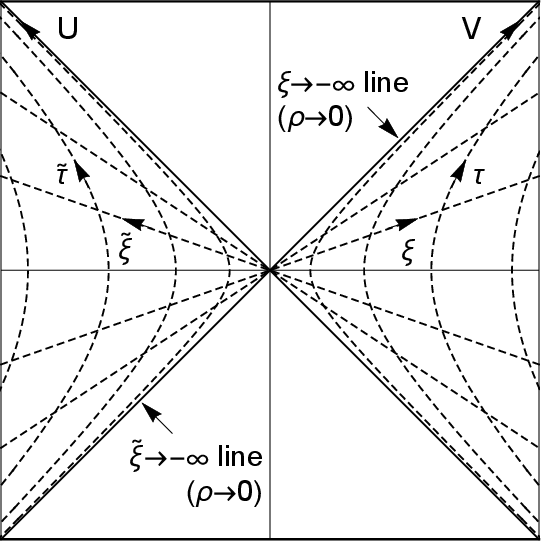} 
\hspace{1.0mm}
\includegraphics[clip,width=4.5cm,angle=0]{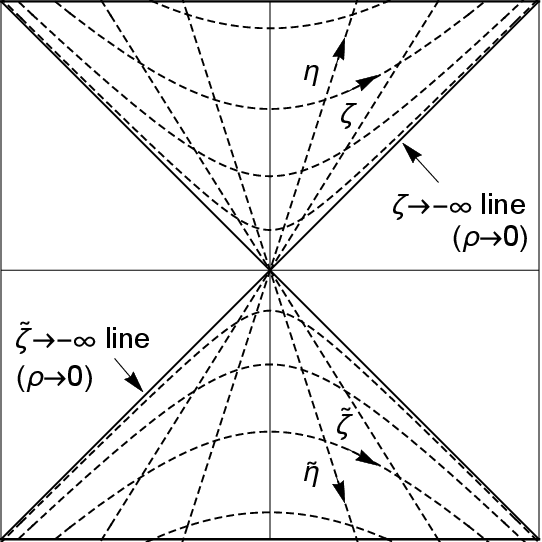}
\end{center}
\vspace{-4.0mm}
\caption{
These figures represent the Rindler coordinates on the Minkowski coordinates. 
The coordinates in each region are defined in (\ref{astew}) and (\ref{dres});
on the left, those of LRW and RRW are represented, and in the right, those of the FRW and PRW  are represented.
$(U,V)$ are the light-cone coordinates defined in (\ref{rebds}), and ``the $\xi \to -\infty$ line'' (and other three) are defined 
in Sec.\,\ref{iytjeb}, which are generally referred to as the Killing horizon.} 
\label{wsdd57}
\end{figure}

We begin with the $4$-dimensional Minkowski spacetime:
\begin{eqnarray}\label{vfdbur}
ds^2=dt^2-\sum_{i=1}^3 (dx^i)^2.
\end{eqnarray}
For the Minkowski coordinates in (\ref{vfdbur}), 
we consider the following four coordinate transformations:
\begin{subequations}\label{vrbwo0}
\begin{align}
\label{vrbwo1}
\cdot \quad
x^0 =& \,\,  a^{-1}e^{a\tilde{\xi}} \sinh a\tilde{\tau}, \quad \hspace{4.0mm}
x^1 =       -a^{-1}e^{a\tilde{\xi}} \cosh a\tilde{\tau},  
\\*[1.0mm]
\label{vrbwo2}
\cdot \quad
x^0 =& \,\,  a^{-1}\,e^{ a\xi} \sinh a\tau, \quad \hspace{3.5mm}
x^1 =       +a^{-1}\,e^{ a\xi} \cosh a\tau,
\\*[3.0mm]
\label{vrbwo3}
\cdot \quad
x^0 =& \, +a^{-1}e^{a\eta} \cosh a\zeta, \quad \hspace{-1.0mm}
x^1 =      a^{-1}e^{a\eta} \sinh a\zeta,  
\\*[1.0mm]
\label{vrbwo4}
\cdot \quad
x^0 =& \, -a^{-1}e^{a\tilde{\eta}} \cosh a\tilde{\zeta}, \quad \hspace{-1.0mm}
x^1 =      a^{-1}e^{a\tilde{\eta}} \sinh a\tilde{\zeta},
\end{align}
\end{subequations}
where
\begin{itemize}
\item 
the regions which are transformed to by (\ref{vrbwo1}) and (\ref{vrbwo2}) are the left and right Rindler-wedges (LRW and RRW), respectively, 
and the regions transformed to by (\ref{vrbwo3}) and (\ref{vrbwo4}) are the future and past Rindler-wedges (FRW and PRW), respectively. 

On the left and right in Fig.\ref{wsdd57}, the LRW/RRW and FRW/PRW are shown, respectively.

\item
We have used $(\tau,\xi)$, $(\tilde{\tau},\tilde{\xi})$, $(\zeta,\eta)$ and $(\tilde{\zeta},\tilde{\eta})$ 
as the coordinates in the LRW, RRW, FRW and PRW respectively, as shown in Fig.\ref{wsdd57}. 

\item
In the Rindler coordinates, $a$ is considered to be fixed.

\item
In the LRW and RRW, there are world-lines of the constant accelerated motion; 
$\xi$ and $\tilde{\xi}$ label those, and $\tau$ and $\tilde{\tau}$ parametrize them.

On the other hand, in the FRW and PRW, it is assumed that there is no object performing constant accelerated motion.
However, in the FRW and PRW, the lines labeled by $\eta$ and $\tilde{\eta}$ and parametrized by $\zeta$ and $\tilde{\zeta}$ 
exist as lines analogous to the world-lines in the LRW and RRW, which are plotted in Fig.\ref{wsdd57}.
\end{itemize}

Applying (\ref{vrbwo0}) to (\ref{vfdbur}), $ds^2$ in each region can be obtained as follows:
\begin{eqnarray}\label{astew}
ds^2 =
\left\{ 
\begin{array}{ll}
\! e^{2a\tilde{\xi}} (d\tilde{\tau}^2 -d\tilde{\xi}^2) - (dx^\perp)^2     & \!\! \textrm{in the LRW}, \\[1.0mm] 
\! e^{2a\xi} (d\tau^2 -d\xi^2) - (dx^\perp)^2                             & \!\! \textrm{in the RRW}, \\[1.0mm]    
\! e^{2a\eta} (d\eta^2 -d\zeta^2) - (dx^\perp)^2                          & \!\! \textrm{in the FRW}, \\[1.0mm]  
\! e^{2a\tilde{\eta}} (d\tilde{\eta}^2 -d\tilde{\zeta}^2) - (dx^\perp)^2  & \!\! \textrm{in the PRW}, 
\end{array} 
\right.
\end{eqnarray} 
where $x^\perp \equiv (x^2,x^3)$. 

Now, let us introduce $\rho$ defined as follows: 
\begin{eqnarray}\label{dvyjd}
\rho \equiv 
\left\{
\begin{array}{lcll}
\! a^{-1}e^{a\tilde{\xi}} & \textrm{and} & a^{-1}e^{a\xi}           & \!\textrm{in the LRW and RRW,} \\[1.0mm]  
\! a^{-1}e^{a\eta}        & \textrm{and} & a^{-1}e^{a\tilde{\eta}}  & \!\textrm{in the FRW and PRW.}  
\end{array}
\right.
\end{eqnarray}
Considering this $\rho$, $ds^2$ in (\ref{astew}) can be given as follows:
\begin{eqnarray}\label{dres}
ds^2 = 
\left\{
\begin{array}{ll}
\! a^2\rho^2 d\tilde{\tau}^2-d\rho^2-(dx^\perp)^2            & \!\textrm{in the LRW,} \\*[1.0mm] 
\! a^2\rho^2 d\tau^2-d\rho^2-(dx^\perp)^2                    & \!\textrm{in the RRW,} \\*[1.0mm] 
\!           d\rho^2 -a^2\rho^2 d\zeta^2-(dx^\perp)^2        & \!\textrm{in the FRW,} \\*[1.0mm] 
\!           d\rho^2 -a^2\rho^2 d\tilde{\zeta}^2-(dx^\perp)^2 & \!\textrm{in the PRW.} 
\end{array}
\right.
\end{eqnarray}
From the above expressions, it can be seen that the $\tau$- and $\tilde{\tau}$-directions are the Killing vectors in the RRW and LRW, and the $\zeta$- and $\tilde{\zeta}$-directions are the Killing vectors in the FRW and PRW. 
Therefore, in the RRW and LRW, the Hamiltonians are defined for the $\tilde{\tau}$- and $\tau$-constant surfaces and in the FRW and PRW, the Hamiltonians are defined for the $\zeta$- and $\tilde{\zeta}$-constant surfaces. 
Therefore, in the Rindler coordinates in the RRW and LRW, $\tilde{\tau}$ and $\tau$ coordinate play the roles of the time, and in the Rindler coordinates in the FRW and PRW, 
the $\tilde{\zeta}$ and $\zeta$ coordinates play the roles of time.

As mentioned above, since $\zeta$ and $\tau$ play the role of time in the FRW and RRW, 
respectively, we perform the Euclideanization as $\tau \to -i\tau$ and $\zeta \to -i\zeta$ in (\ref{dres}). 
As a result, the $ds^2$ of the RRW  and FRW in (\ref{dres}) are written as follows:
\begin{eqnarray}\label{hertrt}
ds_{\rm E}^2 = 
\left\{
\begin{array}{ll}
\! a^2\rho^2 d\tau^2+d\rho^2+(dx^\perp)^2   & \!\textrm{in the RRW,} \\*[1.0mm] 
\! d\rho^2 +a^2\rho^2 d\zeta^2+(dx^\perp)^2 & \!\textrm{in the FRW,} \\ 
\end{array}
\right.
\end{eqnarray}
where we have put as $-ds^2 \to ds_{\rm E}^2$ for the Euclideanization, $\tau \to -i\tau$. 
It can be seen in (\ref{hertrt}) that the $\tau$- and $\zeta$-directions are periodic by $\beta = 2\pi/a$, 
the inverse of which agrees with the Unruh temperature in the constant accelerated system with $a$. 
In the LRW and PRW, the Euclideanization is performed in the same way as for the RRW and FRW, and the same Unruh temperature can be obtained.

\subsection{The solutions in the LRW (PRW) from the solutions in the FRW (PRW)}
\label{vnbreo}

If the solutions in the FRW and RRW are known, solutions in the LRW and PRW, which are parity-symmetric, can be immediately obtained. 
In this subsection, this is discussed.

If a vector solution in the FRW is known, denoting it as $f^\mu (\zeta, \eta)$,
it can be seen from (\ref{vrbwo0}) and Fig.\ref{wsdd57} that 
the vector solution in the PRW being in a party-symmetric relation with $f^\mu (\zeta, \eta)$ for the $x^0=0$ line can be obtained 
simply by exchanging $(\zeta, \eta)$ to $(\tilde{\zeta}, \tilde{\eta})$ leaving the forms of the functions of the solutions in the following way:
\begin{flushleft}
\textrm{\underline{\bf The changing of the solution from the FRW to the PRW:}}
\end{flushleft}
\vspace*{-1.5mm}
\begin{eqnarray}\label{vfaerb}
\bullet\quad\!\!
&(f^0,f^1,f^\perp) \to (-f^0,f^1,f^\perp) \quad \!\!\textrm{with $(\zeta, \eta) \to (\tilde{\zeta}, \tilde{\eta})$}. 
\end{eqnarray}

In the same way, denoting as $g^\mu (\tau, \xi)$ for a given vector solution in the RRW, the party-symmetric vector solutions in the LRW for $g^\mu (\zeta, \eta)$ for the $x^1=0$ line 
can be obtained simply by exchanging $(\tau, \xi)$ to $(\tilde{\tau}, \tilde{\xi})$ as:
\begin{flushleft}
\textrm{\underline{\bf The changing of the solution from the RRW to the LRW:}}
\end{flushleft}
\vspace*{-1.5mm}
\begin{eqnarray}\label{jsrjmu}
\bullet\quad\!\!
&(g^0,g^1,g^\perp) \to (g^0,-g^1,g^\perp) \quad \!\!\textrm{with $(\tau, \xi) \to (\tilde{\tau}, \tilde{\xi})$}. 
\end{eqnarray}

\subsection{Light-cone coordinates}
\label{vtiss4}

We define the light-cone coordinates as:
\begin{eqnarray}\label{rebds}
U \equiv x^0-x^1, \quad V \equiv x^0+x^1,
\end{eqnarray}
where $x^0$ and $x^1$ are the Minkowski coordinates in (\ref{vfdbur}).
We show these in Fig.{\ref{wsdd57}}, from which we can see that
\begin{eqnarray}\label{rvdsd}
U
\left\{
\begin{array}{ll}
\! > 0 & \!\textrm{in the LRW and FRW,} \\*[1.0mm] 
\! < 0 & \!\textrm{in the RRW and PRW,} 
\end{array}
\right. 
\quad
V
\left\{
\begin{array}{ll}
\! < 0 & \!\textrm{in the LRW and FRW,} \\*[1.0mm] 
\! > 0 & \!\textrm{in the RRW and PRW.} 
\end{array}
\right. 
\end{eqnarray} 
Based on (\ref{vrbwo1}) and (\ref{vrbwo2}), 
we can represent $U$ and $V$ in terms of the LRW and RRW coordinates ($\tilde{\tau}$, $\tilde{\xi}$) and ($\tau$, $\xi$) as follows:
\begin{eqnarray}\label{rerwod}
(U,V) =
\left\{
\begin{array}{lcl}
\! a^{-1}( e^{a\tilde{u}},\, -e^{-a\tilde{v}}) & \! 
\textrm{in the LRW with} \! &
(\tilde{u},\tilde{v}) \equiv (\tilde{\tau}+\tilde{\xi}, \tilde{\tau}-\tilde{\xi}), \\*[1.0mm] 
\! a^{-1} (-e^{-au},\, e^{av}) & \!
\textrm{in the RRW with} \! & 
(u,v) \equiv (\tau-\xi, \tau+\xi), \\*[1.0mm] 
\! a^{-1} (e^{-as},\, e^{at}) & \!
\textrm{in the FRW with} \! & 
(s,t) \equiv (-\zeta+\eta, \zeta+\eta),  \\*[1.0mm] 
\! -a^{-1} (e^{-a\tilde{s}},\, e^{a\tilde{t}}) & \!
\textrm{in the PRW with} \! & 
(\tilde{s},\tilde{t}) \equiv (\tilde{\zeta}+\tilde{\eta}, -\tilde{\zeta}+\tilde{\eta}). 
\end{array}
\right.
\end{eqnarray}

\subsection{The lines asymptotically approached to by taking $\xi$, $\eta$, $\tilde{\xi}$ and $\tilde{\eta}$ closer to $-\infty$}
\label{iytjeb}

\subsubsection{The lines at $\xi$ closer to $-\infty$}
\label{iytjeb1}

Let us consider taking $\xi$ in (\ref{vrbwo2}) closer to $-\infty$.
Then, let us take $\vert\tau\vert$ to $\infty$
and refer to the lines asymptotically approached to at that time as ``the $\xi \to -\infty$ line''. 
Note that the exact limit taking $\xi$ to $-\infty$ is excluded by the definition of the Rindler coordinates.

Then, at any points on $x^0>0$ or $x^0<0$ on the $\xi \to -\infty$ line, it can be seen that $\tau$ and $\xi$ in (\ref{vrbwo2}) should be related as 
\begin{eqnarray}\label{bekwfq}
\vert\tau\vert \sim \vert \xi \vert, 
\end{eqnarray}
otherwise, on the $\xi \to -\infty$ line, neither $t$ nor $x^1$ can be finite 
(therefore, the motion of the object on the $\xi \to -\infty$ line in the finite $\tau$ is all packed into the neighborhood of $x^0=0$). 
This can be seen from the forms of $x^0$ and $x^1$ in (\ref{vrbwo2})\footnote{
$x^0$ and $x^1$ in (\ref{vrbwo2}) can be roughly written as
\begin{subequations}\label{wegfy}
\begin{align}
x^0 &\sim e^{a\xi} \sinh a\tau \sim e^{a(\xi+\tau)}-e^{-a(-\xi+\tau)},\\[1.0mm]
x^1 &\sim e^{a\xi} \cosh a\tau \sim e^{a(\xi+\tau)}+e^{-a(-\xi+\tau)}.
\end{align}
\end{subequations}
Then, on the $\xi \to -\infty$ line, if (\ref{bekwfq}) is satisfied, $x^0$ can take finite values behaves as follows:
\begin{eqnarray}
\displaystyle
x^0 \sim 
\left\{
\begin{array}{ll}
e^{a(\xi+\tau)}-e^{-2a\infty}   & \textrm{for $\tau \sim -\xi$,}\\[1.0mm] 
e^{-2a\infty}-e^{-a(-\xi+\tau)} & \textrm{for $\tau \sim +\xi$,} 
\end{array}
\right.
\end{eqnarray}
where ``$\xi \sim \mp\tau$'' mean that $\vert\xi \vert$ and $\vert\tau \vert$ are in the same order.
$x^1$ also can take finite values  on the $\xi \to -\infty$ line, if (\ref{bekwfq}) is satisfied, by the same logic.
}. 

Therefore, on the $\xi \to -\infty$ line, since $\vert\tau\vert$ approaches $\infty$, $x^0/x^1=\tanh(a\tau)$ approaches $\pm 1$.
The lines obtained by the $x^0/x^1=\pm 1$ are the $\pm 45$-degree straight diagonal lines in the $x^1 \ge 0$ region in Fig.\ref{wsdd57}. 

These $\pm 45$-degree straight diagonal lines are generally referred to as the past and future Killing horizons for $x^0<0$ and $x^0>0$, respectively.
In this paper, we refer to the lines obtained by taking $\xi$ closer to $-\infty$, 
which asymptote to those $\pm 45$-degree straight diagonal lines, as ``the $\xi \to -\infty$ line'', 
as noted on the lefthand side of Fig.\ref{wsdd57}. 

\subsubsection{The lines at $\eta$ closer to $-\infty$}
\label{iytjeb2}

By the same logic in Sec.\,\ref{iytjeb1}, when $\eta$ in (\ref{vrbwo3}) is taken closer to $-\infty$, 
it can be seen that $\eta$ and $\zeta$ should be related as follows: 
\begin{eqnarray}\label{bfdhj}
\vert \eta \vert \sim \vert  \zeta \vert,
\end{eqnarray}
and $x^0/x^1=\tanh(a\zeta)$ asymptotes to $\pm 1$.

The lines given by these $x^0/x^1=\pm 1$ are the $\pm 45$-degree straight diagonal lines in the $x^0 \ge 0$ region in Fig.\ref{wsdd57}, 
where $-1$ and $+1$ correspond to the left and right future Killing horizons respectively.

In this paper, we refer to the lines obtained by taking $\eta$ closer to $-\infty$ as ``the $\eta \to -\infty$ line'', 
which asymptote to those $\pm 45$-degree straight diagonal lines, 
as noted on the righthand side of Fig.\ref{wsdd57}.

\subsubsection{The lines at $\tilde{\tau}$ or $\tilde{\eta}$ closer to $-\infty$}
\label{iytjeb3}

By the same logic in Secs.\ref{iytjeb1} and \ref{iytjeb2}, we refer to the lines obtained 
by taking $\tilde{\tau}$ and $\tilde{\eta}$ closer to $-\infty$ as ``the $\tilde{\tau} \to -\infty$ line'' and ``the $\tilde{\eta} \to -\infty$ line'', respectively, 
as noted in both sides of Fig.\ref{wsdd57}.
 
\section{The Lorentz-covariant U(1) gauge field Lagrangians in the Rindler coordinates}
\label{vesbv}

In this section, the Lagrangian densities (referred to as the Lagrangian) of the U(1) gauge field 
in the Lorentz-covariant gauge in the four regions in the Rindler coordinates are obtained. 
Since the analysis to obtain the Lorentz-covariant U(1) gauge field Lagrangian in the LRW and PRW can be performed in the same way as those in the FRW and RRW, 
this section mainly focuses on the analysis concerning the FRW and RRW.
The result is noted in (\ref{etsiph}). 
\newline
 
First, let us begin with the actions of the U(1) gauge field in each of the four regions in the Rindler coordinates:
\begin{eqnarray}\label{uieqp}
S=\int \! d^4x \,\sqrt{-g}\,{\cal L}_{\rm U(1)}, \quad 
{\cal L}_{\rm U(1)}
= -F_{\mu\nu} F^{\mu\nu}/4, 
\end{eqnarray}
where the integral region is the RRW, FRW, LRW or PRW given by the coordinates in (\ref{dres}), 
and $\nabla_\mu$ are covariant derivatives given by the metrices (\ref{dres}), 
and $F_{\mu\nu}=\nabla_\mu A_\nu-\nabla_\nu A_\mu=\partial_\mu A_\nu-\partial_\nu A_\mu$. 

The Lagrangian in the Lorentz-covariant gauge given by $\partial_\mu A^\mu=0$ in the Minkowski coordinates is known to be given as
\begin{eqnarray}\label{ebtwf}
\tilde{{\cal L}}^{(M)}_{\rm U(1)} =
{\cal L}_{\rm U(1)}
+B \, \partial_\mu A^\mu + B^2/2
+i\,\bar{c} \,\partial_\mu \partial^\mu \, c,  
\end{eqnarray}
where $B$ is the auxiliary field (the $B$-field) and $c$, $\bar{c}$ are the ghost fields.
From this, we are able to derive the Lagrangian in the Lorentz-covariant gauge in each of the four regions in the Rindler coordinates
by replacing the differentials with the gravitational covariant derivatives as 
\begin{eqnarray}\label{etsiph}
\tilde{{\cal L}}^{(R)}_{\rm U(1)} =
{\cal L}_{\rm U(1)}
+B \, \nabla_\mu A^\mu + B^2/2
+i\,\bar{c} \,\nabla_\mu \nabla^\mu \, c,
\end{eqnarray}
where the coordinates in the one above refer to those in (\ref{dres}). 

However, since (\ref{ebtwf}) is derived through the Lagrangian gauge-fixed by the Coulomb gauge (a non-covariant gauge),
we will check if (\ref{etsiph}) can be obtained in the Rindler coordinates in the following subsection.  
As noted earlier, in the following subsections, we mainly focus on the analysis concerning the FRW and RRW, 
and the Lagrangians of the U(1) gauge field in the Lorentz-covariant gauge in the PRW and LRW are addressed in Sec.\,\ref{xbmyv}.
\newline

Upon performing the analysis to obtain (\ref{etsiph}) in the following subsections, 
we note the values of the Christoffel's in each region in the Rindler coordinates:
\begin{subequations}\label{rmwg}
\begin{align}
\label{rmwg2}
\cdot \quad \Gamma^{0}_{01} &=\Gamma^{0}_{10}= \rho^{-1},\quad
\Gamma^{1}_{00}  = a^2\rho, \quad \!\textrm{others $=0$ in the RRW and LRW,}
\\[1.0mm]
\label{rmwg1}
\cdot \quad \Gamma^{0}_{11} &= a^2\rho,\quad
\Gamma^{1}_{01} = \Gamma^{1}_{10} = \rho^{-1}, \quad \!\textrm{others $=0$ in the FRW and PRW.}
\end{align}
\end{subequations}
 
In this study, the U(1) gauge field is addressed. 
In such a case, the Faddeev-Popov determinant (FPd) does not include the gauge field (as seen in (\ref{eartd})). 
As a result, the partition functions of the ghost and the gauge fields decouple each other as $Z=Z_{A_\mu,B} \, Z_{c,\bar{c}}$.  
Therefore, we may address only $A_\mu$ and $B$.
With this in mind, we will focus on $A_\mu$ and $B$ in the subsequent discussion.
  
\subsection{The Hamiltonian in the RRW and FRW}
\label{ts6dt5}

Beginning with ${\cal L}_{\rm U(1)}$ in (\ref{uieqp}), the dynamical variables are $A^\mu$, 
and $x^0$ and $x^1$ play the role of the time-coordinate in the FRW and RRW, respectively, as mentioned under (\ref{dres}). 
Therefore, the canonical conjugate momenta in the RRW and FRW are derived as follows:
\begin{subequations}\label{zulst}
\begin{align}
\label{zulst1}
&
\cdot \quad \!\!
\frac{\partial {\cal L}_{\rm U(1)}}{\partial (\partial_0 A^k)}
= -F^0{}_k = g^{00} F_{k0} \equiv \pi_k
\quad\textrm{in the RRW},
\\*[1.0mm]
\label{zulst2}
& 
\cdot \quad \!\!
\frac{\partial {\cal L}_{\rm U(1)}}{\partial (\partial_1 A^k)}
=-F^1{}_k=g^{11} F_{k1} \equiv \pi_k 
\quad\textrm{in the FRW}, 
\end{align}
\end{subequations}
where $k=1,\perp$ in RRW and $k=0,\perp$ in the FRW (in what follows, the indices written in lower case Latin alphabets refer to this). 
When $\pi_k$ are defined as above, $\pi^k$ are given as follows:
\begin{subequations}\label{s69j5}
\begin{align}
\label{s69j51}
&
\cdot \quad \!\!
\pi^k=F^{k0}=\frac{\partial {\cal L}_{\rm U(1)}}{\partial (\partial_0 A_k)}
\quad\textrm{in the RRW},
\\*[1.0mm]
\label{s69j52}
& 
\cdot \quad \!\!
\pi^k= F^{k1} =\frac{\partial {\cal L}_{\rm U(1)}}{\partial (\partial_1 A_k)}
\quad \textrm{in the FRW}.
\end{align}
\end{subequations}
$\pi_k$ and $\pi^k$ have physical meaning as the electric field: 
\begin{subequations}\label{fsgio}
\begin{align}
& \cdot \quad \!\! \pi_k =E_k, \quad \pi^k =E^k \quad\textrm{in the RRW,} 
\\*[1.0mm]
& \cdot \quad \!\! \pi_k =E_k, \quad \pi^k =E^k \quad\textrm{in the FRW.} 
\end{align}
\end{subequations}
As the electric fields have been defined, let us define the magnetic field in each region:
\begin{subequations}\label{khnme}
\begin{align}
\label{khnme1}
\cdot \quad \!\! 
B^k &\equiv -\varepsilon^{ijk}F_{ij}/2, \quad 
B_k =        \varepsilon_{ijk}F^{ij}/2 \quad \textrm{in the RRW,}
\\*[1.0mm]
\label{khnme2}
\cdot \quad \!\!
B^k &\equiv -\varepsilon^{ijk}F_{ij}/2, \quad 
B_k =
\left\{ 
\begin{array}{ll}
\displaystyle -\varepsilon_{ij0}F^{ij}/2 & \textrm{for $k=0$} \\[1.0mm] 
\displaystyle +\varepsilon_{ij \perp}F^{ij}/2 & \textrm{for $k=\perp$}
\end{array}
\right. 
\quad \!\!\! \textrm{in the FRW,}
\end{align}
\end{subequations}
where $k$ in $B_k$ is the same as that in $B^k$ in (\ref{khnme1}) and (\ref{khnme2}), respectively. 
Since $i,j,k=1, \perp$ and $=0,\perp$ in the RRW and FRW respectively, 
it can be seen that $\varepsilon^{ijk}=-\varepsilon_{ijk}$ in the RRW, and $\varepsilon^{ijk}=\varepsilon_{ijk}$ in the FRW.

Regarding (\ref{zulst})-(\ref{khnme}), the Lagrangian (\ref{uieqp}) can be written as
\begin{subequations}\label{bv64r}
\begin{align}
\label{bv64r1}
\cdot \quad \!\! 
{\cal L}_{\rm U(1)}
=& 
-E^k (\nabla_k A_0-\nabla_0 A_k)-(B_k^2-g_{00}E_k^2)/2 
\quad \!\!
\textrm{for the RRW},
\\*[1.0mm]
\label{bv64r2}
\cdot \quad \!\! 
{\cal L}_{\rm U(1)}
=&  
-E^k     (\nabla_k A_1-\nabla_1 A_k)
-(B_k^2-g_{11}E_k^2)/2
\quad\!\!
\textrm{for the FRW},
\end{align}
\end{subequations}
where $B_k^2=B_k B^k$ ($E_k^2$, $E_0^2$ and $E_\perp^2$ are likewise), and $g_{00}$ and $g_{11}$ refer to those in (\ref{dres}).

The first term in the r.h.s. of (\ref{bv64r1}) can be rewritten as
\begin{subequations}\label{ewosdv}
\begin{align}
\label{ewosdv1}
\cdot \quad
\int_{\rm RRW} \! d^4x \sqrt{-g}\,E^k \nabla_k A_0
&= 
-\int_{\rm RRW} \! d^4x \sqrt{-g}\,\partial_k E^k A_0, 
\\*[1.0mm] 
\label{ewosdv2}
\cdot \quad
\int_{\rm FRW} \! d^4x \sqrt{-g}\,E^k \nabla_k A_1
&=
-\int_{\rm FRW} \! d^4x \sqrt{-g}\,\partial_k E^k A_1, 
\end{align}
\end{subequations}
where in the calculation (\ref{ewosdv1}), expressing $E^k$ as $F^{k0}$, we have used $\nabla_k A^0=(\sqrt{-g})^{-1} \partial_k (\sqrt{-g} A^0)$ 
and assumed the boundary condition that the fields vanish at the infinite far region which is given by $\xi$ or $\eta$ to $\infty$.  
By that boundary condition, $\int_{\rm RRW} d^4x \, \partial_k (F^k{}_0 \sqrt{-g} A^0)$ vanishes 
(note that as can be seen from (\ref{astew}), the coordinates of the RRW and FRW reduce to the Minkowski coordinates by $a \to 0$, but do not by $\xi$ or $\eta$  to $\infty$).  
The calculation of (\ref{ewosdv2}) has been performed likewise.
We perform these rewritings of (\ref{ewosdv}) to change the coefficients of $A_0$ and $A_1$ to numbers for the convenience in the calculation (\ref{vvwavh}).

With (\ref{ewosdv1}) and (\ref{ewosdv2}), 
\begin{subequations}\label{b65urfb}
\begin{align} 
\label{b65urfb1}
& \cdot \quad \!\!
\textrm{(\ref{bv64r1})}
= 
E^k \nabla_0 A_k
+\partial_k E^k A_0
-(B_k^2-g_{00}E_k^2)/2,
\\*[1.0mm] 
\label{b65urfb2}
& \cdot \quad \!\!
\textrm{(\ref{bv64r2})}
= 
E^k \nabla_1 A_k
+\partial_k E^k A_1
-(B_k^2-g_{11}E_k^2)/2. 
\end{align}
\end{subequations}
With (\ref{b65urfb}), the Hamiltonian densities (referred to as the Hamiltonian) can be obtained as follows:
\begin{subequations}\label{bvery}
\begin{align}
\label{bvery1}
\cdot \quad \!\!
{\cal H}
= 
\nabla_0 A_\mu\,\pi^\mu-{\cal L}_{\rm U(1)} 
=
\nabla_0 A^0\,\pi_0
-\partial_k E^k A_0
+(B_k^2-g_{00}E_k^2)/2
\quad \!\! \textrm{for the RRW,}
\\*[1.0mm]
\label{bvery2}
\cdot \quad \!\!
{\cal H}
= 
\nabla_1 A_\mu\,\pi^\mu-{\cal L}_{\rm U(1)}
=
\nabla_1 A^1\,\pi_1
-\partial_k E^k A_1
+(B_k^2-g_{11}E_k^2)/2
\quad \!\! \textrm{for the FRW,}
\end{align}
\end{subequations}
where $\pi_0$ in the RRW and $\pi_1$ in the FRW are vanished  by the constraint conditions in Sec.\,\ref{tduk24}.

\subsection{The constitution of path-integral in RRW and FRW}
\label{tduk24}

In the system with the Hamiltonians (\ref{bvery}), there are two constraint conditions in each region, which we denote as $\phi^{(i)}$ $(i=1,2)$:
\begin{subequations}\label{bef67} 
\begin{align}
\label{bef671} 
\cdot \quad \!\!
\phi^{(1)} &\equiv \pi_0 =0,\quad
\phi^{(2)} \equiv \partial_k E^k = 0 
\quad \textrm{for the RRW,}
\\*[1.0mm] 
\label{bef672} 
\cdot \quad \!\!
\phi^{(1)} &\equiv \pi_1 =0,\quad
\phi^{(2)} \equiv \partial_k E^k = 0
\quad \textrm{for the FRW,}
\end{align}
\end{subequations}
where $\pi_0$ and $\pi_1$ above are defined in (\ref{zulst}), which are always vanished as can be seen from (\ref{zulst}); 
the value of $\phi^{(2)}$, as given above, can be seen from the condition that the time-development of $\phi^{(1)}$ generated 
by the Hamiltonians (\ref{bvery}) should be vanished\footnote{ 
The Hamiltonians (\ref{bvery1}) and (\ref{bvery2}) can be written using $\phi^{(2)}$ as
\begin{subequations}\label{tvrtj}
\begin{align}
\label{tvrtj1}
\cdot \quad \!\!
\textrm{(\ref{bvery1})}
=& \,
\nabla_0 A^0\,\pi_0
-\phi^{(2)} A_0
+(B_k^2-g_{00}E_k^2)/2
,\\*[1.0mm]
\label{tvrtj2}
\cdot \quad \!\!
\textrm{(\ref{bvery2})}
=& \,
\nabla_1 A^1\,\pi_1
-\phi^{(2)} A_1
+(B_k^2-g_{00}E_k^2)/2.
\end{align} 
\end{subequations}
The terms $\phi^{(2)} A_0$ and $\phi^{(2)} A_1$ contribute to the time-developments of $\phi^{(1)}$ generated 
by the Hamiltonians (\ref{bvery1}) and (\ref{bvery2}) in the RRW and FRW, respectively.
}.  

For (\ref{bef67}), we take the Coulomb gauge, and denote $\chi^{(i)}$ $(i=1,2)$, as follows:
\begin{subequations}\label{vrue}
\begin{align}
\label{vrue1}
\cdot \quad \!\!
\chi^{(1)} &\equiv A^0=0, \quad 
\chi^{(2)}  \equiv \nabla_k A^k=0 \quad \textrm{for the RRW,}
\\*[1.0mm] 
\label{vrue2}
\cdot \quad \!\!
\chi^{(1)} &\equiv A^1=0, \quad 
\chi^{(2)}  \equiv \nabla_k A^k=0 \quad \textrm{for the FRW,}
\end{align} 
\end{subequations}
where the gauge considered as the Coulomb gauge in the Minkowski coordinates is $\partial_\mu A^\mu$, and $\chi^{(2)}$ in the ones above are the extension of this to the Rindler coordinates;  
therefore, $\nabla_\mu$ in $\chi^{(2)}$ are the gravitational ones, which do not include the gauge field. 

Denoting (\ref{bef67}) and (\ref{vrue}) together as $\tilde{\phi} \equiv \{\phi^{(1)},\phi^{(2)},\chi^{(1)},\chi^{(2)}\} \equiv \{\tilde{\phi}^{(i)} \}$ ($i=1,2,3,4$), 
the Poisson bracket for $\tilde{\phi}$ can be derived for each $x^0$ in the RRW or $x^1$ in the FRW as follows:
\begin{equation}\label{ivw3rk}
\left[
\begin{array}{c}
\{\tilde{\phi}^{(i)}(x),\tilde{\phi}^{(j)}(y)\}_{\rm P.B.} 
\end{array}
\right]
=
\left[
\begin{array}{cccc}
0 & 0 & -1 & 0                    \\
0 & 0 & 0  & -\partial^k \nabla_k \\
1 & 0 & 0  & 0                    \\
0 & \nabla_k \partial^k & 0  & 0 
\end{array}
\right]
\delta^3(\vec{x}-\vec{y}),  
\end{equation}
where the Poisson bracket and $\delta^3(\vec{x}-\vec{y})$ are given as follows:
\begin{eqnarray}
\{X(t,\vec{x}),Y(t,\vec{y})\}_{{\rm P.B.}}
\!\!\! &=& \!\!\!
\left(\frac{\partial X(t,\vec{x})}{\partial A^\mu(t,\vec{x})} \frac{\partial Y(t,\vec{y})}{\partial \pi_\mu(t,\vec{y})}
- \frac{\partial X(t,\vec{x})}{\partial \pi_\mu(t,\vec{y})} \frac{\partial Y(t,\vec{x})}{\partial A^\mu(t,\vec{y})}\right)
\delta^3(\vec{x}-\vec{y}), 
\nonumber \\*[1.0mm]
\delta^3(\vec{x}-\vec{y})
\!\!\! &=& \!\!\!
\left\{ 
\begin{array}{ll}
\displaystyle \delta(x^1-y^1)\,\delta^2(x^\perp-y^\perp) & \textrm{for the RRW,} \\*[1.0mm] 
\displaystyle \delta(x^0-y^0)\,\delta^2(x^\perp-y^\perp) & \textrm{for the FRW.}
\end{array}
\right.
\nonumber
\end{eqnarray}
$X$ and $Y$ represent fields, and $t$ means $x^0$ and $x^1$ for the RRW and the FRW, respectively. 
Since ${\rm Det} \big[\{\tilde{\phi}(x),\tilde{\phi}(y)\}_{\rm P.B.}\big]$ is non-vanishing, 
$\tilde{\phi}$ forms a second-class constraint. 

When the constraints in (\ref{bef67}) and (\ref{vrue}) are imposed in the phase space with $(A^\mu,\pi_\mu)$ as its coordinates, 
the probability amplitude from the infinite-past ground state to the infinite-future ground state in the RRW and FRW can be written 
by the following path-integrals:
\begin{subequations}\label{iedub}
\begin{align}
\label{iedub1}
\cdot \quad \!\!
T_{\rm RRW}
&\equiv 
\int \! {\cal D}\!A \, {\cal D}\pi 
\prod_{x \in {\textrm{RRW}}}
\big[
\delta[\phi^{(1)}]\,\delta[\phi^{(2)}]\,\delta[\chi^{(1)}]\,\delta[\chi^{(2)}]
\big]\cdot
\prod_{x^0} 
M_c
\exp \big[ 
i \int_{\textrm{RRW}} \! d^4x \,\sqrt{-g}\, {\cal L}_{\rm path}
\big], 
\\*[1.0mm]
\label{iedub2}
\cdot \quad \!\!
T_{\rm FRW}
&\equiv 
\int \! {\cal D}\!A \, {\cal D}\pi 
\prod_{x \in {\textrm{FRW}}}
\big[
\delta[\phi^{(1)}]\,\delta[\phi^{(2)}]\,\delta[\chi^{(1)}]\,\delta[\chi^{(2)}]
\big]\cdot
\prod_{x^1} 
M_c
\exp \big[ 
i \int_{\textrm{FRW}} \! d^4x \,\sqrt{-g}\, {\cal L}_{\rm path}
\big], 
\end{align}
\end{subequations}
where 
\begin{equation}
{\cal L}_{\rm path}\equiv
\left\{
\begin{array}{l}
\nabla_0 A^\mu\,\pi_\mu-{\cal H}={\cal L}_{\rm U(1)} \quad \textrm{for the RRW,} \\*[1.0mm]
\nabla_1 A^\mu\,\pi_\mu-{\cal H}={\cal L}_{\rm U(1)} \quad \textrm{for the FRW.}  
\end{array}
\right.
\nonumber
\end{equation}
${\cal H}$ is defined in (\ref{bvery1}) and (\ref{bvery2}),  respectively, and ${\cal L}_{\rm U(1)}$ is given in (\ref{b65urfb1}) and (\ref{b65urfb2}) respectively. 
$M_c$ and ${\cal D}\!A \, {\cal D}\pi $ are defined as
\begin{subequations}
\begin{align}
M_c 
&\equiv 
{\rm Det} \big[ \{\phi^{(i)}(x),\chi^{(j)}(y)\}_{\rm P.B.} \big]
=
{\rm Det} \big[
\left[
\begin{array}{cccc}
-1 & 0                    \\
0  & -\partial^k \nabla_k 
\end{array}
\right]
\delta^3(\vec{x}-\vec{y})
\big]
, \quad
\nonumber
\\*[1.0mm]
{\cal D}\!A \, {\cal D}\pi 
&= 
\displaystyle {\cal C}\,\prod_{\mu=0}^3 \,\prod_{x \in {\rm RRW/FRW}} dA^\mu(x)\,d\pi_\mu(x),
\nonumber
\end{align}
\end{subequations}
where ${\cal C}$ means some constant and $x$ refers to the coordinates in the RRW or FRW. 
${\rm Det}$ is the functional determinant of the functions on the three-dimensional space at constant time in the RRW or FRW.

\subsection{Analysis of path-integrals to obtain the Lagrangian in the Lorentz-covariant gauge in the RRW and FRW}
\label{teeah54}

Let us integrate out $\pi_0$ in the RRW and $\pi_1$ in the FRW in (\ref{iedub}).
This can be performed readily as $\phi^{(1)}=\pi_0$ and $\pi_1$ in the RRW and FRW,  respectively, as in (\ref{bef67}), and as a result, (\ref{iedub}) can be written as
\begin{subequations}\label{gsrtx}
\begin{align}
\label{gsrtx1}
\cdot \quad \!\!
T_{\textrm{RRW}}
=& 
\int \! {\cal D}\!A \, {\cal D}\pi_k  
\prod_{x \, \in \, \textrm{RRW}}
\delta[\phi^{(2)}]\,
\delta[\chi^{(1)}]\,
\delta[\chi^{(2)}]
\cdot 
\prod_{x^0} M_c
\,\exp \big[
i \int_{\textrm{RRW}} \! d^4x \,\sqrt{-g}\, {\cal L}_{\rm U(1)}\vert_{\pi_0=0}
\big],
\\*[1.0mm]
\label{gsrtx2}
\cdot \quad \!\!
T_{\textrm{FRW}}
=& 
\int \! {\cal D}\!A \, {\cal D}\pi_k 
\prod_{x \, \in \, \textrm{FRW}}
\delta[\phi^{(2)}]\,
\delta[\chi^{(1)}]\,
\delta[\chi^{(2)}]
\cdot 
\prod_{x^1} M_c \,
\exp \big[
i \int_{\textrm{FRW}} \! d^4x \,\sqrt{-g}\, {\cal L}_{\rm U(1)}\vert_{\pi_1=0}
\big].
\end{align}
\end{subequations}
Since ${\cal L}_{\rm U(1)}$ in (\ref{gsrtx1}) and (\ref{gsrtx2}) do not include $\pi_0$ and $\pi_1$, respectively, as can be seen in (\ref{b65urfb}), 
we may write ${\cal L}_{\rm U(1)}|_{\pi_0=0}$ and ${\cal L}_{\rm U(1)}|_{\pi_1=0}$ as ${\cal L}_{\rm U(1)}$. 
Therefore, in what follows, we write those as ${\cal L}_{\rm U(1)}$.

Now, introducing a functional variable $\eta=\eta(x)$, we give $\delta[\phi^{(2)}]$ by a functional integral:
\begin{eqnarray}\label{bvfer} 
 \delta[\phi^{(2)}]
=\int {\cal D}\eta \, \exp \big[ i \int_{\textrm{RRW$/$FRW}} \! d^4x \,\sqrt{-g}\,\eta \, \phi^{(2)} \big]. 
\end{eqnarray}
Then, (\ref{gsrtx1}) and (\ref{gsrtx2}) can be written as 
\begin{subequations}\label{gastrl}
\begin{align}
\label{gastrl1}
\cdot \quad \!\!
\textrm{(\ref{gsrtx1})}
=& 
\int \! {\cal D}\!A \, {\cal D}\pi_k
\int {\cal D} \eta \, 
\prod_{x \in \textrm{RRW}}
\delta[\chi^{(1)}]\,
\delta[\chi^{(2)}]
\cdot 
\prod_{x^0} M_c \,
\exp \big[
i \int_{\textrm{RRW}} \! d^4x \,\sqrt{-g}\, 
\big\{{\cal L}_{\rm U(1)}+\eta \,\phi^{(2)}\big\}
\big],
\\*[1.0mm]
\label{gastrl2}
\cdot \quad \!\!
\textrm{(\ref{gsrtx2})}
=& 
\int \! {\cal D}\!A \, {\cal D}\pi_k
\int {\cal D} \eta \, 
\prod_{x \in \textrm{FRW}}
\delta[\chi^{(1)}]\,
\delta[\chi^{(2)}]
\cdot 
\prod_{x^1} M_c \,
\exp \big[
i \int_{\textrm{FRW}} \! d^4x \,\sqrt{-g}\, 
\big\{{\cal L}_{\rm U(1)}+\eta \,\phi^{(2)}\big\}
\big],
\end{align}
\end{subequations}
where 
\begin{subequations}\label{vvwavh}
\begin{align}
\label{vvwavh1}
\cdot \quad \!\!
{\cal L}_{\rm U(1)}+\eta \,\phi^{(2)}
=&\cdots+(A_0+\eta ) \phi^{(2)}+\cdots \quad \textrm{for the RRW,} 
\\*[1.0mm]
\label{vvwavh2}
\cdot \quad \!\!
{\cal L}_{\rm U(1)}+\eta \,\phi^{(2)}
=&\cdots+(A_1+\eta)\phi^{(2)} +\cdots  \quad \textrm{for the FRW.} 
\end{align}
\end{subequations}
In the equations above, we have written only the part which is important in the following analysis. 
Now we can integral out $A^0$ for the RRW in (\ref{gastrl1}).    
Then, since $\chi^{(1)}=A^0$ as in (\ref{vrue1}), the term concerning $A_0$ in (\ref{vvwavh1}) disappears. 
However, we can take $\eta$ as $A^0$, by which the term which has just disappeared is now reintroduced.
In the same way, in the FRW in (\ref{gastrl2}), we integrate out $A^1$ and take $\eta$ as $A^1$.
As a result, (\ref{gastrl1}) and (\ref{gastrl2}) can be given as follows:
\begin{subequations}\label{vtiof}
\begin{align}
\label{vtiof1}
\cdot \quad \!\!
\textrm{(\ref{gastrl1})}
=& 
\int \! {\cal D}\!A \, {\cal D}\pi_k \!
\prod_{x \in \textrm{RRW}}
\delta[\chi^{(2)}] \cdot
\prod_{x^0} M_c 
\cdot\exp \big[
i \int_{\textrm{RRW}} \! d^4x \,\sqrt{-g}\, {\cal L}_{\rm U(1)}
\big],
\\*[1.0mm]
\label{vtiof2}
\cdot \quad \!\!
\textrm{(\ref{gastrl2})}
=& 
\int \! {\cal D}\!A \, {\cal D}\pi_k \!
\prod_{x \in \textrm{FRW}}
\delta[\chi^{(2)}] \cdot
\prod_{x^1} M_c
\cdot\exp \big[
i \int_{\textrm{FRW}} \! d^4x \,\sqrt{-g}\, {\cal L}_{\rm U(1)}
\big]. 
\end{align}
\end{subequations}

Considering (\ref{b65urfb1}) and (\ref{b65urfb2}) as the expression of ${\cal L}_{\rm U(1)}$ in (\ref{vtiof1}) and (\ref{vtiof2}), respectively, 
${\cal L}_{\rm U(1)}$ includes the terms $\partial_k E^k A_0$ and $\partial_k \pi^k A_1$, respectively. 
We can restate those terms back by rewriting (\ref{ewosdv1}) and (\ref{ewosdv2}) in reverse. 
As a result, the component of ${\cal L}_{\rm U(1)}$ in (\ref{vtiof1}) and (\ref{vtiof2}) are provided, respectively, as
\begin{subequations}\label{dusea1}
\begin{align}
\label{dusea11}
\cdot \quad \!\!
\textrm{${\cal L}_{\rm U(1)}$ in (\ref{vtiof1})}
=& \,\,
F_{0k}\,\pi^k 
-(B_k^2-g_{00}\,\pi_k^2)/2 
\nonumber\\*[1.0mm]
=& \,
-\frac{g_{00}}{2}\sum_{k=1,\perp}(
(\pi_k+g^{00}F_{0k})(\pi_k+g^{00}F_{0k})
-(g_{00})^2F_{0k}F_{0k})
-B_k^2/2
\nonumber\\*[1.0mm]
=& \,
-\frac{g_{00}}{2}\sum_{k=1,\perp}(\pi_k+g^{00}F_{0k})(\pi_k+g^{00}F_{0k})+{\cal L}_{\rm U(1)},
\\*[1.0mm]
\label{dusea12}
\cdot \quad \!\!
\textrm{${\cal L}_{\rm U(1)}$ in (\ref{vtiof2})}
=& \,\,
F_{1k}\,\pi^k 
-(B_k^2-g_{11}\,\pi_k^2)/2 
\nonumber\\*[1.0mm]
=& \,\,
\frac{g_{11}}{2}
\Big(
\hspace{4mm}
(\pi_0+g^{11}F_{1k})
(\pi_0+g^{11}F_{1k})
-(g^{11})^2 F_{10} F_{10}
\nonumber\\*[0.0mm]
&
\hspace{8mm}
-(\pi_\perp+g^{11}F_{1k})
(\pi_\perp+g^{11}F_{1k})
+(g^{11})^2 F_{1\perp} F_{1\perp}
\Big)
-B_k^2/2
\nonumber\\*[1.0mm]
=& \,\,
\frac{g_{11}}{2}
(
 (\pi_k+g^{11}F_{10})(\pi_k+g^{11}F_{10})
-(\pi_\perp+g^{11}F_{1\perp})(\pi_\perp+g^{11}F_{1\perp})
)+{\cal L}_{\rm U(1)},
\end{align}
\end{subequations}
where $B_k^2=B_k B^k$ and $\pi_k^2=\pi_k\pi^k$ as mentioned in (\ref{bv64r}).
If $A^\mu$ and $\pi_\mu$ were not independent of each other and were in the relations (\ref{zulst1}) and (\ref{zulst2}) respectively, 
the r.h.s. of (\ref{dusea11}) and (\ref{dusea12}) reduce to $0+{\cal L}_{\rm U(1)}$. 

Then, $\pi_k$ in (\ref{vtiof1}) and (\ref{vtiof2}) can be integrated out (by redefining $\pi_k$). As a result, 
\begin{subequations}\label{tedrha}
\begin{align}
\label{tedrha1}
\textrm{(\ref{vtiof1})}
=& 
\int \! {\cal D}\! A 
\prod_{x \in \textrm{RRW}}
\delta[\chi^{(2)}] \cdot
\prod_{x^0} M_c
\cdot
\exp \big[
i \int_{\textrm{RRW}} d^4x \,\sqrt{-g}\, {\cal L}_{\rm U(1)}
\big],
\\*[1.0mm]
\label{tedrha2}
\textrm{(\ref{vtiof2})}
=& 
\int \! {\cal D}\! A 
\prod_{x \in \textrm{FRW}}
\delta[\chi^{(2)}] \cdot
\prod_{x^1} M_c
\cdot
\exp \big[
i \int_{\textrm{FRW}} d^4x \,\sqrt{-g}\, {\cal L}_{\rm U(1)}
\big].
\end{align}
\end{subequations}

Once we have obtained the form $\int DA \,\Omega_{\textrm{Coulomb}}\, e^{i\int d^4x {\cal L}_{\rm U(1)}}$ 
as seen in (\ref{tedrha})
($\Omega_{\textrm{Coulomb}}$ denotes the parts $\prod [\delta(\chi^{(2)})]\cdot \prod M_c$ in the ones above), 
by proceeding in the same way as the case of the Minkowski coordinates, 
we can replace the part of $\Omega_{\textrm{Coulomb}}$ with the one by the Lorentz-covariant gauge, 
and obtain the Lagrangian in Lorentz-covariant gauge in the RRW and FRW, which is (\ref{etsiph}).
Since that analysis in the Minkowski coordinates is well-known, 
and is performed not explicitly depending on the coordinates 
unlike the analysis in which (\ref{tedrha1}) and (\ref{tedrha2}) are obtained,
we proceed with the analysis from (\ref{tedrha}) to (\ref{etsiph}) in Appendix.\ref{buobhs}. 

\subsection{The Lorentz-covariant U(1) gauge field Lagrangians in the LRW and PRW}
\label{xbmyv}

Since the metrices of the FRW (RRW) and PRW (LRW) are mathematically the same, 
as seen in (\ref{dres}), 
if the path-integrals in the FRW and RRW can be given as in (\ref{tedrha}), 
the path-integrals which correspond to those in the PRW and LRW can be given as
\begin{subequations}\label{sknuek}
\begin{align}
\label{sknuek1} 
T_{\rm LRW}
=& 
\int \! {\cal D}\! A 
\prod_{x \in \textrm{LRW}}
\big[\delta(\chi^{(2)})\big] \cdot
\prod_{x^0} M_c
\cdot
\exp \big[
i \int_{\textrm{LRW}} d^4x \,\sqrt{-g}\, {\cal L}_{\rm U(1)}
\big],
\\*[1.0mm]
\label{sknuek2}
T_{\rm PRW}
=& 
\int \! {\cal D}\! A 
\prod_{x \in \textrm{PRW}}
\big[\delta(\chi^{(2)})\big] \cdot
\prod_{x^1} M_c
\cdot
\exp \big[
i \int_{\textrm{PRW}} d^4x \,\sqrt{-g}\, {\cal L}_{\rm U(1)}
\big],
\end{align}
\end{subequations}
where the coordinates in (\ref{sknuek1}) and (\ref{sknuek2}) refer to the LRW and PRW given by (\ref{dres}); 
therefore, $x^0$ and $x^1$ mean $\tilde{\tau}$ and $\tilde{\zeta}$ in (\ref{dres}), respectively. 
Then, performing the analysis following (\ref{tedrha1}) and (\ref{tedrha2}) 
in the same way for (\ref{sknuek1}) and (\ref{sknuek2}),
the Lagrangians of the U(1) gauge field in the Lorentz-covariant gauge in the LRW and PRW can be obtained as in (\ref{etsiph}).

\section{The solutions of the U(1) gauge field and the $B$-field in each region in the Rindler coordinates}
\label{f2v4t}

In the previous section, the gauge-fixed Lagrangian was obtained as given in (\ref{ebtwf}).
In this section, by solving the equations of motion obtained from that Lagrangian, 
the mode-solutions of the U(1) gauge field and the $B$-field in the RRW and FRW are obtained. 

The results are noted in (\ref{ba54nwy}) and (\ref{b543es}), 
and the forms of fields as the solution are noted in (\ref{w4aeea}) and (\ref{weh28a}). 
The method for solving the equations of motion is noted under (\ref{weh28a}). 
The solution obtained in this study is not the general solution, as mentioned at the end of Sec.\,\ref{sy5sst}.

The forms of the U(1) gauge field and the $B$-field as the solutions in the LRW and PRW and their mode-solutions are given in Sec.\,\ref{nnklf}, 
and the normalization constants of the mode-solutions in the four regions are given in Sec.\,\ref{brpnevt}.

\subsection{The solution of the U(1) gauge field and the $B$-field in the RRW and FRW}
\label{sy5sst}
 
From the Lagrangian (\ref{etsiph}), the equations of motion can be obtained as follows\footnote{
$\int d^4x \sqrt{-g}\,\tilde{{\cal L}}^{(R)} = \int d^4x \sqrt{-g}\,(\cdots -A^\mu\, \partial_\mu B + \cdots)$ 
and 
$\partial_\mu (\sqrt{-g}\,F^{\mu \nu})=\sqrt{-g} \,\nabla_\mu F^{\mu \nu}$.
}:
\begin{subequations}\label{e3ier}
\begin{align}
\label{e3ier1}
-\nabla_\mu F^{\mu \nu}+\partial^\nu B &=0, \\*[1.0mm]
\label{e3ier2}
\nabla_\mu A^\mu + B &=0.
\end{align}
\end{subequations} 
From these, the equations of motion that the fields should satisfy in the RRW are derived as follows: 
\begin{subequations}\label{ebraue}
\flushleft{\textrm{\underline{\bf The equations of motion in the RRW:}}}
\begin{align}
\label{ebraue1}
(g^{\mu\nu}\partial_\mu \partial_\nu-3\rho^{-1}\partial_1)A^0 
&= - 2a^{-2}\rho^{-3}\partial_0 A^1, 
\\*[1.0mm]
\label{ebraue2}
(g^{\mu\nu}\partial_\mu \partial_\nu-3\rho^{-1}\partial_1-\rho^{-2})A^1 
&= 2\rho^{-1}(B+\partial_\perp A^\perp), \\*[1.0mm]
\label{ebraue3}
(g^{\mu\nu}\partial_\mu \partial_\nu-\rho^{-1}\partial_1)A^\perp &=0, 
\\*[1.0mm]
\label{ebraue4}
\partial_0 A^0 +\partial_1 A^1+\partial_\perp A^\perp+\rho^{-1}A^1 +B &=0,
\\*[1.0mm]
\label{ebraue5}
(g^{\mu\nu}\partial_\mu \partial_\nu-\rho^{-1}\partial_1)B &=0,
\end{align}
\end{subequations}
where, combining (\ref{e3ier1}) and (\ref{e3ier2}), $\nabla_\nu\nabla^\nu A^\mu=0$ can be obtained, from which (\ref{ebraue1})-(\ref{ebraue3}) can be obtained 
(in obtaining (\ref{ebraue2}), (\ref{ebraue4}) is used). 
From (\ref{e3ier2}), (\ref{ebraue4}) can be obtained.
Multiplying the entire (\ref{e3ier1}) by $\nabla_\nu$, $\nabla_\mu \nabla^\mu B=0$  can be obtained, from which (\ref{ebraue5}) can be obtained.
Non-zero $\Gamma^\mu_{\nu \lambda}$ are noted in (\ref{rmwg}).
The $0$- and $1$-directions correspond to $\tau$- and $\rho$-directions, and $\perp$ correspond to  $2,3$-directions. 

In the same way, the equations of motion for the fields in the FRW can be derived from (\ref{e3ier}) as follows:
\begin{subequations}\label{eeylh}
\flushleft{\textrm{\underline{\bf The equations of the motion in FRW:}}}
\begin{align}
\label{eeylh1}
(g^{\mu\nu}\partial_\mu \partial_\nu+3\rho^{-1}\partial_0)A^1 
&= 2a^{-2}\rho^{-3}\partial_1 A^0,
\\*[1.0mm]
\label{eeylh2}
(g^{\mu\nu}\partial_\mu \partial_\nu+3\rho^{-1}\partial_0+\rho^{-2})A^0 
&= -2\rho^{-1}(B+\partial_\perp A^\perp), \\*[1.0mm]
\label{eeylh3}
(g^{\mu\nu}\partial_\mu \partial_\nu+\rho^{-1}\partial_0)A^\perp &=0, 
\\*[1.0mm]
\label{eeylh4}
\partial_0 A^0 +\partial_1 A^1+\partial_\perp A^\perp+\rho^{-1}A^0 +B &=0, 
\\*[1.0mm]
\label{eeylh5}
(g^{\mu\nu}\partial_\mu \partial_\nu+\rho^{-1}\partial_0)B &=0, 
\end{align}
\end{subequations}
where the $0$- and $1$-directions refer to the $\rho$- and $\zeta$-directions, respectively.

Since the coordinate systems in the RRW and FRW are homogeneous for the ($1$, $\perp$)- and ($0$, $\perp$)-directions, respectively, 
as seen in (\ref{dres}), the forms of fields in the RRW and FRW can be supposed by the following Fourier-expansions:
\begin{subequations}\label{wyea}
\flushleft{\textrm{\underline{\bf The Fourier-expanded fields in the RRW:}}}
\begin{align}
\label{wyea1}
A^\mu(\tau,\rho,x^\perp) 
&=
\int_{-\infty}^\infty \! dk_0 \int_{-\infty}^\infty \! d^{2}k_\perp\,
\tilde{{\cal N}}^{(\mu)}_{k}\,e^{-ikx}\,\tilde{A}^{(\mu)}_{k}(\rho), 
\\*[1.0mm]
\label{wyea2}
B(\tau,\rho,x^\perp) 
&=
\int_{-\infty}^\infty \! dk_0 \int_{-\infty}^\infty \! d^{2}k_\perp\,
\tilde{{\cal N}}^{(B)}_{k}\,e^{-ikx}\,\tilde{B}_{k}(\rho), 
\end{align}
\end{subequations}
\begin{subequations}\label{rxwwae}
\flushleft{\textrm{\underline{\bf The Fourier-expanded fields in the FRW:}}}
\begin{align}
\label{rxwwae1}
A^\mu(\rho,\zeta,x^\perp) 
&=
\int_{-\infty}^\infty \! dk_1 \int_{-\infty}^\infty \! d^{2}k_\perp\,
\tilde{{\cal N}}^{(\mu)}_{k}\,e^{-ikx}\,\tilde{A}^{(\mu)}_{k}(\rho), 
\\*[1.0mm]
\label{rxwwae2}
B(\rho,\zeta,x^\perp) 
&=
\int_{-\infty}^\infty \! dk_1 \int_{-\infty}^\infty \! d^{2}k_\perp\,
\tilde{{\cal N}}^{(B)}_{k}\,\tilde{B}_{k}(\rho), 
\end{align}
\end{subequations}
where
\begin{itemize}
\item[$\cdot$]
$k$ in the subscripts and $kx$ in the shoulder of $e$ in the r.h.s. of (\ref{wyea}) are abbreviations of ``$k_0,k_\perp$'' and ``$k_0\tau-k_\perp x^\perp$'', respectively.

On the other hand, 
$k$ in the subscripts and $kx$ in the shoulder of $e$ in the r.h.s. of (\ref{rxwwae}) are abbreviations of ``$k_1,k_\perp$'' and ``$-k_1\zeta-k_\perp x^\perp$'', respectively. 

\item[$\cdot$]
$A^\mu(\tau,\rho,x^\perp)$ and $B(\tau,\rho,x^\perp)$ in this study are assumed to be real.
 
\item[$\cdot$]
$\tilde{{\cal N}}^{(\mu)}_{k}$ and $\tilde{{\cal N}}^{(B)}_{k}$ are constants of each mode, which can take complex numbers. 
These will be denoted by decomposing them as
\begin{eqnarray}\label{rwfyg53}
\tilde{{\cal N}}^{(\mu)}_{k}=\, {\cal N}^{(\mu)}_{k}\,\bm{a}^{\mu}_{k}, \quad
\tilde{{\cal N}}^{(B)}_{k}  =\, {\cal N}^{(B)}_{k}\,\bm{b}_{k},
\end{eqnarray}
where 
${\cal N}^{(\mu)}_{k}$ and ${\cal N}^{(B)}_{k}$ are the normalization constants to be determined by the mode-solutions, 
$\tilde{A}^{(\mu)}_{k}(\rho)$ and $\tilde{B}_{k}(\rho)$, (as noted in (\ref{e3rrjei})); 
and $\bm{a}^{\mu}_{k}$ and $\bm{b}_{k}$ are numbers which become annihilation operators 
when the canonical quantization is performed (as seen in Sec.\,\ref{bywbd}).
\end{itemize}

Applying (\ref{wyea}) to (\ref{ebraue}), the equations of motion for each mode in the RRW can be derived as follows:
\begin{subequations}\label{r4rg2}
\flushleft{\textrm{\underline{\bf The equations of motion for each mode in the RRW:}}}
\begin{align}
\label{r4rg21}
(
(a\rho)^{-2}k_0^2
+\partial_1^2
-k_\perp^2
+3\rho^{-1}\partial_1)     \,\tilde{\cal N}^{(0)}_{k}\tilde{A}^0_{k}
&=- 2a^{-2}\rho^{-3}\,ik_0 \,\tilde{\cal N}^{(1)}_{k}\tilde{A}^1_{k}, 
\\*[1.0mm]
\label{r4rg22}
(
\rho^{-2}(a^{-2}k_0^2+1)
+\partial_1^2
-k_\perp^2
+3\rho^{-1}\partial_1
) \,\tilde{\cal N}^{(1)}_{k}\tilde{A}^1_{k}
&= -2\rho^{-1}(\tilde{{\cal N}}^{(B)}_{k} \tilde{B}_{k}+ik_\perp \,
\tilde{\cal N}^{(\perp)}_{k} \tilde{A}^\perp_{k}),  
\\*[1.0mm]
\label{r4rg23}
((a\rho)^{-2}k_0^2+\partial_1^2-k_\perp^2+\rho^{-1}\,\partial_1)
\,\tilde{\cal N}^{(\perp)}_{k}\tilde{A}^\perp_{k}
&=0, 
\\*[1.0mm]
\label{r4rg25}
-ik_0 \,\tilde{\cal N}^{(0)}_{k}\tilde{A}^0_{k} 
+ (\partial_1+\rho^{-1})\,\tilde{\cal N}^{(1)}_{k}\tilde{A}^1_{k}
&=
-(\tilde{\cal N}^{(B)}_{k_0}\tilde{B}_{k}
+ik_\perp \,\tilde{\cal N}^{(\perp)}_{k_0}\tilde{A}^\perp_{k}),  
\\*[1.0mm]
\label{r4rg24}
((a\rho)^{-2}k_0^2+\partial_1^2-k_\perp^2+\rho^{-1}\,\partial_1)
\,\tilde{\cal N}^{(B)}_{k}\tilde{B}_{k}
&=0,
\end{align}
\end{subequations}
where $\partial_1^2$ means $ \partial_1 \partial_1$. 
Since different directions of $\tilde{A}^{(\mu)}_{k} $ and $B$ are mixed in each equation above, 
these equations of motion are given including the normalization constants as shown above. 

Similarly, applying (\ref{rxwwae}) to (\ref{eeylh}), the equations of motion for each mode in the FRW are obtained as follows:
\begin{subequations}\label{arwrlb}
\flushleft{\textrm{\underline{\bf The equations of motion for each mode in the FRW:}}}
\begin{align}
\label{arwrlb1}
(
(a\rho)^{-2}k_1^2
+\partial_0^2
+k_\perp^2
+3\rho^{-1}\partial_0)        \,\tilde{{\cal N}}^{(1)}_{k}\tilde{A}^1_{k}
&= 2a^{-2}\rho^{-3}\,ik_1 \,\tilde{{\cal N}}^{(2)}_{k}\tilde{A}^0_{k}, 
\\*[1.0mm]
\label{arwrlb2}
(
\partial_0^2
+\rho^{-2}(a^{-2}k_1^2+1)
+k_\perp^2
+3\rho^{-1}\partial_0) \,\tilde{{\cal N}}^{(0)}_{k}\tilde{A}^0_{k}
&= 
-2\rho^{-1}
( 
\tilde{{\cal N}}^{(B)}_{k} \tilde{B}_{k}
+ik_\perp \, \tilde{{\cal N}}^{(\perp)}_{k} \tilde{A}^\perp_{k}
),  
\\*[1.0mm]
\label{arwrlb3}
(
(a\rho)^{-2}k_1^2
+\partial_0^2
+k_\perp^2
+\rho^{-1}\,\partial_0)
\,\tilde{{\cal N}}^{(\perp)}_{k}\tilde{A}^\perp_{k}
&=0, 
\\*[1.0mm]
\label{arwrlb5}
ik_1 \,\tilde{{\cal N}}^{(1)}_{k}\tilde{A}^1_{k} 
+ (\partial_0+\rho^{-1})\tilde{{\cal N}}^{(0)}_{k}\tilde{A}^0_{k}
&=
-(\tilde{{\cal N}}^{(B)}_{k}\tilde{B}_{k}
+ik_\perp \,\tilde{{\cal N}}^{(\perp)}_{k}\tilde{A}^\perp_{k}), 
\\*[1.0mm]
\label{arwrlb4}
(
(a\rho)^{-2}k_1^2
+\partial_0^2
+k_\perp^2
+\rho^{-1}\,\partial_0)
\,\tilde{{\cal N}}^{(B)}_{k}\tilde{B}_{k}
&=0,
\end{align}
\end{subequations}
where $\partial_0^2$ means $\partial_0 \partial_0$. 
Saving the explanation of how we have solved (\ref{r4rg2}) and (\ref{arwrlb}) for later, 
we first show the results of the mode-solutions in the RRW and FRW obtained by solving them in the following:
\begin{subequations}\label{ba54nwy}
\flushleft{\textrm{\underline{\bf The mode-solutions in the RRW:}}}
\begin{align}
\label{ba54nwy1}
\bullet\quad \!\!
\tilde{A}^{(\perp)}_{k} 
&= 
K_{i\alpha}(b\rho),
\\*[1.0mm]
\label{ba54nwy3}
\bullet\quad \! 
\tilde{A}^{(1)}_{k} 
&= 
\rho^{-1}K_{i\alpha}(b \rho),
\\*[1.0mm]
\label{ba54nwy4}
\bullet\quad \!
\tilde{A}^{(0)} _{k}
&= 
-\frac{i}{k_0\,\rho}\,\partial_1K_{i\alpha}(b \rho)
=\frac{ib}{2k_0\,\rho}\,(K_{-1+i\alpha}(b \rho)+K_{1+i\alpha}(b \rho)),
\\*[1.0mm]
\label{ba54nwy5}
\bullet\quad \,\,
\tilde{B}_{k} 
&=  
-ik_\perp \,\tilde{A}^{(\perp)}_{k},
\end{align}
\end{subequations}
\begin{subequations}\label{b543es} 
\flushleft{\textrm{\underline{\bf The mode-solutions in the FRW:}}}
\begin{align}
\label{b543es1}
\bullet\quad \!\!
\tilde{A}^{(\perp)}_{k} 
&= 
J_{i\alpha}(b\rho), 
\\*[1.0mm]
\label{b543es3}
\bullet\quad \!
\tilde{A}^{(0)}_{k} 
&= 
\rho^{-1}J_{i\alpha}(b \rho),
\\*[1.0mm]
\label{b543es4}
\bullet\quad \!
\tilde{A}^{(1)}_{k}
&= 
\frac{i}{k_0\,\rho}\,\partial_0 J_{i\alpha}(b \rho)
=\frac{ib}{2k_0\,\rho}\,(J_{-1+i\alpha}(b \rho)-J_{1+i\alpha}(b \rho)),
\\*[1.0mm]
\label{b543es5}
\bullet\quad \,\,
\tilde{B}_{k} 
&=  
-ik_\perp \,\tilde{A}^{(\perp)}_{k},
\end{align}
\end{subequations}
where
\begin{itemize}

\item[$\cdot$]
The meaning of $k$ in the subscripts of $\tilde{A}^{(\mu)}_{k}$ and $\tilde{B}_{k} $ is given under (\ref{wyea}) and (\ref{rxwwae}).

\item[$\cdot$]
Both $\partial_1$ and $\partial_0$ in (\ref{ba54nwy4}) and (\ref{b543es4}) can be denoted as ${\partial}/{\partial \rho}$.
$\alpha$ in (\ref{ba54nwy}) and (\ref{b543es}) are respectively defined as $\alpha \equiv k_0/a$ and $\alpha \equiv k_1/a$, 
and $b$ is defined as $b \equiv \sqrt{k_2^2+k_3^2}$ both in (\ref{ba54nwy}) and (\ref{b543es}).
$K_{i\alpha}(b\rho)$ is the modified Bessel function of the second kind, and $J_{i\alpha}(b\rho)$ is the Bessel function of the first kind. 

\item[$\cdot$]
$\tilde{{\cal N}}^{(\perp)}_k = \tilde{{\cal N}}^{(B)}_k$ is supposed in the process of obtaining $\tilde{B}_{k_0,k_\perp}$ (specifically, seen (\ref{svdo})).
From this, using (\ref{rwfyg53}), the following condition is introduced:
\begin{eqnarray}\label{vras5g1}
{\cal N}^{(\perp)}_k \, \bm{a}^{\perp}_k = {\cal N}^{(B)}_{k}\, \bm{b}_k.
\end{eqnarray}

Here, since $\tilde{B}_{k}$ in the RRW and FRW are obtained as $ik_\perp\,\tilde{A}^\perp_k$ as seen in (\ref{ba54nwy5}) and (\ref{b543es5}),
$\tilde{B}_{k}$ and $\tilde{A}^\perp_k$ are different only by the constant multiplication in the RRW and FRW.
Therefore, denoting as $\chi_{1,k} \equiv {\cal N}^{(B)}_k\tilde{B}_k$ and $\chi_{2,k} \equiv {\cal N}^{(\perp)}_k\tilde{A}^\perp_k$, 
it can be deduced from (\ref{aerts1}) and (\ref{aerts2}):
\begin{eqnarray}\label{vhcsdo}
(\chi_{1,k},\chi_{1,k'})_{\rm KG}=(\chi_{2,k},\chi_{2,k'})_{\rm KG}.
\end{eqnarray} 
From this, it can be seen that $\chi_{1,k}=\chi_{2,k}$ is identically maintained in the RRW and FRW.
From this, it is turned out that ${\cal N}^{(B)}_k$ and ${\cal N}^{(\perp)}_k$ in the RRW and FRW are related as follows: 
\begin{eqnarray}\label{vae47th}
{\cal N}^{(B)}_k=(ik_\perp)^{-1}\,{\cal N}^{(\perp)}_k. 
\end{eqnarray}

With this ${\cal N}^{(B)}_k$, it is concluded from (\ref{vras5g1}) that there is the following relation 
between the coefficients $\bm{a}^{(\perp)}_k$ and $\bm{b}_k$: $ik_\perp\,\bm{a}^{(\perp)}_k =\bm{b}_k$, at the classical level, 
which is a conclusion led from the supposition of $\tilde{{\cal N}}^{(\perp)}_k = \tilde{{\cal N}}^{(B)}_k$. 

\item[$\cdot$]
Next, in obtaining $\tilde{A}^{(1)}_{k_0,k_\perp} $ in the FRW and $\tilde{A}^{(0)}_{k_0,k_\perp} $ in the RRW, 
$\tilde{{\cal N}}^{(1)}_k = \tilde{{\cal N}}^{(0)}_k$ 
is supposed (see (\ref{hwsjv})). 
Using (\ref{rwfyg53}), this leads to
\begin{eqnarray}\label{vras5g2}
{\cal N}^{(1)}_k \, \bm{a}^{1}_k = 
{\cal N}^{(0)}_k \, \bm{a}^{0}_k
\end{eqnarray}
in the RRW and FRW. 

However, in the RRW and FRW, since $\tilde{A}^{(0)}_{k_0,k_\perp}$ and $\tilde{A}^{(1)}_{k_0,k_\perp}$ are obtained as seen in (\ref{ba54nwy4}) and (\ref{b543es4}), 
$\tilde{A}^{(0)}_{k_0,k_\perp}$ and $\tilde{A}^{(1)}_{k_0,k_\perp}$ are not in the linear relation  in the RRW and FRW. 
Therefore, an identity such as $\chi_{1,k}=\chi_{2,k}$ under (\ref{vhcsdo}) is unclear in the RRW and FRW.

However, in the RRW and FRW, it is turned out that ${\cal N}^{(1)}_k$ and ${\cal N}^{(0)}_k$ are equivalent to each other as seen in (\ref{ly3es2}) and (\ref{4rwerh2}). 
Therefore, it can be found from (\ref{vras5g2}) that $\bm{a}^{1}_k= \bm{a}^{0}_k$ at the classical level in the RRW and FRW, 
which is a conclusion which leads from the supposition of $\tilde{{\cal N}}^{(1)}_k = \tilde{{\cal N}}^{(0)}_k$. 

\item[$\cdot$]
From what is mentioned above, 
it can be seen that ${{\cal N}}^{(\perp)}_{k}$ and ${{\cal N}}^{(B)}_{k}$ (${{\cal N}}^{(1)}_{k}$ and ${{\cal N}}^{(0)}_{k}$) are not independent each other,
therefore these can be denoted using a notation. 
Also, $\bm{a}^{\perp}_{k}$ and $\bm{b}_{k}$ ($\bm{a}^{1}_{k}$ and $\bm{a}^{0}_{k}$) can be denoted using  a notation.

However, after the canonical quantization is performed, $\bm{a}^{\perp}_{k}$ and $\bm{b}_{k}$ ($\bm{a}^{1}_{k}$ and $\bm{a}^{0}_{k}$) 
become annihilation operators vanishing the excitations in different directions. 
In this sense, $\bm{a}^{(\perp)}_{k}$ and $\bm{b}_{k}$ ($\bm{a}^{1}_{k}$ and $\bm{a}^{0}_{k}$) are physically distinct from each other. 
Therefore, we distinctively denote those as $\bm{a}^{(\perp)}_{k}$ and $\bm{b}_{k}$ ($\bm{a}^{1}_{k}$ and $\bm{a}^{0}_{k}$).

Corresponding to this, we distinctively  denote ${\cal N}^{(\perp)}_{k}$ and ${\cal N}^{(B)}_{k}$ (${\cal N}^{(1)}_{k}$ and ${\cal N}^{(0)}_{k}$) 
as ${\cal N}^{(B)}_{k}$ and ${\cal N}^{(\perp)}_{k}$ (${\cal N}^{(1)}_{k}$ and ${\cal N}^{(0)}_{k}$), as well.
\end{itemize} 

Now that the mode-solutions have been obtained as given in (\ref{ba54nwy}) and (\ref{b543es}), let us write as:
\begin{subequations}\label{ds3vws}
\begin{align}
\label{ds3vws1}
\cdot \quad \!\!
A^\mu(\tau,\rho,x^\perp) 
&= 
\int_{-\infty}^\infty \! dk_0 \int_{-\infty}^\infty \! d^{2}k_\perp\,
{\cal N}^{(\mu)}_{k}\,\bm{a}^{\mu}_{k}\,\tilde{A}^{(\mu)}_{k}(\rho) \, e^{-ikx}
\nonumber\\*[1.0mm]
&= 
\Big(
\int_{-\infty}^0 \! dk_0 \,
+\int_0^\infty \! dk_0 \,
\Big)
\int_{-\infty}^\infty \! d^{2}k_\perp \,
{\cal N}^{(\mu)}_{k}\,\bm{a}^{\mu}_{k}\,\tilde{A}^{(\mu)}_{k}(\rho) \, e^{-ikx}
\quad\textrm{for RRW,}
\\*[1.0mm]
\label{ds3vws2}
\cdot \quad \!\!
A^\mu(\rho,\zeta,x^\perp) 
&=
\int_{-\infty}^\infty \! dk_1 \int_{-\infty}^\infty \! d^{2}k_\perp\,
{\cal N}^{(\mu)}_{k}\,\bm{a}^{\mu}_{k}\,\tilde{A}^{(\mu)}_{k}(\rho) \, e^{-ikx}
\nonumber\\*[1.0mm]
&=
\Big(
\int_{-\infty}^0 \! dk_1 
+\int_0^\infty \! dk_1 
\Big)
\int_{-\infty}^\infty \! d^{2}k_\perp \,
{\cal N}^{(\mu)}_{k}\,\bm{a}^{\mu}_{k}\,\tilde{A}^{(\mu)}_{k}(\rho) \, e^{-ikx}
\quad\textrm{for FRW.}
\end{align}
\end{subequations}
Next, let us consider flipping ($k_0$, $k_\perp$) to ($-k_0$, $-k_\perp$) in (\ref{ds3vws1}), and ($k_1$, $k_\perp$) to ($-k_1$, $-k_\perp$) in (\ref{ds3vws2}). 

\begin{itemize}
\item[$\cdot$]
There is symmetry for the $k_\perp$-direction in our system 
(actually, $\tilde{A}^{(\mu)}_{k}(\rho)$ and ${\cal N}^{(\mu)}_{k}$ are obtained independently of $k_\perp$ as in (\ref{ba54nwy}), (\ref{b543es}), (\ref{4rwerh}) and (\ref{ly3es})). 
In addition, $\bm{a}^{(\mu)}_{k_0,-k_\perp}=\bm{a}^{(\mu)}_{k_0,k_\perp}$ and $\bm{a}^{(\mu)}_{k_1,-k_\perp}=\bm{a}^{(\mu)}_{k_1,k_\perp}$ 
can be assumed in (\ref{ds3vws1}) and (\ref{ds3vws2}) respectively.

\item[$\cdot$]
Next, for the flip of $k_0$ to $-k_0$, ${\rm Re}(K_{-i\alpha}(b \rho))={\rm Re}(K_{i\alpha}(b \rho))$ and ${\rm Im}(K_{i\alpha}(b \rho))=0$, 
and $J_{-i\alpha}(b \rho)=J_{i\alpha}^*(b \rho)$, where $\tilde{A}^{(\mu)}_{k}(\rho)$ in (\ref{ds3vws1}) and (\ref{ds3vws2}) are $K_{i\alpha}(b \rho)$ and $J_{i\alpha}(b \rho)$ respectively.

In addition, 
$\bm{a}^{(\mu)}_{-k_0,k_\perp}=\bm{a}^{(\mu)*}_{k_0,k_\perp}$ and $\bm{a}^{(\mu)}_{-k_1,k_\perp}=\bm{a}^{(\mu)*}_{k_1,k_\perp}$ can be assumed in (\ref{ds3vws1}) and (\ref{ds3vws2}) respectively.
\end{itemize}
From this, the first terms in the r.h.s. of (\ref{ds3vws1}) and (\ref{ds3vws2}) can be written as
\begin{subequations}\label{dwe4d}
\begin{align}
\cdot \quad \!\!
\textrm{First term in r.h.s. of (\ref{ds3vws1})}
=&
\int_{0}^\infty \! dk_0 \int_{-\infty}^\infty \! d^{2}k_\perp\,
{\cal N}^{(\mu)}_{-k_0,k_\perp}\,\bm{a}^{\mu*}_{k}\,\tilde{A}^{(\mu)}_{k}(\rho) \, e^{ikx}, 
\\*[1.0mm]
\cdot \quad \!\!
\textrm{First term in r.h.s. of (\ref{ds3vws2})}
=&
\int_{0}^\infty \! dk_1 \int_{-\infty}^\infty \! d^{2}k_\perp\,
{\cal N}^{(\mu)}_{-k_1,k_\perp}\,\bm{a}^{\mu*}_{k}\,\tilde{A}^{\mu*}_{k}(\rho) \, e^{ikx}. 
\end{align}
\end{subequations}
As the issue of the notation, we may denote ${\cal N}^{(\mu)}_{-k_0,k_\perp}$ and ${\cal N}^{(\mu)}_{-k_1,k_\perp}$ in the equations above
as ${\cal N}^{(\mu)}_{k_0,k_\perp}$ and ${\cal N}^{(\mu)}_{k_1,k_\perp}$ respectively. 
(concrete expressions of these are determined in Sec.\,\ref{brpnevt}.)

In conclusion, the fields as the solutions in the FRW and RRW can be expressed as follows:
\begin{subequations}\label{w4aeea}
\flushleft{\textrm{\underline{\bf The solutions in the RRW:}}}
\begin{align}
\label{w4aeea1}
A^\mu(\tau,\rho,x^\perp) 
&=
\int_{0}^\infty \! dk_0 \int_{-\infty}^\infty \! d^{2}k_\perp\,
{\cal N}^{(\mu)}_{k}(
\bm{a}^{\mu}_{k}\, e^{-ikx}
+\bm{a}^{\mu*}_{k} \, e^{ikx})
\,\tilde{A}^{(\mu)}_{k}(\rho), 
\\*[1.0mm]
\label{w4aeea2}
B(\tau,\rho,x^\perp) &=
\int_{0}^\infty \! dk_0 \int_{-\infty}^\infty \! d^{2}k_\perp\,
{\cal N}^{(B)}_{k}(
\bm{b}_{k}\,e^{-ikx}
+\bm{b}_{k}^*\, \,e^{ikx}
)\,\tilde{B}_{k}(\rho), 
\end{align}
\end{subequations}
\begin{subequations}\label{weh28a}
\flushleft{\textrm{\underline{\bf The solutions in the FRW:}}}
\begin{align}
\label{weh28a1}
A^\mu(\rho,\zeta,x^\perp) 
&=
\int_{0}^\infty \! dk_1 \int_{-\infty}^\infty \! d^{2}k_\perp\,
{\cal N}^{(\mu)}_{k}(
\bm{a}^{\mu}_{k}\,\tilde{A}^{(\mu)}_{k}(\rho)\, e^{-ikx}
+\bm{a}^{\mu*}_{k}\,\tilde{A}^{\mu*}_{k}(\rho) \, e^{ikx}), 
\\*[1.0mm]
\label{weh28a2}
B(\rho,\zeta,x^\perp) &=
\int_{0}^\infty \! dk_1 \int_{-\infty}^\infty \! d^{2}k_\perp\,
{\cal N}^{(B)}_{k}(
\bm{b}_{k}\,\tilde{B}_{k}(\rho)\,e^{-ikx}
+\bm{b}_{k}^*\,\tilde{B}^*_{k}(\rho)\, \,e^{ikx}
), 
\end{align}
\end{subequations}
where the mode parts $\tilde{A}^{(\mu)}_{k}(\rho)$ and $\tilde{B}_{k}(\rho)$ are given by (\ref{ba54nwy}) and (\ref{b543es}). 
The forms of (\ref{w4aeea}) and (\ref{weh28a}) guarantee that $A^\mu$ and $B$ in the RRW and FRW are real.
~\newline

We will explain how we have solved (\ref{r4rg2}) and obtained (\ref{ba54nwy}). 
\begin{itemize}
\item[$\cdot$]
First, we obtain the solutions of $\tilde{A}^\perp_{k}$ in the FRW and RRW.
\begin{itemize}
\item[$-$]
(\ref{r4rg23}) can be solved using the Mathematica. 
As a result, the solution of $\tilde{A}^\perp_{k}$ in the RRW is obtained as 
$J_{i\alpha}(-ib\rho)+Y_{i\alpha}(-ib\rho)$. 
However, as $K_{i\alpha}(b\rho)$ can satisfy (\ref{r4rg23}), 
we considered $K_{i\alpha}(b\rho)$ instead of that, and (\ref{ba54nwy1}) was obtained.
\vspace{1.5mm}

\item[$-$]
Also, (\ref{arwrlb3}) can be solved using the Mathematica. 
As a result, the solution of $\tilde{A}^\perp_{k}$ in the FRW is obtained as $J_{i\alpha}(b\rho)+Y_{i\alpha}(b\rho)$. 
From this result, we finally considered  $J_{i\alpha}(b\rho)$ as the solution of $\tilde{A}^\perp_{k}$ and (\ref{b543es1}) was obtained.
\end{itemize}

\item[$\cdot$]
The solutions of $\tilde{B}_{k}$ satisfy (\ref{r4rg24}) and (\ref{arwrlb4}), 
which are the same equations as (\ref{r4rg23}) and (\ref{arwrlb3}). 
Therefore, the solutions of $\tilde{B}_{k}$ in the RRW and FRW are proportional to $\tilde{A}^\perp_{k}$ in (\ref{ba54nwy1}) and (\ref{b543es1}).
We fixed the overall coefficients of $\tilde{B}_{k}$ in such a way that 
\begin{subequations}\label{svdo}
\begin{align}
\label{svdo1}
\cdot \quad \!\!
\textrm{r.h.s. of (\ref{r4rg22})}&=
\tilde{{\cal N}}^{(B)}_{k}\tilde{B}_{k}
+ik_\perp \,\tilde{{\cal N}}^{(\perp)}_{k}\tilde{A}^\perp_{k}=0, 
\\*[1.0mm]
\label{svdo2}
\cdot \quad \!\!
\textrm{r.h.s. of (\ref{arwrlb2})}&=
\tilde{{\cal N}}^{(B)}_{k} \tilde{B}_{k}
+ik_\perp \, \tilde{{\cal N}}^{(\perp)}_{k} \tilde{A}^\perp_{k}=0.
\end{align}
\end{subequations}
From (\ref{svdo1}) and (\ref{svdo2}), using the solutions of $\tilde{A}^\perp_{k}$ in (\ref{ba54nwy1}) and (\ref{b543es1}) 
and supposing $\tilde{{\cal N}}^{(B)}_{k}=\tilde{{\cal N}}^{(\perp)}_{k}$ (we discuss this in (\ref{vras5g1})), 
$\tilde{B}_{k}$ in the RRW and FRW can be obtained as in (\ref{ba54nwy5}) and (\ref{b543es5}), respectively. 
These (\ref{ba54nwy5}) and (\ref{b543es5}) can satisfy (\ref{r4rg24}) and (\ref{arwrlb5}).
  
\item[$\cdot$]
Since we have supposed the r.h.s. of (\ref{r4rg22}) and (\ref{arwrlb2}) as 0 as seen in (\ref{svdo}), 
we can set the l.h.s. of (\ref{r4rg22}) and (\ref{arwrlb2}) as 0. 
Solving those, $\tilde{A}^1_{k}$ in the RRW can be obtained as in (\ref{ba54nwy3}), 
and $\tilde{A}^0_{k}$ in the FRW can be obtained as in  (\ref{b543es3}).

\item[$\cdot$]
The r.h.s. of (\ref{r4rg25}) and (\ref{arwrlb5}) are essentially the same as the r.h.s. of (\ref{r4rg22}) and (\ref{arwrlb2}), 
which are now being taken to $0$ as seen in (\ref{svdo}).
Therefore, we can set the r.h.s. of (\ref{r4rg25}) and (\ref{arwrlb5}) as  $0$.
From these, the following equations are obtained:
\begin{subequations}\label{hwsjv}
\begin{align}
\label{hwsjv1}
\cdot \quad \!\!
\textrm{r.h.s. of (\ref{r4rg25})} 
&=
-ik_0 \,\tilde{{\cal N}}^{(0)}_{k}\tilde{A}^0_{k} 
+ (\partial_1+\rho^{-1})\,\tilde{{\cal N}}^{(1)}_{k}\tilde{A}^1_{k}
=
-ik_0 \,\tilde{{\cal N}}^{(0)}_{k}\tilde{A}^0_{k} 
+\rho^{-1}\,\partial_1 K_{i\alpha}(b \rho)
=0,
\\*[1.0mm]
\label{hwsjv2}
\cdot \quad \!\!
\textrm{r.h.s. of (\ref{arwrlb5})} 
&=
-ik_1 \,\tilde{{\cal N}}^{(1)}_{k}\tilde{A}^1_{k} 
+ (\partial_0+\rho^{-1})\,\tilde{{\cal N}}^{(0)}_{k}\tilde{A}^0_{k}
=-ik_1 \,\tilde{{\cal N}}^{(1)}_{k}\tilde{A}^1_{k} 
+ \rho^{-1}\,\partial_0 J_{i\alpha}(b \rho)=0,
\end{align}
\end{subequations}
where 
the mode-solutions of $\tilde{A}^1_{k}$ given in (\ref{ba54nwy3}) and $\tilde{A}^0_{k}$ given in (\ref{b543es3}) have been used, 
and $\tilde{{\cal N}}^{(0)}_{k}=\tilde{{\cal N}}^{(1)}_{k}$ have been supposed (we discuss this in (\ref{vras5g2})).
From (\ref{hwsjv1}) and (\ref{hwsjv2}), 
$\tilde{A}^0_{k}$ in the RRW can be obtained as in (\ref{ba54nwy4}),
and $\tilde{A}^1_{k}$  in the FRW can be obtained as in (\ref{b543es4}).

\item[$\cdot$]
Although (\ref{r4rg21}) and (\ref{arwrlb1}) are not used in the process above, it can be verified that 
(\ref{r4rg21}) is satisfied by the solutions $\tilde{A}^0_{k}$ and $\tilde{A}^1_{k}$ in  (\ref{ba54nwy4}) and (\ref{ba54nwy3}), 
and (\ref{arwrlb1}) is satisfied by the solutions $\tilde{A}^0_{k}$ and $\tilde{A}^1_{k}$ in (\ref{b543es3}) and (\ref{b543es4}).

The reason for the appearance of the non-used equation is that 
the two equations (\ref{r4rg25}) and (\ref{r4rg24}), and (\ref{arwrlb5}) and (\ref{arwrlb4}), 
are not independent of each other, which can be seen in the description under (\ref{ebraue}). 
\end{itemize}

In the process above, the ansatz (\ref{svdo}) has been set.
However, from the perspective of the general solution, 
it is okay if only the both sides of (\ref{r4rg22}) or (\ref{arwrlb2}) are equivalent to each other, 
and the ansatz (\ref{svdo}) is just one situation where the equations of motion (\ref{r4rg2}) and (\ref{arwrlb}) are held. 
In this sense, the solutions (\ref{ba54nwy}) and (\ref{b543es}) represent solutions, but not general solutions.

\subsection{The solution of the U(1) gauge field and the $B$-field in the LRW and PRW}
\label{nnklf}

In the previous subsection, the mode-solutions in the RRW and FRW have been obtained in (\ref{ba54nwy}) and (\ref{b543es}), 
and the fields as the solutions have been given in (\ref{w4aeea}) and (\ref{weh28a}).
In the Rindler coordinates, if a solution in one region is known, 
the parity-symmetric solution for that in another side can be known according to (\ref{vfaerb}) and (\ref{jsrjmu}).
Therefore, in this subsection, the mode-solutions and the fields as the solutions in the LRW and PRW are given 
from what have been been obtained in the previous subsection with (\ref{vfaerb}) and (\ref{jsrjmu}). 
The normalization constants of the mode-solutions in the LRW and PRW are obtained in Sec.\,\ref{brpnevt}.
\newline

According to (\ref{vfaerb}), the forms of the fields in the PRW, which are parity-symmetric solutions for the fields in FRW of (\ref{weh28a}) for the $x^0=0$ line,  
can be given from the fields in FRW of (\ref{weh28a}) as follows: 
\begin{subequations}\label{hoaag}
\flushleft{\textrm{\underline{\bf The solutions in the PRW:}}}
\begin{align}
\label{hoaag1}
A^\mu(\tilde{\zeta},\rho,x^\perp) 
&=
\int_{0}^\infty \! dk_1 \int_{-\infty}^\infty \! d^{2}k_\perp\,
{\cal N}^{(\mu)}_{k}(
\bm{a}^{\mu}_{k}\,\tilde{A}^{(\mu)}_{k}(\rho)\, e^{-ikx}
+\bm{a}^{\mu*}_{k}\,\tilde{A}^{\mu*}_{k}(\rho) \, e^{ikx}), 
\\*[1.0mm]
\label{hoaag2}
B(\tilde{\zeta},\rho,x^\perp) &=
\int_{0}^\infty \! dk_1 \int_{-\infty}^\infty \! d^{2}k_\perp\,
{\cal N}^{(B)}_{k}(
\bm{b}_{k}\,\tilde{B}_{k}(\rho)\,e^{-ikx}
+\bm{b}_{k}^*\,\tilde{B}^*_{k}(\rho)\, \,e^{ikx}
), 
\end{align}
\end{subequations}
where the coordinates in (\ref{hoaag}) refer to those of the PRW in (\ref{dres}), 
and $k$ and $kx$ in the subscripts and shoulder of $e$ are abbreviations of ``$k_1,k_\perp$'' and ``$-k_1\tilde{\zeta}-k_\perp x^\perp$'' respectively. 
${\cal N}^{(\mu)}_{k}$ and ${\cal N}^{(B)}_{k}$ are the normalization constants (which are obtained in Sec.\,\ref{brpnevt}). 
\begin{eqnarray}\label{b5edv}
\label{b5edv1}
\bullet\quad \!\!
\tilde{A}^\perp_{k} = J_{i\alpha}(b\rho), 
\quad
\tilde{A}^0_{k} = -\rho^{-1}J_{i\alpha}(b \rho),
\quad
\tilde{A}^1_{k} = i(k_0\,\rho)^{-1}\,\partial_0 J_{i\alpha}(b \rho),
\quad
\tilde{B}_{k} = -ik_\perp \,\tilde{A}^\perp_{k},
\end{eqnarray}
where the definitions of $\partial_0$, $\alpha$, $b$ and $K_{i\alpha}$ are the same as those in (\ref{b543es}). 
$\rho$ in (\ref{b5edv}) are given with $\tilde{\eta}$ as noted in (\ref{dvyjd}). 
\newline

In the same way, the fields and the mode-solutions in the LRW can be given from those in (\ref{w4aeea}) and (\ref{ba54nwy}) as follows: 
\begin{subequations}\label{w8atd}
\flushleft{\textrm{\underline{\bf The solutions in LRW:}}}
\begin{align}
\label{w8atd1}
A^\mu(\tilde{\tau},\rho,x^\perp) 
&=
\int_{0}^\infty \! dk_0 \int_{-\infty}^\infty \! d^{2}k_\perp\,
{\cal N}^{(\mu)}_{k}(
\bm{a}^{\mu}_{k}\, e^{-ikx}
+\bm{a}^{\mu*}_{k} \, e^{ikx})
\,\tilde{A}^{(\mu)}_{k}(\rho), 
\\*[1.0mm]
\label{w8atd2}
B(\tilde{\tau},\rho,x^\perp) &=
\int_{0}^\infty \! dk_0 \int_{-\infty}^\infty \! d^{2}k_\perp\,
{\cal N}^{(B)}_{k}(
\bm{b}_{k}\,e^{-ikx}
+\bm{b}_{k}^*\, \,e^{ikx}
)\,\tilde{B}_{k}(\rho), 
\end{align}
\end{subequations}
where
\begin{eqnarray}\label{bbmry}
\label{bbmry1}
\bullet\quad \!\!
\tilde{A}^\perp_{k} 
= 
K_{i\alpha}(b\rho), 
\quad
\tilde{A}^1_{k} 
= 
-\rho^{-1}K_{i\alpha}(b \rho),
\quad
\tilde{A}^0 _{k}
= 
-i(k_0\,\rho)^{-1}\,\partial_1K_{i\alpha}(b \rho),
\quad
\tilde{B}_{k} 
=  
-ik_\perp \,\tilde{A}^\perp_{k}.
\end{eqnarray}
The definitions of $\alpha$, $b$ and $K_{i\alpha}$ are the same as those in (\ref{ba54nwy}).
The coordinates in (\ref{w8atd}) refer to those of the LRW in (\ref{dres}), 
and $k$ and $kx$ in the subscripts and shoulder of $e$ are abbreviations of ``$k_0,k_\perp$'' and ``$-k_0\tilde{\tau}-k_\perp x^\perp$'' respectively. 
${\cal N}^{(\mu)}_{k}$ and ${\cal N}^{(B)}_{k}$ are the normalization constants, which are obtained in Sec.\,\ref{brpnevt}. 

\subsection{The normalization constants of the mode-solutions in each region}
\label{brpnevt}

In this subsection, the normalization constants in all directions of the mode-solutions of the U(1) gauge field and the $B$-field in all four regions 
in the Rindler coordinates are obtained.
First, the normalization constants of the mode-solutions in the FRW and RRW are obtained 
from the explicit calculation based on the Klein-Gordon (KG) inner-product. 
These results are given in (\ref{ly3es}) and (\ref{4rwerh}).

Then, from the fact that the coordinates and the mode-solutions in the PRW and LRW can be considered to be the same as those in the FRW and RRW,
the normalization constants of the mode-solutions in the FRW and RRW are similarly obtained as those in the FRW and RRW. 
Therefore, we immediately obtain the normalization constants in the PRW and LRW from the results of those in the FRW and RRW. 
These results are noted in (\ref{s3rtr}).
\newline

We begin by defining the KG inner-product.
Then, in general, an integral with regard to a vector on a three-dimensional (3D) hypersurface in four-dimensional (4D) spacetime can be written as 
\begin{align}\label{d35wufa}
\int_{\Sigma} V^\mu \, d\Sigma_\mu,
\end{align}
where $V^\mu$ represents some vector, and $d\Sigma_\mu$ are the components of the area element on the 3D hypersurface. 

Let us suppose the form of $ds^2$ as $ds^2 =g_{ii}(dx^i)^2+g_{jk}dx^jdx^k$
($i$ means one direction and the summation is not taken for $i$; 
on the other hand $j,k$ mean all the directions except for the $i$-direction and the summations except for $i$ are taken for $j,k$).
Then, if we take the 3D hypersurface as a $x^i$-constant, the $i$-direction is perpendicular to the 3D hypersurface, and $V^\mu \, d\Sigma_\mu$ in (\ref{d35wufa}) is given as
\begin{eqnarray}\label{erhnt2}
V^\mu \,d\Sigma_\mu = g^{ii}\,V_i \, \sqrt{g_{ii} \,\det{(g_{jk})}} \, d^3x, 
\end{eqnarray}
where $j,k$ take the 4D directions except for the $i$-direction.

Considering the Rindler coordinates defined by (\ref{dres}), 
let us take the $\zeta$- or $\tau$-constant hypersurface in the FRW or RRW 
(our Rindler coordinates are defined in (\ref{dres})) as the 3D hypersurface discussed in above.
At this time, (\ref{d35wufa}) are given as
\begin{eqnarray}\label{dewvja}
\cdot \quad \!\! 
\int_\infty^0 \, d\rho \int_{-\infty}^\infty \! d^2x^\perp \, (a\rho)^{-1} \, V_1 
\quad \! \textrm{for the FRW,}
\quad
\int_\infty^0 \, d\rho \int_{-\infty}^\infty \! d^2x^\perp \, (a\rho)^{-1} \, V_0
\quad \! \textrm{for the RRW.}
\end{eqnarray}

Next, let us define the conserved current as follows:
\begin{align}\label{dberw}
J_\mu^{(f_A,g_B)}(x) \equiv i f_A^\ast(x) \overleftrightarrow{\nabla}\!_\mu \, g_B(x),
\end{align}
where $f_A^\ast \overleftrightarrow{\nabla}\!_\mu \, g_B \equiv f_A^\ast \,\nabla_\mu g_B - g_B \,\nabla_\mu f_A^\ast$, 
and $f_A$ and $g_B$ are some solutions of equations of motion. 

From (\ref{dewvja}) and (\ref{dberw}), the KG inner-products in the FRW and RRW are defined as follows:
\begin{subequations}\label{stwgi}
\begin{align}
\label{stwgi1}
\cdot \quad \!\! 
(f_A,g_B)_{\rm KG} &\equiv 
\int_\infty^0 \, d\rho \int_{-\infty}^\infty \! d^2x^\perp \, (a\rho)^{-1} J^{(f_A,g_B)}_0
\quad\textrm{for the FRW,}
\\*[1.0mm]
\label{stwgi2}
\cdot \quad \!\! 
(f_A,g_B)_{\rm KG} &\equiv 
\int_\infty^0 \, d\rho \int_{-\infty}^\infty \! d^2x^\perp \, (a\rho)^{-1} J^{(f_A,g_B)}_1
\quad\textrm{for the RRW.}
\end{align}
\end{subequations}
With (\ref{stwgi}), we, in what follows, determine the normalization constants ${\cal N}^{(\mu)}_{k}$ by the following demand:
\begin{subequations}\label{e3rrjei}
\begin{align}
\label{e3rrjei1}
\cdot \quad \!\! 
({\cal N}_{k}^{(\mu)}  \tilde{A}^{(\mu)}_{k},\, {\cal N}_{k'}^{(\mu)} \tilde{A}^{(\mu)}_{k'})_{\rm KG}&=
\delta(k_0-k_0')\,\delta^2(k_\perp-k_\perp')\quad\textrm{for the FRW,}
\\*[1.0mm]
\label{e3rrjei2}
\cdot \quad \!\! 
({\cal N}_{k}^{(\mu)}  \tilde{A}^{(\mu)}_{k},\, {\cal N}_{k'}^{(\mu)} \tilde{A}^{(\mu)}_{k'})_{\rm KG}&=
\delta(k_1-k_1')\,\delta^2(k_\perp-k_\perp')\quad\textrm{for the RRW,}
\end{align}
\end{subequations}
where $k$ and $k'$ in the subscripts are defined under (\ref{wyea}).
$\tilde{A}^{(\mu)}_{k}$ are given in (\ref{ba54nwy}) and (\ref{weh28a}). 
Saving the explanation for how we have obtained the normalization constants for later, in conclusion, we can obtain these normalization constants as follows: 
\begin{subequations}\label{ly3es}
\flushleft{\textrm{\underline{\bf The normalization constants in the FRW:}}}
\begin{align}
\label{ly3es1}
& \bullet\quad \!\!  
{\cal N}_{k}^{(\perp)}
=ik_\perp\,{\cal N}_{k}^{(B)}
=\frac{1}{\sqrt{2(2\pi)^2a\sinh (\pi \alpha)}}
,\\*[1.0mm]
\label{ly3es2}
& \bullet\quad \!\!  
{\cal N}_{k}^{(0)}=
{\cal N}_{k}^{(1)}= 
\frac
{\alpha}
{2\pi b \sqrt{a\sinh (\pi \alpha)}},
\end{align}
\end{subequations}
\begin{subequations}\label{4rwerh}
\flushleft{\textrm{\underline{\bf The normalization constants in the RRW:}}}
\begin{align}
\label{4rwerh1}
& \bullet\quad \!\!  {\cal N}_{k}^{(\perp)}
=ik_\perp\,{\cal N}_{k}^{(B)}
= 
\frac{1}{2\pi^2}
\sqrt{\frac{\sinh (\pi \alpha)}{a}},\\*[1.0mm]
\label{4rwerh2}
& \bullet\quad \!\!  {\cal N}_{k}^{(0)}=
{\cal N}_{k}^{(1)}= 
\frac{\alpha}{2\pi^2b} \sqrt{\frac{2\sinh (\pi\alpha)}{a}},
\end{align}
\end{subequations}
where $\alpha$ and $b$ are defined under (\ref{ba54nwy}). 
The relation between $ {\cal N}_{k}^{(\perp)}$ and ${\cal N}_{k}^{(B)}$ in the FRW and RRW is the one given in (\ref{vae47th}), 
which is obtained from the condition supposed upon obtaining the solution of $\tilde{B}_{k}$.
\newline

As mentioned in the beginning of this subsection, the normalization constants of the mode-solutions in the PRW and LRW are the same as those in (\ref{ly3es}) and (\ref{4rwerh}); 
therefore, we can give those normalization constants as follows:
\begin{subequations}\label{s3rtr}
\flushleft{\textrm{\underline{\bf The normalization constants in the PRW and LRW:}}}
\begin{align}
\label{s3rtr1}
\bullet &\quad \!\! {\cal N}_{k}^{(\perp)}
=ik_\perp\,{\cal N}_{k}^{(B)}
=\textrm{(\ref{ly3es1})},  \quad\quad \!\! 
{\cal N}_{k}^{(0)}={\cal N}_{k}^{(1)}=\textrm{(\ref{ly3es2})} \quad \textrm{for the PRW,}\\*[1.0mm]
\label{s3rtr2}
\bullet &\quad \!\! {\cal N}_{k}^{(\perp)}
=ik_\perp\,{\cal N}_{k}^{(B)}
=\textrm{(\ref{4rwerh1})}, \quad\quad \!\! 
{\cal N}_{k}^{(0)}={\cal N}_{k}^{(1)}=\textrm{(\ref{4rwerh2})} \quad \textrm{for the LRW.}
\end{align}
\end{subequations}
~\newline

In what follows, we note the equations that we have calculated to obtain ${\cal N}_{k}^{(\perp)}$, ${\cal N}_{k}^{(0)}$ and ${\cal N}_{k}^{(1)}$ in (\ref{ly3es}) and (\ref{4rwerh}). 
As for ${\cal N}_{k}^{(\perp)}$, we show the equations appearing in the calculation process.  
The other ${\cal N}_{1}^{(\perp)}$ and ${\cal N}_{0}^{(\perp)}$ can be calculated in the same way as those. 
The definition of the KG inner-product is given in (\ref{stwgi}), and the demands to determine ${\cal N}_{k}^{(\mu)}$ are given in (\ref{e3rrjei}).
\begin{subequations}\label{aerts}
\flushleft{\textrm{\underline{\bf The calculation to determine ${\cal N}_{k}^{(\perp)}$:}}}
\begin{align}
\label{aerts1}
\cdot \quad\! 
(
{\cal N}_{k}^{(\perp)} \tilde{A}^{\perp}_{k},\,
{\cal N}_{k'}^{(\perp)}\tilde{A}^{\perp}_{k'}
)_{\rm KG}  
=& \,\,
({\cal N}_{k}^{(\perp)})^2 \,\int_\infty^0 \! d\rho 
\int_{-\infty}^\infty \! \frac{d^2x^\perp}{a\rho} \,g_{\perp\perp}\,
i(A_k^{\perp*}\nabla_1 A_{k'}^\perp-A_{k'}^\perp \nabla_1 A_k^{\perp*})
\nonumber\\*[1.0mm]
=& \,\,
({\cal N}_{k}^{(\perp)})^2 \,(2\pi)^2 \delta^2(k_\perp-k'_\perp)\,e^{i(k_1-k'_1)\tau} \, 
(\alpha+\alpha')
\int_0^\infty \! \frac{d\rho}{b\rho}\, J_{i\alpha}(b\rho)J_{i\alpha'}(b\rho)
\nonumber\\*[1.0mm]
=& \,\, 
({\cal N}_{k}^{(\perp)})^2\, 2(2\pi^2)^2 \, a \sinh (\pi \alpha)
\,\delta(k_0-k'_0)\,\delta^2(k_\perp-k'_\perp)
\!\!\quad\textrm{for the FRW,}
\\*[3.0mm]
\label{aerts2}
\cdot \quad\!
(
{\cal N}_{k}^{(\perp)} \tilde{A}^{\perp}_{k},\,
{\cal N}_{k'}^{(\perp)}\tilde{A}^{\perp}_{k'}
)_{\rm KG}
=& \,\,
({\cal N}_{k}^{(\perp)})^2 \,\int_\infty^0 \! d\rho 
\int_{-\infty}^\infty \! \frac{d^2x^\perp}{a\rho} \,g_{\perp\perp}\,
i(A_k^{\perp*}\nabla_0 A_{k'}^\perp-A_{k'}^\perp \nabla_0 A_k^{\perp*})
\nonumber\\*[1.0mm]
=& \,\,
({\cal N}_{k}^{(\perp)})^2 \,(2\pi)^2 \delta^2(k_\perp-k'_\perp)\,e^{i(k_0-k'_0)\tau} \, 
(\alpha+\alpha')
\int_0^\infty \! \frac{d\rho}{\rho}\, K_{i\alpha}(b\rho)K_{i\alpha'}(b\rho)
\nonumber\\*[1.0mm]
=& \,\, 
({\cal N}_{k}^{(\perp)})^2
\frac{a \, (2\pi^2)^2}{\sinh (\pi \alpha)}
\,\delta(k_0-k'_0)\,\delta^2(k_\perp-k'_\perp)
\!\!\quad\textrm{for the RRW,}
\end{align}
\end{subequations}
where in (\ref{aerts1}), in the second line, $\nabla_1 A_k^{\perp}=\partial_1 A_k^{\perp}+\Gamma^\perp_{1\mu}A_k^{\mu}$ is just $\partial_1 A_k^{\perp}$;
from the second to third lines, we put $b'=b$ based on the appearances of $\delta(k_0-k'_0)$ and $\delta^2(k_\perp-k'_\perp)$ in the equation; 
in the third line, we used (\ref{ehktyd11}) in Appendix.\ref{r3g67kb} to perform the integral. 
From the last line, (\ref{ly3es1}) can be obtained. 
(\ref{aerts2}) is calculated in the same way as (\ref{aerts1}), and (\ref{4rwerh1}) can be obtained as well.

Next, we note the concrete expressions of
$({\cal N}_{k}^{(1)}\tilde{A}^{1\ast}_{k},\,{\cal N}_{k'}^{(1)}\tilde{A}^1_{k'})_{\rm KG}$ and
$({\cal N}_{k}^{(0)}\tilde{A}^{0\ast}_{k},\,{\cal N}_{k}^{(0)}\tilde{A}^0_{k'})_{\rm KG}$
that we have computed to determine ${\cal N}_{k}^{(1)}$ and ${\cal N}_{k}^{(0)}$ in (\ref{4rwerh}) and (\ref{s3rtr}). 
\vspace{0.5mm}
\begin{subequations}\label{ev4re}
\flushleft{\textrm{\underline{\bf The calculation to determine ${\cal N}_{k}^{(1)}$ in the FRW and RRW:}}}
\begin{align}
\label{ev4re1}
\cdot \quad\!
& \,\,
(
{\cal N}_{k}^{(1)}\tilde{A}^{1\ast}_{k},\,
{\cal N}_{k'}^{(1)}\tilde{A}^1_{k'}
)_{\rm KG}
=
({\cal N}_{k}^{(1)})^2 
\int_\infty^0 \!\! d\rho 
\int_{-\infty}^\infty \! \frac{d^2x^\perp}{a\rho} \,
i\,g_{11}\,\big(
\tilde{A}_k^{1*}\partial_1 \tilde{A}_{k'}^1
+\Gamma^1_{10}\,\tilde{A}_k^{1*}\tilde{A}_{k'}^0-(k' \leftrightarrow k)^*
\big),  
\\*[3.0mm]
\label{ev4re2}
\cdot \quad\!
& \,\,
(
{\cal N}_{k}^{(1)}\tilde{A}^{1\ast}_{k},\,
{\cal N}_{k}^{(1)}\tilde{A}^1_{k'})_{\rm KG}
=
({\cal N}_{k}^{(1)})^2 
\int_\infty^0 \!\! d\rho 
\int_{-\infty}^\infty \! \frac{d^2x^\perp}{a\rho} \,
i\,g_{11}\,\big(\tilde{A}_k^{1*}\partial_0 \tilde{A}_{k'}^1
+\Gamma^1_{00}\tilde{A}_k^{1*}\tilde{A}_{k'}^0-(k' \leftrightarrow k)^*
\big),
\end{align}
\end{subequations}
for the FRW and RRW, respectively.
The calculation from (\ref{ev4re1}) and (\ref{ev4re2}) can be performed in the same way as (\ref{aerts}), 
and ${\cal N}_{k}^{(1)}$ can be finally determined as noted in (\ref{ly3es2}) and (\ref{4rwerh2}). 
To integrate out the $\rho$-integrals, (\ref{ehktyd14}), (\ref{ehktyd13}), (\ref{wreh14}) and (\ref{wreh13}) in Appendix.\ref{r3g67kb} were used.

\vspace{0.5mm}
\begin{subequations}\label{fsv45re}
\flushleft{\textrm{\underline{\bf The calculation to determine ${\cal N}_{k}^{(0)}$ in the FRW and RRW:}}}
\begin{align}
\label{fsv45re1}
\cdot \quad\!
& \,\,
(
{\cal N}_{k}^{(0)}\tilde{A}^{0\ast}_{k},\,
{\cal N}_{k'}^{(0)}\tilde{A}^0_{k'}
)_{\rm KG}
=
({\cal N}_{k}^{(0)})^2 
\int_\infty^0 \!\! d\rho 
\int_{-\infty}^\infty \! \frac{d^2x^\perp}{a\rho} \,
i\,g_{00}\,\big(
 \tilde{A}_k^{0*}\partial_1 \tilde{A}_{k'}^0
+\Gamma^0_{11}\tilde{A}_k^{0*}\tilde{A}_{k'}^1-(k' \leftrightarrow k)^*
\big),
\\*[3.0mm]
\label{fsv45re2}
\cdot \quad\!
& \,\,
(
{\cal N}_{k}^{(0)}\tilde{A}^{0\ast}_{k},\,
{\cal N}_{k}^{(0)}\tilde{A}^0_{k'})_{\rm KG}
=
({\cal N}_{k}^{(0)})^2 
\int_\infty^0\!\! d\rho 
\int_{-\infty}^\infty \! \frac{d^2x^\perp}{a\rho} \,
i\,g_{00}\,\big(\tilde{A}_k^{0*}\partial_0 \tilde{A}_{k'}^0
+\Gamma^0_{01}\tilde{A}_k^{0*}\tilde{A}_{k'}^1-(k' \leftrightarrow k)^*
\big),
\end{align}
\end{subequations}
for the FRW and RRW, respectively.
From (\ref{fsv45re1}) and (\ref{fsv45re2}), ${\cal N}_{k}^{(0)}$ in (\ref{ly3es2}) and (\ref{4rwerh2}) can be obtained, 
where (\ref{ehktyd12}), (\ref{ehktyd14}), (\ref{wreh12}) and (\ref{wreh14}) in Appendix.\ref{r3g67kb} are used to integrate out the $\rho$-integrals.

\section{The equal-time canonical quantization of the U(1) gauge field in the Rindler coordinates}
\label{bywbd}

In this section, 
formulating the equal-time canonical commutation relations (referred to as CCR in what follows, omitting the term ``equal-time'') of the U(1) gauge field in the FRW and RRW,
it is shown that the coefficients of each mode of the U(1) gauge field in each region have physical meaning as creation/annihilation operators.  
Since we have explicitly obtained the mode-expanded solution of the U(1) gauge field in the FRW and RRW 
including the normalization constants as seen in (\ref{ba54nwy}), (\ref{b543es}), (\ref{ly3es}) and  (\ref{4rwerh}), we can formulate the CCR without ambiguity or speculation. 
Then, by proceeding with calculation following the definition, we can obtain the results (\ref{aredv}) and (\ref{roere}).
\newline

\begin{flushleft}
\textrm{\underline{\bf The CCR in the FRW:}}
\end{flushleft}
Let us formulate the CCR of the fields in $\tilde{{\cal L}}^{(R)}_{\rm U(1)}$ in (\ref{etsiph}) 
in the FRW as follows:
\begin{subequations}\label{rntsr}
\begin{align}
\label{rntsr1}
[A^i(\rho,\zeta,x^\perp),\pi_j(\rho',\zeta,x'{}^\perp)] &
= \, -(a\rho)^{-1} \, i\delta^i_j\delta(\rho-\rho')\delta^{2}(x^\perp-x'{}^\perp),
\\*[1.0mm]
\label{rntsr2}
[A^1(\rho,\zeta,x^\perp),\pi_{1}(\rho',\zeta,x'{}^\perp)] &
= \, -(a\rho)^{+1} \, i\delta(\rho-\rho')\delta^{2}(x^\perp-x'{}^\perp),
\\*[1.0mm]
\label{rntsr3} 
[A^\mu(\rho,\zeta,x^\perp),A^\nu(\rho',\zeta,x'{}^\perp)] &= 0, 
\\*[3.0mm]
\label{rntsr4}
[B(\rho,\zeta,x^\perp),\pi^{(B)}(\rho',\zeta,x'{}^\perp)]&
=\, -(a\rho)^{+1} \, i\delta(\rho-\rho')\delta^{2}(x^\perp-x'{}^\perp), 
\\*[1.0mm]
\label{rntsr5}
[B(\rho,\zeta,x^\perp),B(\rho',\zeta,x'{}^\perp)]&=0.
\end{align}
\end{subequations}
where $\zeta$-coordinates have been commonly taken as the $\zeta$-coordinate plays the role of the time in the FRW, 
and $\pi_1$ and $\pi^{(B)}$ in the FRW are given in terms of $A^\mu$ as follows:
\begin{eqnarray}\label{iytkv}
\pi_i =-F^1{}_i, \quad 
\pi_1 =\frac{\partial {\tilde{{\cal L}}^{(R)}_{\rm U(1)}}}{\partial (\partial_1 A^1)}=B=-\nabla_\mu A^\mu, \quad
\pi^{(B)} =\frac{\partial {\tilde{{\cal L}}^{(R)}_{\rm U(1)}}}{\partial (\partial_1 B)} =-A^1,
\end{eqnarray}
where $\pi_i$ is the same as the one already given in (\ref{zulst2}).
In $\pi_1$, (\ref{e3ier2}) is used.
$\pi^{(B)}$ is obtained by changing the term $B \, \nabla_\mu A^\mu$ in $\tilde{{\cal L}}^{(R)}_{\rm U(1)}$ to $-\partial_\mu B \, A^\mu$ 
using $\nabla_\mu A^\mu=(\sqrt{-g})^{-1} \partial_\mu (\sqrt{-g} A^\mu)$ and assuming $\int_{\rm FRW} d^4x \, \partial_\mu (\sqrt{-g} \,B\, A^\mu)=0$. 
The ghost fields are not addressed for the reason mentioned in Sec.\,\ref{ts6dt5}. 

When $a$ is taken to $0$, $a \rho$ in (\ref{rntsr}) become $1$ ($\rho$ is defined in (\ref{dvyjd})) and (\ref{rntsr}) can agree with the CCR in the Minkowski coordinates 
except for the difference of the sign. 
The difference of the sign can be considered to be due to the difference between the Killing vector at $a=0$ in the FRW and the Killing vector in the Minkowski coordinates. 

With (\ref{iytkv}), (\ref{rntsr1}), (\ref{rntsr2}) and (\ref{rntsr4}) can be given as follows:
\begin{subequations}\label{zrstnm}
\begin{align}
\label{zrstnm1}
[A^i(\tau,\rho,x^\perp),-\partial^1 {A}_j(\tau,\rho',x'{}^\perp)] &
= \, -(a\rho)^{-1} \,i\delta^i_j\delta(\rho-\rho')\delta^{2}(x^\perp-x'{}^\perp),
\\*[1.0mm]
\label{zrstnm2}
[A^1(\tau,\rho,x^\perp),-\partial_1{A}^1(\tau,\rho',x'{}^\perp)] &
= \, -(a\rho)^{+1} \,i\delta(\rho-\rho')\delta^{2}(x^\perp-x'{}^\perp), 
\\*[1.0mm]
\label{zrstnm4}
[B(\tau,\rho,x^\perp),-A^1(\tau,\rho',x'{}^\perp)]&
=\, -(a\rho)^{+1} \,i\delta(\rho-\rho')\delta^{2}(x^\perp-x'{}^\perp). 
\end{align}
\end{subequations}
Since $B=-\nabla_\mu A^\mu$ as seen in (\ref{iytkv}), it can be seen that (\ref{zrstnm2}) and (\ref{zrstnm4}) are equivalent to each other, 
therefore let us look at (\ref{zrstnm1}) and (\ref{zrstnm2}) as follows.

Now, $-\partial^1 {A}_j$ in (\ref{zrstnm1})  can be rewritten as $-g^{11}g_{jj}\partial_1 {A}^j$.
Here, $g^{11}=-(a\rho)^{-2}$, and $g_{00}=+1$ and $g_{\perp\perp}=-1$.
Considering these, (\ref{zrstnm1}) can be written as 
\begin{eqnarray}\label{aqtnr}
[A^i(\tau,\rho,x^\perp),\partial_1 {A}^j(\tau,\rho',x'{}^\perp)] 
= 
\left\{ 
\begin{array}{ll}
-(a\rho)^{+1} \,i\delta(\rho-\rho')\delta^{2}(x^\perp-x'{}^\perp) & \!\! \textrm{for $i=j=0$,} \\[1.0mm] 
+(a\rho)^{+1} \,i\delta(\rho-\rho')\delta^{2}(x^\perp-x'{}^\perp) & \!\! \textrm{for $i=j=\perp$,} \\[1.0mm]
0 & \!\! \textrm{for $i \not= j$.}  
\end{array} 
\right.
\end{eqnarray} 
Writing (\ref{zrstnm1}) and (\ref{zrstnm2}) together as one relation, 
the CCR of the U(1) gauge field in the FRW can be written as follows:
\begin{eqnarray}\label{reawv}
\bullet  \quad \!\!
[A^\mu(\tau,\rho,x^\perp),\partial_1 A^\nu(\tau,\rho',x'{}^\perp)] 
= -(a\rho)^{+1} g^{{\rm (M)}\mu\nu}\,i\delta(\rho-\rho')\delta^{2}(x^\perp-x'{}^\perp),
\end{eqnarray}
where $g^{{\rm (M)}\mu\nu}={\rm diag}(+,-,-,-)$ and $(a\rho)^{+1}=\sqrt{\vert g^{11} \vert }^{\,-1}$.
\newline

\begin{flushleft}
\textrm{\underline{\bf The CCR in the RRW:}}
\end{flushleft}
The CCR of the fields in $\tilde{{\cal L}}^{(R)}_{\rm U(1)}$ in the RRW is also formulated as follows:
\begin{subequations}\label{ssrro} 
\begin{align}
\label{ssrro1}
[A^i(\tau,\rho,x^\perp),\pi_j(\tau,\rho',x'{}^\perp)] &
= \, +(a\rho)^{-1} \, i\delta^i_j\delta(\rho-\rho')\delta^{2}(x^\perp-x'{}^\perp),
\\*[1.0mm]
\label{ssrro2}
[A^0(\tau,\rho,x^\perp),\pi_{0}(\tau,\rho',x'{}^\perp)] &
= \, +(a\rho)^{+1} \, i\delta(\rho-\rho')\delta^{2}(x^\perp-x'{}^\perp),
\\*[1.0mm]
\label{ssrro3}
[A^\mu(\tau,\rho,x^\perp),A^\nu(\tau,\rho',x'{}^\perp)] &= 0, 
\\*[3.0mm]
\label{ssrro4}
[B(\tau,\rho,x^\perp),\pi^{(B)}(\tau,\rho',x'{}^\perp)]&
=\, +(a\rho)^{+1} \, i\delta(\rho-\rho')\delta^{2}(x^\perp-x'{}^\perp), 
\\*[1.0mm]
\label{ssrro5}
[B(\tau,\rho,x^\perp),B(\tau,\rho',x'{}^\perp)]&=0,
\end{align}
\end{subequations}
where $\tau$-coordinates have been commonly taken as $\tau$-coordinate plays the role of the time in the RRW, 
and $\pi_i$, $\pi_0$ and $\pi^{(B)}$ in the RRW above are given as follows:
\begin{eqnarray}\label{ft3krv}
\pi_i =-F^0{}_i, \quad 
\pi_0 =B=-\nabla_\mu A^\mu, \quad
\pi^{(B)} =-A^0,
\end{eqnarray}
where $\pi_i$ is the same as the one already given in (\ref{zulst1}), 
and $\pi_0$ and $\pi^{(B)}$ are known from (\ref{etsiph}) in the same way as the case of the FRW above. 
When $a$ is taken to $0$, (\ref{ssrro}) can agree with the CCR in the Minkowski coordinates.

Using (\ref{ft3krv}), (\ref{ssrro1}), (\ref{ssrro2}) and (\ref{ssrro4}) can be written as 
\begin{subequations}\label{r6isdv}
\begin{align}
\label{r6isdv1}
[A^i(\tau,\rho,x^\perp),-\partial^0 {A}_j(\tau,\rho',x'{}^\perp)] &
= \, +(a\rho)^{-1} \,i\delta^i_j\delta(\rho-\rho')\delta^{2}(x^\perp-x'{}^\perp),
\\*[1.0mm]
\label{r6isdv2}
[A^0(\tau,\rho,x^\perp),-\partial_0{A}^0(\tau,\rho',x'{}^\perp)] &
= \, +(a\rho)^{+1} \,i\delta(\rho-\rho')\delta^{2}(x^\perp-x'{}^\perp), 
\\*[1.0mm]
\label{r6isdv4}
[B(\tau,\rho,x^\perp),-A^0(\tau,\rho',x'{}^\perp)]&
=\, +(a\rho)^{+1} \,i\delta(\rho-\rho')\delta^{2}(x^\perp-x'{}^\perp), 
\end{align}
\end{subequations} 
Since (\ref{r6isdv2}) and (\ref{r6isdv4}) are equivalent to each other by the relation concerning $B$ in (\ref{ft3krv}), we look at (\ref{r6isdv1}) and (\ref{r6isdv2}) in what follows.

Then, $-\partial^0 {A}_j$  in (\ref{r6isdv1}) can be written as $-g^{00}g_{jj}\partial_0 {A}^j$ ($g^{00}=(a\rho)^{-2}$ and $g_{jj}=-1$ for all $j$), and (\ref{r6isdv1}) can be written as 
\begin{eqnarray}\label{tu6klu}
[A^i(\tau,\rho,x^\perp),\partial_0 {A}^j(\tau,\rho',x'{}^\perp)] 
= 
\left\{ 
\begin{array}{ll}
(a\rho)^{+1} \,i\delta(\rho-\rho')\delta^{2}(x^\perp-x'{}^\perp) &  \textrm{for $\, i=j$,} \\[1.0mm] 
0                                                                &  \textrm{for $\, i \not= j$.}  
\end{array} 
\right.
\end{eqnarray}
Therefore, (\ref{r6isdv1}) and (\ref{r6isdv2}) can be written together into one relation as follows:
\begin{eqnarray}\label{aeba}
\label{aeba1}
\bullet  \quad \!\!
[A^\mu(\tau,\rho,x^\perp),\partial_0 A^\nu(\tau,\rho',x'{}^\perp)] 
= -(a\rho)^{+1} g^{{\rm (M)}\mu\nu}\,i\delta(\rho-\rho')\delta^{2}(x^\perp-x'{}^\perp),
\end{eqnarray}
where $g^{{\rm (M)}\mu\nu}={\rm diag}(+,-,-,-)$ as well as (\ref{reawv}) and $(a\rho)^{+1}=\sqrt{g^{00}}^{\,-1}$. 
\newline

When $A^\mu$ satisfies the commutation relations (\ref{reawv}) or (\ref{aeba}), the following commutation relation can be satisfied for arbitrary $f_{A}$ 
($A$ denotes some indices or labels) in each region:
\begin{subequations}\label{rnomt}
\begin{align}
\label{rnomt1}
\cdot \quad\!
& [(f_A(x), A^\mu(x))_{\rm KG},(A^\nu(y), f_B(y))_{\rm KG}]\nonumber\\*[1.0mm]
=&
- \int \! d^3x \! \int \! d^3y 
\sqrt{g^{11}(x)}\sqrt{g^{11}(y)}
\,(
f_A^*(x)  \partial_\zeta f_B(y)  \,[\partial_\zeta A^\mu(x),A^\nu(y)]
+f_B(y) \partial_\zeta f_A^*(x) \,[A^\mu(x),\partial_\zeta A^\nu(y)]
) 
\nonumber\\*[1.0mm]
=& -g^{{\rm (M)}\mu\nu}(f_A(x), f_B(x))_{\rm KG}
\quad\textrm{for the FRW,}
\\*[3.0mm]
\label{rnomt2}
\cdot \quad\!
& [(f_A(x), A^\mu(x))_{\rm KG},(A^\nu(y), f_B(y))_{\rm KG}]
\nonumber\\*[1.0mm]
=&
- \int \! d^3x \! \int \! d^3y 
\sqrt{g^{00}(x)}\sqrt{g^{00}(y)}
\,(
f_A^*(x)  \partial_\tau f_B(y)  \,[\partial_\tau A^\mu(x),A^\nu(y)]
+f_B(y) \partial_\tau f_A^*(x) \,[A^\mu(x),\partial_\tau A^\nu(y)]
) 
\nonumber\\*[1.0mm]
=& 
- g^{{\rm (M)}\mu\nu}(f_A(x), f_B(x))_{\rm KG}
\quad\textrm{for the RRW,}
\end{align}
\end{subequations}
where the KG inner-product defined in (\ref{stwgi}) has been used in the equations above.
\newline

Here, in general, let us consider $\phi=\sum_i({\bm a}_i f_i+{\bm a}_i^\dagger f_i^*)$, where $\phi$ means a real function and $f_i$ is supposed to satisfy
$
  (f_i  ,f_j  )_{\rm KG} 
=-(f_j^*,f_i^*)_{\rm KG}
= \delta_{ij}$ and
$
  (f_i^*,f_j)_{\rm KG} 
=-(f_j,f_i^*)_{\rm KG}=0$. 
These $\phi$ and $f_i$ satisfy the following relations: 
\begin{subequations}\label{rtrsy} 
\begin{align}
  (f_i, \phi)_{\rm KG} 
=& \, 
\sum_j  (f_i,{\bm a}_j f_j)_{\rm KG}
=\sum_j {\bm a}_j (f_i,f_j)_{\rm KG}
= {\bm a}_i, \\*[1.0mm]
  (\phi, f_i)_{\rm KG} 
=& \,
\sum_j   ({\bm a}_j f_j,f_i)_{\rm KG}
= \sum_j {\bm a}_j^\dagger  (f_j,f_i)_{\rm KG}
= {\bm a}_i^\dagger, \\*[1.0mm]
  (\phi, f_i^*)_{\rm KG} 
=& \,
\sum_j {\bm a}_j (f_j^*,f_i^*)_{\rm KG}
= -{\bm a}_i.
\end{align}
\end{subequations}
Then, denoting (\ref{rnomt1}) and (\ref{rnomt2}) respectively as 
\begin{subequations}\label{rearwsr}
\begin{align}
\cdot \quad\!
[(f_A,A^\mu)_{\rm KG},(A^\nu,f_B)_{\rm KG}]
=& \,
- g^{{\rm (M)}\mu\nu}\,(f_A,f_B)_{\rm KG},
\\*[1.0mm]
\cdot \quad\!
[(f_A,A^\mu)_{\rm KG},(A^\nu,f_B^\ast)_{\rm KG}]
=& \,
- g^{{\rm (M)}\mu\nu}\,(f_A,f_B^\ast)_{\rm KG}, 
\end{align}
\end{subequations}
when $f_A$ are given by the mode-functions (\ref{ba54nwy}) or (\ref{b543es}) normalized by (\ref{4rwerh}) or (\ref{ly3es}) for the RRW or FRW respectively, 
it can be seen from the general relations (\ref{rtrsy}) that the coefficient of each mode $\bm{a}^\mu_{k_1,k_\perp}$ and $\bm{a}^\mu_{k_0,k_\perp}$
in (\ref{weh28a1}) and (\ref{w4aeea1}) satisfy the following commutation relations:
\begin{flushleft}
\textrm{\underline{\bf In the FRW:}}
\end{flushleft}
\vspace*{-2mm}
\begin{eqnarray}\label{aredv}
\bullet  \quad \!
\displaystyle
[\bm{a}^\mu_{k_1,k_\perp},\bm{a}^{\nu \dagger}_{k_1',k_\perp'}] 
= - g^{{\rm (M)}\mu\nu}\delta(k_1-k_1')\,\delta^2(k_\perp-k_\perp'),
\quad
[\bm{a}^\mu_{k_1,k_T},\bm{a}^\nu_{k_1',k_\perp'}] 
= [\bm{a}^{\mu \dagger}_{k_1,k_\perp},\bm{a}^{\nu \dagger}_{k_1',k_\perp'}]=0,
\end{eqnarray}
\vspace*{-6mm}
\begin{flushleft}
\textrm{\underline{\bf In the RRW:}}
\end{flushleft}
\vspace*{-2mm}
\begin{eqnarray}\label{roere}
\bullet \quad \!
\displaystyle
[\bm{a}^\mu_{k_0,k_\perp},\bm{a}^{\nu \dagger}_{k_0',k_\perp'}] 
= - g^{{\rm (M)}\mu\nu}\delta(k_0-k_0')\,\delta^2(k_\perp-k_\perp'), 
\quad
[\bm{a}^\mu_{k_0,k_T},\bm{a}^\nu_{k_0',k_\perp'}] 
= [\bm{a}^{\mu \dagger}_{k_0,k_\perp},\bm{a}^{\nu \dagger}_{k_0',k_\perp'}]=0.
\end{eqnarray}

From these, it can be seen that $\bm{a}^\mu_{k_1,k_\perp}$ and $\bm{a}^\mu_{k_1,k_\perp}$ have physical meaning as annihilation operators. 
\newline

The CCR of the U(1) gauge field and the $B$-field in the PRW and LRW are formulated 
in the same way as (\ref{rntsr}) of the FRW and (\ref{ssrro}) of the RRW, 
as the metrices of the PRW and LRW are mathematically the same as those in the FRW and RRW as can be seen in (\ref{dres}) 
and the Lagrangian (\ref{etsiph}) is common in the four regions in the Rindler coordinates. 
Therefore, from the results of (\ref{aredv}) and (\ref{roere}), the coefficient of each mode of the U(1) gauge field in the PRW and LRW  in (\ref{hoaag1}) and (\ref{w8atd1}), 
$\bm{a}^\mu_{k_1,k_\perp}$ and $\bm{a}^\mu_{k_1,k_\perp}$, 
satisfy the following commutation relations:
\begin{flushleft}
\textrm{\underline{\bf In the PRW:}}
\end{flushleft}
\vspace*{-2mm}
\begin{eqnarray}\label{akl3dy} 
\bullet \quad \!
[\bm{a}^\mu_{k_1,k_\perp},\bm{a}^{\nu \dagger}_{k_1',k_\perp'}] 
= - g^{{\rm (M)}\mu\nu}\delta(k_1-k_1')\,\delta^2(k_\perp-k_\perp'),
\quad
[\bm{a}^\mu_{k_1,k_T},\bm{a}^\nu_{k_1',k_\perp'}] 
= [\bm{a}^{\mu \dagger}_{k_1,k_\perp},\bm{a}^{\nu \dagger}_{k_1',k_\perp'}]=0,
\end{eqnarray}
\vspace*{-6mm}
\begin{flushleft}
\textrm{\underline{\bf In the LRW:}}
\end{flushleft}
\vspace*{-2mm}
\begin{eqnarray}\label{xuk2re}
\bullet \quad \!
[\bm{a}^\mu_{k_0,k_\perp},\bm{a}^{\nu \dagger}_{k_0',k_\perp'}] 
= - g^{{\rm (M)}\mu\nu}\delta(k_0-k_0')\,\delta^2(k_\perp-k_\perp'), 
\quad
[\bm{a}^\mu_{k_0,k_T},\bm{a}^\nu_{k_0',k_\perp'}] 
= [\bm{a}^{\mu \dagger}_{k_0,k_\perp},\bm{a}^{\nu \dagger}_{k_0',k_\perp'}]=0,
\end{eqnarray}
and have physical meaning as the annihilation operator, as well.

\section{Polarization vectors}
\label{yervd}

In the previous section, it was shown that 
the coefficient of each mode-solution in each direction of the U(1) gauge field in each region of Rindler coordinates
has the physical meaning as the creation/annihilation operators. 
Usually, the creation/annihilation operators in each direction are treated by decomposing those in the polarization directions.
Therefore, in this section, polarization vectors in each region in the Rindler coordinates are given. 
As a result, the values of the coordinates of $\eta$ and $\xi$ in the FRW and RRW and $\tilde{\eta}$ and $\tilde{\xi}$ in the PRW and LRW are restricted, 
as the region where the norms of the 1-particle states in the scalar and longitudinal polarization directions are less than zero.
Regarding the origin of those restrictions, it can be concluded that 
those can be attributed to the non-covariance in the canonical quantization (\ref{reawv}).
\newline

Let us decompose the annihilation operators in each region with the polarization vectors as follows:
\begin{subequations}\label{rvenya}
\begin{align}
\label{rvenya1}
\cdot \quad \!\!
\bm{a}^\mu_{k_1,k_\perp}
=& \sum_{\sigma=S,L,\pm} \varepsilon^{(\sigma)\mu}\,\bm{a}^{(\sigma)}_{k_1,k_\perp} \quad\textrm{for the FRW and PRW}, \\*
\label{rvenya2}
\cdot \quad \!\!
\bm{a}^\mu_{k_0,k_\perp}
=& \sum_{\sigma=S,L,\pm} \varepsilon^{(\sigma)\mu}\,\bm{a}^{(\sigma)}_{k_0,k_\perp} \quad\textrm{for the RRW and LRW}.
\end{align}
\end{subequations}
We take the polarization directions as the scalar, longitudinal, positive and negative helicity directions, 
and $S$, $L$ and $\pm$ in the indices of the summations refer to those.
With (\ref{rvenya1}) and (\ref{rvenya2}), the relations in (\ref{aredv})-(\ref{xuk2re}) can be written as 
\begin{subequations}\label{rsytr}
\begin{align}
\label{rsytr1}
\cdot \quad \!\!
\sum_{\sigma,\sigma'}\varepsilon^{(\sigma)\mu}\varepsilon^{(\sigma')\nu\ast}\,
[\bm{a}^{(\sigma)}_{k_1,k_\perp},\bm{a}^{(\sigma')\dagger}_{k'_1,k'_\perp}] 
=& - g^{{\rm (M)}\mu\nu}\,\delta(k_1-k_1')\,\delta^2(k_\perp-k_\perp')
\quad\!\textrm{for the FRW and PRW,} 
\\*[1.0mm]
\label{rsytr2} 
\cdot \quad \!\!
\sum_{\sigma,\sigma'}\varepsilon^{(\sigma)\mu}\varepsilon^{(\sigma')\nu\ast}\,
[\bm{a}^{(\sigma)}_{k_0,k_\perp},\bm{a}^{(\sigma')\dagger}_{k'_0,k'_\perp}] 
=& - g^{{\rm (M)}\mu\nu}\,\delta(k_0-k_0')\,\delta^2(k_\perp-k_\perp') 
\quad\!\textrm{for the RRW and LRW,}
\end{align}
\end{subequations}
where $g^{\rm (M)\mu\nu}$ is defined under (\ref{reawv}).

Now, let us introduce the matrix $\eta^{(\sigma\sigma')}$ as follows:
\begin{subequations}\label{rvbrek}
\begin{align}
\label{rvbrek1}
\cdot \quad \!\!
[\bm{a}^{(\sigma)}_{k_1,k_\perp},\bm{a}^{(\sigma')\dagger}_{k'_1,k'_\perp}] 
=& \,\eta^{(\sigma\sigma')}\delta(k_1-k_1')\,\delta^2(k_\perp-k'_\perp)\quad\textrm{for the FRW and PRW,}  \\*[1.0mm] 
\label{rvbrek2}
\cdot \quad \!\!
[\bm{a}^{(\sigma)}_{k_0,k_\perp},\bm{a}^{(\sigma')\dagger}_{k'_0,k'_\perp}] 
=& \,\eta^{(\sigma\sigma')}\delta(k_0-k_0')\,\delta^2(k_\perp-k'_\perp)\quad\textrm{for the RRW and LRW.} 
\end{align}
\end{subequations}
$\eta^{(\sigma\sigma')}$ plays the role of the metric of the norms of the 1-particle states in each of the polarization directions of the U(1) gauge field.
By applying (\ref{rvbrek1}) and (\ref{rvbrek2}) to (\ref{rsytr1}) and (\ref{rsytr2}),
$g^{\rm (M)\mu\nu}$ can be written as:
\begin{eqnarray}\label{trntr1}
g^{\rm (M)\mu\nu} 
=
-\sum_{\sigma,\sigma'}\varepsilon^{(\sigma)\mu}\,
\varepsilon^{(\sigma')\nu\ast}\,\eta^{(\sigma\sigma')}.
\end{eqnarray}
Multiplying this by $\varepsilon^{(\rho)\ast}_\mu \, \varepsilon^{(\rho')}_\nu$, the following relation can be obtained in common for all the four regions:
\begin{eqnarray}\label{tkoelin}
\varepsilon^{(\rho)\ast}_\mu \, \varepsilon^{(\rho')}_\nu \, g^{\rm (M)\mu\nu} 
=
- \sum_{\sigma,\sigma'} \,
(g^{\rm (R)}_{\mu\lambda} \,\varepsilon^{(\sigma)\mu}   \varepsilon^{(\rho )\lambda*})\,
(g^{\rm (R)}_{\nu\tau}    \,\varepsilon^{(\sigma')\nu *}\varepsilon^{(\rho')\tau})\,
\eta^{(\sigma\sigma')},
\end{eqnarray}
where $g^{\rm (R)\mu\nu}$ means the metric in the FRW, PRW, RRW or LRW in (\ref{dres}).
~\newline

Let us express $k_\mu$, the four-dimensional momentum of the U(1) gauge field, as follows:
\begin{subequations}\label{reprv}
\begin{align}
\label{reprv1}
&
\begin{array}{ll}
\cdot & \!\!\!
\textrm{
$k^\mu=(k,\omega,0,0)$
\! with \,\! 
$ \displaystyle
\omega
= \frac{\vert\vec{k}\vert}{\sqrt{\vert g_{11}\vert}}
= \frac{\vert\vec{k}\vert}{\vert a  \rho \vert}
$ 
\,\! and \,\!
$
\vec{k}=(
\underbrace{\!\!k_{\,}\!\!}_{0},
\underbrace{0,0}_{\perp}
)$ \,\! for the FRW and PRW,} 
\end{array}
\\*[1.0mm]
\label{reprv2}
&
\begin{array}{ll}
\cdot & \!\!\!
\textrm{
$k^\mu=(\omega,k,0,0)$
\! with \,\! 
$ \displaystyle
\omega
= \frac{\vert\vec{k}\vert}{\sqrt{\vert g_{00}\vert}}
= \frac{\vert\vec{k}\vert}{\vert a  \rho \vert}
$ 
\,\! and \,\!
$
\vec{k}=(
\underbrace{\!\!k_{\,}\!\!}_{1},
\underbrace{0,0}_{\perp}
)$ \,\! for the FRW and PRW,}
\end{array}
\end{align}
\end{subequations}
where $\vec{k}$ is the three-dimensional vector obtained from $k^\mu$ removing $\omega$ from $k^\mu$, 
and the indices under the under-breaths mean the directions that the components the under-breaths attach to refer to.
These $k_\mu$ satisfy the conditions as a massless field $0=k_\mu k^\mu$. 

Let us discuss how the $(S,L,\pm)$-directions of the polarization vectors correspond to the directions in the Rindler coordinates.
First, from $0=k_\mu k^\mu$, we can see that the direction of the world-line of the U(1) gauge field (the direction of travel in the four-dimensional Rindler coordinates) 
is either of parallel or perpendicular to the Killing horizon 
(which is shown in  Fig.\ref{wsdd57} as $\textrm{$\xi$, $\tilde{\xi}$, $\zeta$ or $\tilde{\zeta}$} \to -\infty$ line). 

Then, since $\tau$ and $\tilde{\tau}$ play the role of time in the RRW and LRW, the world-line of the U(1) gauge field in the RRW and LRW should be parametrized 
by $\tau$ and $\tilde{\tau}$ (as noted under (\ref{dres})).
Then, since the direction parallel to the Killing horizon is parametrized by $\tau$ and $\tilde{\tau}$ 
(while the direction perpendicular to the Killing horizon is $\tau$ or $\tilde{\tau}$ constant), 
the direction of the world-line of the U(1) gauge field is supposed as parallel to the Killing horizon. 
In the same way, in the FRW and PRW, the direction of the world-line of the U(1) gauge field is assumed to be parallel to the Killing horizon. 

Then, since the longitudinal direction in the polarization vector is the direction of the world-line of the U(1) gauge field,
the $(S,L)$-directions of the polarization vectors agree with the $(1,0)$-directions in the FRW and PRW, 
and the $(S,L)$-directions of the polarization vectors agree with the $(0,1)$-directions in the RRW and LRW
(the $\pm$ polarization-directions and the $\perp$-directions in the Rindler coordinates agree with each other up to the $O(2)$ rotation around the $L$-direction).

Now, let us assume that, 
\begin{eqnarray}\label{rrba}
\textrm{when the polarization vectors are considered, $\rho$ is taken as constant.}
\end{eqnarray}
The definition of $\rho$ is given in (\ref{dvyjd}), from which it can be seen that this assumption means to take the acceleration of motion as constant in each region 
(to be exact, there is no accelerated motion in the FRW and PRW, and the lines 
that $\rho$ labels in the FRW and PRW 
are lines analogous to the world-lines of constant accelerated motion 
in the LRW and RRW). 
Since we are considering the constant accelerated motion, we may impose this assumption (\ref{rrba}) upon considering the polarization vectors. 

This assumption is crucial in defining the polarization vectors. 
This is because, as can be seen in (\ref{reprv}), $\omega$ depends on a coordinate $\rho$. 
Then, as can be seen later, such a $\omega$ leads to $\varepsilon^{(S)\mu}$ and $\varepsilon^{(L)\mu}$ depending on the coordinate $\rho$, 
which means that $\bm{a}^\mu_{k_1,k_\perp}$ and $\bm{a}^\mu_{k_0,k_\perp}$ depend on the coordinate $\rho$.
Then, the equations of motion become unsatisfied. 
In such a situation, if $\rho$ is fixed by (\ref{rrba}), $\omega$ becomes substantially coordinate-independent.
Accordingly, $\varepsilon^{(S)\mu}$ and $\varepsilon^{(L)\mu}$ becomes substantially coordinate-independent, and the problem mentioned above does not occur. 
\newline

In the situation where $k^\mu$ is given as (\ref{reprv}), 
we will consider the following polarization vector: 
\begin{subequations}\label{sceabu}
\begin{align}
\label{sceabu1}
& \bullet \quad \!\! 
\varepsilon^{(+)\mu} = -(0,0, 1, i)/\sqrt{2}, \quad
\varepsilon^{(-)\mu} = \varepsilon^{(+)\mu *}, 
\\*[1.0mm]
\label{scwehu3}
& \bullet \quad \!\!
\varepsilon^{(L)\mu}= -i k^\mu,
\\*[1.0mm]
\label{scwehu2}
& \bullet \quad \!\!
\varepsilon^{(S)\mu}= 
\left\{ 
\begin{array}{ll}
\! i(-k,\omega,0,0)/2\vert\vec{k}\vert^2 & \textrm{for the FRW and PRW,} \\[1.0mm] 
\! i(\omega,-k,0,0)/2\vert\vec{k}\vert^2 & \textrm{for the RRW and LRW,} 
\end{array}  
\right.
\end{align}
\end{subequations}
where $\varepsilon^{(L)\mu}$ and $\varepsilon^{(S)\mu}$ are common across the four regions.
According to the assumption (\ref{rrba}), this $\varepsilon^{(\sigma)\mu}$ is substantially constant for the Rindler coordinates. 
Below, we mention how (\ref{sceabu}) has been obtained. 
\begin{itemize}
\item[$\cdot$]
If $\vec{k}$ is given as (\ref{reprv}), $\varepsilon^{(\pm)\mu}$ can  be immediately determined as (\ref{sceabu1}) in the four regions in the Rindler coordinates.  

\item[$\cdot$]
Due to the fact that the $L$-polarization direction is parallel to $k^\mu$ in the 4D Rindler coordinates,
$\varepsilon^{(L)\mu}$ should be a constant multiplication of $k^\mu$.
In addition, it should be able to reduce to the one in the Minkowski coordinates, which is $\varepsilon^{(L)\mu}_0$ in (\ref{xeadt})\footnote{
The polarization vectors in the Minkowski coordinates which are referred to in the body text are the following one:
\begin{eqnarray}\label{xeadt}
\label{xeadt1}
&  
\varepsilon^{(+)\mu}_0 = -(0,0, 1, i)/\sqrt{2}, \quad
\varepsilon^{(-)\mu}_0 = \varepsilon^{(+)\mu *}_0, \quad
\varepsilon^{(L)\mu}_0 = -i k^\mu, \quad
\varepsilon^{(S)\mu}_0 = i(\omega,-\vec{k})/2\vert\vec{k}\vert^2,
\end{eqnarray}
where $k^\mu$ here is $(\omega,k,0,0)$ with $\omega= \vert\vec{k}\vert$ 
($\vec{k}$ is the three-dimensional spatial vector, $(k,0,0)$). 
$\varepsilon^{(L)\mu}_0$ should be a  constant multiplication of $k^\mu$.
The reason for the multiplication by $-i$ is that $\eta^{(\sigma\sigma')}$ can be real by that.}, at $a=0$.
From these conditions, $\varepsilon^{(L)\mu}$ is fixed as noted in (\ref{scwehu3}). 

Actually, since $g_{00}$ in the LRW and RRW in (\ref{dres}) reduce to $1$ at $a=0$, $\varepsilon^{(L)\mu}$ in the LRW and RRW reduce to $-i(\vert k \vert, k,0,0)$ at $a=0$, 
which is $\varepsilon^{(L)\mu}_0$ in (\ref{xeadt}). 

On the other hand, 
as for $\varepsilon^{(L)\mu}$ in the FRW and PRW, 
$g_{11}$ in the FRW and PRW reduce to $1$ at $a=0$,
and $\varepsilon^{(L)\mu}$ in the FRW and PRW reduce to $-i(k,\vert k \vert,0,0)$ at $a=0$, which 
do not agree with $\varepsilon^{(L)\mu}_0=-i(\vert k \vert,k,0,0)$ in (\ref{xeadt}). 

As for this disagreement, from comparison of (\ref{vrbwo2}) and (\ref{vrbwo3}), 
it can be seen that, in the FRW, $x^0$ is the direction of the constant accelerated motion, 
and $x^1$ is the time-direction in that constant accelerated motion.
Therefore, in the alignment of the components of FRW vectors, 
the first component has the spatial meaning and the second component has time-like meaning; namely,
\begin{eqnarray}
\lim_{a \to 0}\varepsilon^{(L)\mu}=-i(
\underbrace{\!\!\!\! \big. k \big.             \!\!\!\!}_{\rm space},
\underbrace{\big.\vert k \vert\big.}_{\rm time},
0,0), \quad
\varepsilon^{(L)\mu}_0=-i(
\underbrace{\big. \vert k \vert \big.}_{\rm time},
\underbrace{\!\!\!\! \big. k \big.             \!\!\!\!}_{\rm space},
0,0). 
\nonumber
\end{eqnarray}
Considering this fact, 
the disagreement between $\lim_{a \to 0}\varepsilon^{(L)\mu}$ and $\varepsilon^{(L)\mu}_0$ 
can be simply understood 
as the difference of the alignment of the components 
(and, therefore, can be fixed  by changing  the alignment of the components), 
and
does not pose a problem.
The disagreement in the PRW can be explained in a similar manner.

\item[$\cdot$]
Let us look at $\varepsilon^{(S)\mu}$.  
In the Minkowski case, $\varepsilon^{(S)\mu}$ is normally determined according to one of the equation of motion including $A^\mu$ and $B$  
(as $B$ is a scalar field, the polarization vector for the scalar direction can be fixed from that equation of motion\footnote{
In the Minkowski case, as one of the equations of motion, $\partial_\mu A^\mu+\alpha B=0$ ($\alpha=1$) is obtained.
From this, an equation that $\varepsilon^{(S)\mu}$ should satisfy is obtained as $-k_\mu\varepsilon^{(\sigma)\mu}=\delta^{\sigma S}$, where $k_\perp=0$.
}), which in this study corresponds to (\ref{ebraue4}) and (\ref{eeylh4}), 
which lead to (\ref{r4rg25}) and (\ref{arwrlb5}).

However, in this study, $B$ is obtained not by the whole (\ref{r4rg25}) and (\ref{arwrlb5}), 
but by the r.h.s. of (\ref{r4rg25}) $=0$ and (\ref{arwrlb5}) $=0$ in the RRW and FRW, respectively. 
Then, now that $k_\perp=0$, therefore those equations no longer work as the equations $\varepsilon^{(S)\mu}$ should adhere to.

However, this means that we can freely take $\varepsilon^{(S)\mu}$.  
Therefore, we have taken $\varepsilon^{(S)\mu}$ as noted in (\ref{scwehu2}), 
which can reduce to the Minkowski one $\varepsilon^{(S)\mu}_0$ as mentioned in (\ref{xeadt}). 
(In the FRW and PRW, there is a similar type of the disagreement as the one discussed above in the case of $\varepsilon^{(L)\mu}$; 
however, similar to the discussion mention there, those disagreements do not pose any problem either.)
\end{itemize}
~\newline

\vspace{-5mm}
Applying (\ref{sceabu}) to (\ref{tkoelin}), we can obtain $\eta^{(\sigma\sigma')}$ as follows\footnote{
If we assign as $\rho=\rho'=S$ in (\ref{tkoelin}), 
(\ref{tkoelin}) can be calculated as follows:
\begin{eqnarray}\label{rs1nt}
\varepsilon^{(S)\ast}_\mu \, \varepsilon^{(S')}_\nu \, g^{\rm (M)\mu\nu} 
=
-\sum_{\sigma,\sigma'} \,
(g^{\rm (R)}_{\mu\lambda} \,\varepsilon^{(\sigma)\mu}   \varepsilon^{(S)\lambda*})\,
(g^{\rm (R)}_{\nu\tau}    \,\varepsilon^{(\sigma')\nu *}\varepsilon^{(S)\tau})\,
\eta^{(\sigma\sigma')}
=
-\eta^{(LL)},
\end{eqnarray}
where (\ref{sceabu}) has been used. 
Performing this for each component, the first line in (\ref{dvase}) can be obtained.
The second line in (\ref{dvase}) can be obtained by assigning (\ref{sceabu}) to each component in the first line.
}: 
\begin{eqnarray}\label{dvase}
\eta^{(\sigma\sigma')}  
\!\! &=& \!\!  
-\, \bordermatrix{
& S &  L & + & -\cr
S & 
\varepsilon^{(L)\ast}_\mu \, \varepsilon^{(L)}_\nu \, g^{\rm (M)\mu\nu} & 
\varepsilon^{(L)\ast}_\mu \, \varepsilon^{(S)}_\nu \, g^{\rm (M)\mu\nu} & 0 & 0 \cr  
L & 
\varepsilon^{(S)\ast}_\mu \, \varepsilon^{(L)}_\nu \, g^{\rm (M)\mu\nu} & 
\varepsilon^{(S)\ast}_\mu \, \varepsilon^{(S)}_\nu \, g^{\rm (M)\mu\nu} & 0 & 0 \cr 
+ & 0 & 0 &
\varepsilon^{(+)\ast}_\mu \, \varepsilon^{(+)}_\nu \, g^{\rm (M)\mu\nu} & 
\varepsilon^{(+)\ast}_\mu \, \varepsilon^{(-)}_\nu \, g^{\rm (M)\mu\nu} \cr
- & 0 & 0 &
\varepsilon^{(-)\ast}_\mu \, \varepsilon^{(+)}_\nu \, g^{\rm (M)\mu\nu} & 
\varepsilon^{(-)\ast}_\mu \, \varepsilon^{(-)}_\nu \, g^{\rm (M)\mu\nu} \cr
}
\nonumber \\*[1.0mm]
\!\! &=& \!\!  \hspace{9.5mm}
\left\{
\begin{array}{l}
\left(
\begin{array}{cccc}
(\vert g_{11}\vert-1 )/4k_1^2 & (1+\vert g_{11} \vert)/2      & 0 & 0 \cr 
(1+\vert g_{11} \vert)/2      & (\vert g_{11}\vert-1 )\,k_1^2 & 0 & 0 \cr 
0 & 0 & 1 & 0 \cr
0 & 0 & 0 & 1  
\end{array}
\right)  
\quad \! \textrm{for the FRW and PRW,} \vspace{3.0mm}\cr 
\left(
\begin{array}{cccc}
(1-\vert g_{00} \vert)/4k_0^2 & (1+\vert g_{00} \vert)/2 & 0 & 0 \cr 
(1+\vert g_{00} \vert)/2      & (1-\vert g_{00} \vert)\,k_0^2  & 0 & 0 \cr 
0 & 0 & 1 & 0  \cr
0 & 0 & 0 & 1 
\end{array}
\right)  
\quad \!\textrm{for the RRW and LRW,}
\end{array}
\right.
\end{eqnarray}
where  
$
\varepsilon^{(\sigma)}_\mu
=
g^{\rm (R)}_{\mu\nu}\varepsilon^{(\sigma)\nu} 
$. 

Then, it can be seen from (\ref{rvbrek}) that the norms of the 1-particle states in the $S$- and $L$-directions are given as follows:
\begin{subequations}\label{dvnkl}
\begin{align}
\label{dvnkl1}
&
\cdot \quad \!\!
\begin{array}{l}
\displaystyle
\langle 0_{\rm R}\vert 
\bm{a}^{(S)}_{k_0,k_\perp}
\bm{a}^{(S)\,\dagger}_{k_0,k_\perp} 
\vert 0_{\rm R} \rangle = \,({\vert g_{11}\vert-1})/{4k_1^2},
\\* [3.0mm]
\displaystyle
\langle 0_{\rm R}\vert 
\bm{a}^{(L)}_{k_0,k_\perp}
\bm{a}^{(L)\,\dagger}_{k_0,k_\perp} 
\vert 0_{\rm R} \rangle
= (\vert g_{11}\vert-1)\,k_1^2,   
\end{array}
\quad\!\textrm{for the FRW and PRW,}
\\*[3.0mm]
\label{dvnkl2}
&
\cdot \quad \!\!
\begin{array}{l}
\displaystyle
\langle 0_{\rm R}\vert 
\bm{a}^{(S)}_{k_0,k_\perp}
\bm{a}^{(S)\,\dagger}_{k_0,k_\perp} 
\vert 0_{\rm R} \rangle = \,({1-\vert g_{00}\vert})/{4k_0^2},
\\* [3.0mm] 
\displaystyle
\langle 0_{\rm R}\vert 
\bm{a}^{(L)}_{k_0,k_\perp}
\bm{a}^{(L)\,\dagger}_{k_0,k_\perp} 
\vert 0_{\rm R} \rangle
=(1-\vert g_{00}\vert)\,k_0^2, 
\end{array}
\quad\!\textrm{for the RRW and LRW.}
\end{align}
\end{subequations}

Here, $\eta^{(SL)}=\eta^{(LS)} > 1/2$, 
which is the same situation with the Minkowski case 
in the sense that $\eta^{(SL)} $ and $ \eta^{(LS)}$ are always positive and pose no problem.   

\begin{figure}[H]   
\vspace{0mm} 
\begin{center}
\includegraphics[clip,width=4.5cm,angle=0]{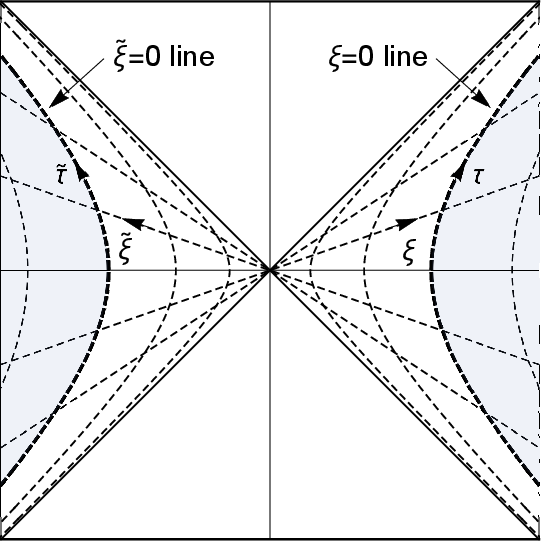} 
\hspace{3mm} 
\includegraphics[clip,width=4.5cm,angle=0]{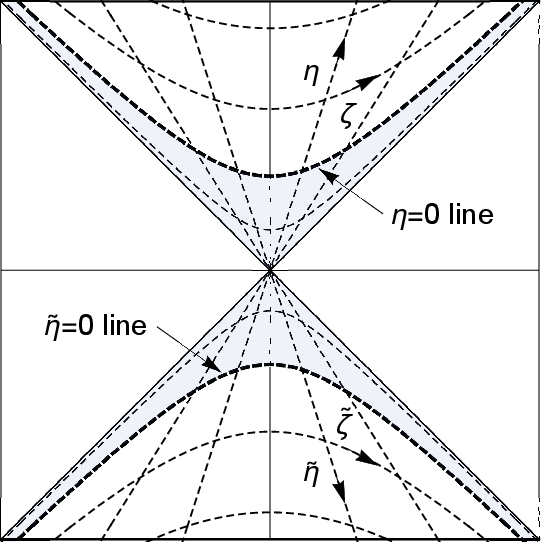}  
\end{center}
\vspace{-5.0mm}
\caption{
In the left- and right-hand figures, 
supposing the bold dashed lines as the $\tilde{\xi}=0$ and $\xi=0$ lines and the $\eta=0$ and $\tilde{\eta}=0$ lines, respectively,
the regions restricted by (\ref{sdvlne}), 
in which the norms of the 1-particle states in the $S$- and $L$- polarization directions are less than 0, are shown as the blue region.}
\label{wdet7}
\end{figure}

To ensure that (\ref{dvnkl1}) and (\ref{dvnkl2}) are less than zero, the following conditions should be satisfied:
\begin{subequations}\label{eael}
\begin{align}
\label{eael1}
\cdot \quad \!\!
\vert g_{11}\vert-1 & \le 0 \quad \!\textrm{for the FRW and PRW,}
\\*[1.0mm]
\label{eael2}
\cdot \quad \!\!
1-\vert g_{00}\vert & \le 0 \quad \!\textrm{for the RRW and LRW.}
\end{align}
\end{subequations}
Since $g_{11}=-a^2\rho^2$ and $\rho=a^{-1}e^{a\eta}$ in the FRW and PRW and $g_{00}=a^2\rho^2$ and $\rho=a^{-1}e^{a\xi}$ in the RRW and LRW, 
(\ref{eael1}) and (\ref{eael2}) can be rewritten as 
\begin{subequations}\label{erbcne}
\begin{align}
\label{erbcne1}
\cdot \quad \!\!
e^{a\eta} \le 1 \quad \!\! \textrm{and}\!\! \quad e^{a\tilde{\eta}} \le 1 \!
\quad &\textrm{for the FRW and PRW, respectively,}\\*[1.0mm]
\label{erbcne2}
\cdot \quad \!\!
1 \le e^{a\xi}  \quad \!\! \textrm{and}\!\! \quad 1 \le e^{a\tilde{\xi}} \! 
\quad &\textrm{for the RRW and LRW, respectively.}
\end{align}
\end{subequations}
From these, the region where the polarization vector (\ref{sceabu}) can be defined in the sense that 
the norms of the 1-particle states in the $S$- and $L$-directions are less than zero is restricted as follows:
\begin{subequations}\label{sdvlne}
\begin{align}
\label{sdvlne1}
\bullet \quad\!
&\textrm{$\eta$, $\tilde{\eta}$} \,
\left\{
\begin{array}{cc}
\le 0 & \textrm{for $a \ge 0$} \vspace{1.5mm} \cr 
> 0   & \textrm{for $a <   0$} 
\end{array}
\right. \,\textrm{in the FRW and PRW,}
\\*[1.0mm]
\bullet \quad\!
&\textrm{$\xi$, $\tilde{\xi}$} \,
\left\{
\begin{array}{cc}
\ge 0 & \textrm{for $a \ge 0$} \vspace{1.5mm} \cr 
< 0   & \textrm{for $a >   0$} 
\end{array}
\right. \,\textrm{in the RRW and LRW,}
\end{align}
\end{subequations}
for $\eta$, $\tilde{\eta}$, $\xi$, $\tilde{\xi}$ $\in (-\infty,\infty)$.  
The regions restricted by (\ref{sdvlne}) are shown in Fig.\ref{wdet7}.
~\newline

\vspace{-8mm}
As can be seen in (\ref{sdvlne}), (\ref{sdvlne}) are constraints which are irrelevant of the relativity.
Below, we discuss why such constraints can appear.

First, (\ref{rsytr1}) and (\ref{rsytr2}) are the equations with regard to the quantities defined in the Rindler coordinates, 
but the metrices in the l.h.s. of those equations are not the Rindler metrices, but the Minkowski metrices; 
therefore, (\ref{rsytr1}) and (\ref{rsytr2}) are some non-covariant equations. 
Therefore, when some equation is derived based on (\ref{rsytr1}) or (\ref{rsytr2}), that equation is irrelevant of the relativity. 
Therefore, since $\eta^{(\sigma\sigma')}$ in (\ref{dvase}) is obtained based on (\ref{rsytr1}) or (\ref{rsytr2}), 
constraints irrelevant of the relativity, such as (\ref{sdvlne}), appears. 
(Therefore, if some polarization vector other than (\ref{sceabu}) is considered, some constraint different from (\ref{sdvlne}) would appear from $\eta^{(\sigma\sigma')}$ at that time.) 

Therefore, since the cause of (\ref{sdvlne}) is the appearance of the Minkowski metrices in (\ref{rsytr1}) and (\ref{rsytr2}), let us consider the origin of those metrices. 
We can see that, in (\ref{reawv}) and (\ref{aeba}), the canonical commutation relations in the Rindler coordinates are defined with the Minkowski metrices; 
once (\ref{reawv}) and (\ref{aeba}) are given, (\ref{rsytr1}) and (\ref{rsytr2}) follow from (\ref{reawv}) and (\ref{aeba}) by proceeding with a calculation following the definition. 
Therefore, as long as (\ref{reawv}) and (\ref{aeba}) are considered, the appearance of the Minkowski metric in (\ref{rsytr1}) and (\ref{rsytr2}) is inevitable. 
Why the Minkowski metric appears in the equation considered in the Rindler coordinates can be considered as a general property of canonical quantization that 
the relativistic covariance is not maintained in canonical quantization.

In conclusion, the appearance of (\ref{sdvlne}) can be attributed to the general property of canonical quantization mentioned above,
and (\ref{sdvlne}) is not a problem of the classical solutions obtained in Sec.\,\ref{f2v4t}-\ref{brpnevt}. 
In this sense, (\ref{sdvlne}) is no more than what specifies the region where the polarization vector (\ref{sceabu}) can be defined in each region in the Rindler coordinates.

Lastly, when (\ref{sdvlne}) is not satisfied, since all four directions are positive norm states, the U(1) gauge field should be massive. 
However, we could not find any effective mass term at that time.
Yet, this is no problem because the relativistic consistency is not maintained in (\ref{sdvlne}) for the reason mentioned above.
Therefore, even if (\ref{sdvlne}) is not satisfied and all four directions become positive norm states, it is not necessary that the effective mass term of the U(1) gauge field must occur.
 
\section{Summary} 
\label{Summary}

To briefly summarize this study.
Since the Unruh temperature has been described around 1975~\cite{Fulling:1972md,Davies:1974th,Unruh:1976db} up until the present, 
while the canonical quantizations of the scalar and spinor fields have been performed, 
that of the gauge field has not been done, in the Rindler coordinates. 
As those three fields are the fundamental fields in the expression of the Lorentz group, 
it is desirable that the canonical quantization of the gauge field are also performed.
Therefore, we have tackled this problem in this study.

It can be considered that 
the integrals appearing in the KG inner-product of the mode-solutions of the U(1) gauge field in the Rindler coordinates have remained unclear 
(those integrals are not even included in \cite{Gradshteyn:1943cpj}). 
This means that the KG inner-product cannot be concretely performed and the normalization constants of the mode-solutions cannot be determined, 
which are crucial in actually performing the canonical quantization of the U(1) gauge field in the Rindler coordinates. 

In this study, calculating those integrals, 
the normalized mode-solutions of the U(1) gauge field have been obtained 
in the Lorentz-covariant gauge in all directions of spacetime in the Rindler coordinates,
and the canonical quantization of the U(1) gauge field in the Rindler coordinates has been performed. 

The mode-solutions obtained in this study are not general solutions, as mentioned at the end of Sec.\,\ref{nnklf}.
However as all directions of the spacetime have been solved by a very explicit manner, 
and no reference is found up until now in which all directions of the gauge field have been solved in such an explicit manner,
it is considered that the mode-solutions given in this study provide a significant contribution to the field. 

Based on those calculations, 
we have shown that 
the coefficients of the normalized mode-solutions of the U(1) gauge field have the role of the creation/annihilation operators 
in the Rindler coordinates. 
Thereby, the U(1) gauge field in the constant accelerated system can be shown 
to feel the Unruh temperature as well~\cite{Takeuchi:2023nxi}.

Regarding the future directions which can be immediately derived from this study, 
extending this study to the non-Abelian gauge theory is considered. 
However, it is well known that in the non-Abelian gauge theory in the Minkowski coordinates,
the BRST invariant Lagrangian can be formulated, but the equations of motion cannot be analytically solved. 
Therefore, in the case that this study is extended to the non-Abelian gauge theory, 
it would be probably possible to formulate the BRST invariant Lagrangian, 
however it would be impossible to obtain the classical solutions of the gauge fields. 
As a study in which the creation/annihilation operators of the U(1) gauge field in the Rindler coordinates are active, 
issues concerning photon antibunching can be considered. 
Currently, many ways to detect the Unruh effect have been investigated, and this would be interesting as one of those new ways.
  
\appendix

\section{Analysis of path-integrals from (\ref{tedrha}) by replacing the Coulomb gauge to a Lorentz-covariant gauge}
\label{buobhs}

Let us replace the Coulomb gauge in (\ref{tedrha}), which is the part $\Omega_{\textrm{Coulomb}}$ in the text under (\ref{tedrha}), to the Lorentz-covariant gauge.  
The Lorentz-covariant gauge which we will consider is 
\begin{eqnarray}\label{ts2rer}
f[A^\mu(x)]={\cal C}(x), \quad  f[A^\mu(x)]\equiv \nabla_\mu A^\mu(x),
\end{eqnarray}
where the gauge to be considered as the Lorentz-covariant gauge in the Minkowski coordinates is usually $\partial_\mu A^\mu$, and this $f$ is the extension of this to the Rindler coordinates;  
therefore $\nabla_\mu$ in this $f$ is the gravitational one, which does not include the gauge field. 
${\cal C}$ is some real function (since ${\cal C}$ is finally integrated out as seen in (\ref{bb54va}), there is no need to specify ${\cal C}$).   

Then, the inverse of the Faddeev-Popov determinant is given as
\begin{eqnarray}\label{sbrtv}
\varDelta_{\rm FP}[A^\mu]^{-1} = \int {\cal D}U \! \prod_{x \in {\rm RRW/FRW}} \delta(f[A^U{}^\mu]-{\cal C}),
\end{eqnarray}
where $A^U{}^\mu$ denotes a gauge transformed $A^\mu$ such as $A^\mu+\partial^\mu U$. 
Under the assumption that $A^U$ satisfying $f={\cal C}$ exists only in the neighborhood of $U=1$, $\varDelta_{\rm FP}[A^\mu]$ can be given as follows:
\begin{eqnarray}\label{eartd}
\varDelta_{\rm FP}[A^\mu] = {\rm Det}\big[\nabla_\mu \partial^\mu  \,\delta^4(x-y)\big],
\end{eqnarray}
where this ${\rm Det}$ is the functional determinant of the functions in the RRW or FRW.
Here, note that since $\nabla_\mu$ in $f$ does not include the gauge fields if the gauge field is the U(1), 
accordingly $\varDelta_{\rm FP}[A]$ does not include the gauge fields if the gauge field is the U(1), 
which leads to a situation where the gauge and ghost fields do not couple each other (as can be seen from (\ref{edsstr})). 
Therefore, we denote $\varDelta_{\rm FP}[A^\mu]$ as $\varDelta_{\rm FP}$ in what follows.

Then, inserting the unity obtained from (\ref{sbrtv}) into (\ref{tedrha}), 
and exploiting the gauge invariance,  
(\ref{tedrha1}) and (\ref{tedrha2}) can be given as follows:
\begin{subequations}\label{erbi75}
\begin{align}
\label{erbi751}
\cdot \quad\!\!
\textrm{(\ref{tedrha1})}
=&
\int \! {\cal D}\!A \,
\Big[
\int \! {\cal D}U \! \prod_{x \in {\rm RRW}} 
\big[\delta(\chi^{(2)})\big] \cdot
\prod_{x^0} M_c
\Big]
\,
\varDelta_{\rm FP}
\prod_{x \in {\rm RRW}}
\big[\delta(f[A^\mu]-{\cal C})\big]  
\exp \big[i\int_{\rm RRW} d^4x \,\sqrt{-g}\,{\cal L}_{\rm U(1)} \big],
\\*[1.0mm]
\label{erbi752}
\cdot \quad\!\!
\textrm{(\ref{tedrha2})}
=&
\int \! {\cal D}\!A \,
\Big[
\int \! {\cal D}U \! \prod_{x \in {\rm FRW}} 
\big[\delta(\chi^{(2)})\big] \cdot
\prod_{x^1} M_c
\Big]
\,
\varDelta_{\rm FP}
\prod_{x \in {\rm FRW}}
\big[\delta(f[A^\mu]-{\cal C})\big]  
\exp \big[i\int_{\rm FRW} d^4x \,\sqrt{-g}\,{\cal L}_{\rm U(1)} \big].
\end{align}
\end{subequations}
$\int \! {\cal D}U \! \prod_{x} [\delta(\chi^{(2)})]$
in (\ref{erbi751}) and (\ref{erbi752}) can be given as
\begin{subequations}
\begin{align}
\label{sbrtv1}
\cdot \quad\!\!
\int \! {\cal D}U \! \prod_{x \in {\rm RRW}} 
\big[\delta(\chi^{(2)})\big]  
&= 
\displaystyle 
\prod_{x^0} 
\Big({\rm Det}
\big[ \nabla_k \partial^k \,\delta^3(\vec{x}-\vec{y})\big]
\Big)^{-1},
\\*[1.0mm]
\label{sbrtv2}
\cdot \quad\!\!
\int \! {\cal D}U \! \prod_{x \in {\rm FRW}} 
\big[\delta(\chi^{(2)})\big]  
&= 
\displaystyle 
\prod_{x^1} 
\Big( {\rm Det}
\big[ \nabla_k \partial^k \,\delta^3(\vec{x}-\vec{y})\big]
\Big)^{-1},
\end{align}
\end{subequations}
where it is assumed that $A^U{}^\mu$ satisfying $\chi^{(2)}=0$ exists only in the neighborhood of $U=1$; 
$\delta^3(\vec{x}-\vec{y})$ are defined in (\ref{ivw3rk}).
Since (\ref{sbrtv1}) and (\ref{sbrtv2}) are the inverse of $\prod_{x^0/x^1} M_c$ given in (\ref{iedub1}) and (\ref{iedub2}) respectively, (\ref{erbi751}) and (\ref{erbi752}) can be given as follows:
\begin{subequations}\label{4reuo4}
\begin{align}
\label{4reuo41}
\cdot \quad\!\!
\textrm{(\ref{erbi751})}
&= 
\int \! {\cal D}\!A \,
\varDelta_{\rm FP}
\prod_{x \in {\rm RRW}}
\big[\delta(f[A^\mu]-{\cal C})\big] \cdot   
\exp \big[i \int_{\rm RRW} d^4x \,\sqrt{-g} \,{\cal L}_{\rm U(1)} \big].
\\*[1.0mm]
\label{4reuo42}
\cdot \quad\!\!
\textrm{(\ref{erbi752})}
&=
\int \! {\cal D}\!A \,
\varDelta_{\rm FP}
\prod_{x \in {\rm RRW}}
\big[\delta(f[A^\mu]-{\cal C})\big] \cdot   
\exp \big[i \int_{\rm RRW} d^4x \,\sqrt{-g} \,{\cal L}_{\rm U(1)} \big].
\end{align}
\end{subequations}

Reweighting $\varDelta_{\rm FP}[A]$ in (\ref{4reuo41}) and (\ref{4reuo42})  
using the ghost and anti-ghost fields $c$ and $\bar{c}$ as 
\begin{eqnarray}\label{edsstr}
\varDelta_{\rm FP}=
\int \! {\cal D}c\,{\cal D}\bar{c} \, 
\exp \, [ \, 
i \! \int_{\rm RRW/FRW} \! d^4x \, \sqrt{-g} \, 
i \,\bar{c} \, \nabla_\mu \partial^\mu  c
],
\end{eqnarray}
we incorporate the contribution of $\varDelta_{\rm FP}$ into the Lagrangian.
Then, inserting the unity defined as
\begin{eqnarray}\label{bb54va}
1= 
\int \! {\cal D}B\, {\cal D}{\cal C}
\exp \big[ \int_{\rm RRW/FRW} \! d^4x \, \sqrt{-g} \, (B\,{\cal C}+ B^2/2) \big], 
\end{eqnarray}
we integrate out ${\cal C}$, where ${\cal C}$ in (\ref{bb54va}) is identified with ${\cal C}$ in the Lorentz-covariant gauge in (\ref{ts2rer}). 
From the form of the path-integral obtained performing these, we can obtain the Lagrangian of the U(1) gauge field in the Lorentz-covariant gauge in the Rindler coordinates given in (\ref{etsiph}).  

\section{Integral formulas}
\label{r3g67kb}

In this Appendix, we give the integral formulas used in Sec.\,\ref{brpnevt} 
to determine the normalization constants in the mode-solutions of the U(1) gauge field in the four regions in the Rindler coordinates. 
The integral formulas in this Appendix are essential in the calculation of the KG inner-product between those mode-solutions. 
We have obtained those in this study to calculate those KG inner-products. 
\newline 

First, we note the integrals and their results in the following:
\begin{subequations}\label{ehktyd1}
\begin{align}
\label{ehktyd11}
\int_0^\infty \! dx \, x^{-1}J_{iu}(x)J_{iv}(x)
=& \,
\frac{\sinh(\pi u)}{u}\delta(u+v),\\*[1.0mm]
\label{ehktyd12}
\int_0^\infty \! dx \, x^{-3}J_{iu}(x)J_{iv}(x)
=& \,
-\frac{\sinh(\pi u)}{2u(1+u^2)}\delta(u+v),\\*[1.0mm]
\label{ehktyd14}
\int_0^\infty \! dx \, x^{-2} J_{iu}(x)(J_{-1+iv}(x)-J_{1+iv}(x))
=& \,\,
-\frac{\sinh(\pi u)}{u(1+u^2)}\delta(u+v),\\*[1.0mm]
\label{ehktyd13}
\int_0^\infty \! dx \, x^{-1}
(J_{-1+iu}(x)-J_{1+iu}(x))
(u \leftrightarrow v)
=& \,
\frac{2u\sinh(\pi u)}{1+u^2}
\delta(u+v),
\end{align}
\end{subequations}
\vspace{-3.0mm}
\begin{subequations}\label{wreh1}
\begin{align}
\label{wreh11}
\int_0^\infty \! dx \, x^{-1}K_{iu}(x)K_{iv}(x)
=& \,
\frac{1}{2u}\frac{\pi^2}{\sinh(\pi u)}\delta(u-v),\\*[1.0mm]
\label{wreh12}
\int_0^\infty \! dx \, x^{-3}K_{iu}(x)K_{iv}(x)
=& \,
\frac{1}{4u(1+u^2)}\frac{\pi^2}{\sinh(\pi u)}\delta(u-v),\\*[1.0mm]
\label{wreh14}
\int_0^\infty \! dx \, x^{-2} K_{iu}(x)(K_{-1+iv}(x)+K_{1+iv}(x))
=& \,\,
-\frac{1}{2u(1+u^2)}\frac{\pi^2}{\sinh(\pi u)}\delta(u-v),\\*[1.0mm]
\label{wreh13}
\int_0^\infty \! dx \, x^{-1}
(K_{-1+iu}(x)+K_{1+iu}(x))
(u \leftrightarrow v)
=& \,
-\frac{u}{1+u^2}
\frac{\pi^2}{\sinh(\pi u)}\delta(u-v),
\end{align}
\end{subequations}
where $J_{iu}(x)$ and $K_{iu}(x)$ are the Bessel function of the first kind and the modified Bessel function of the second kind, respectively. 
(\ref{ehktyd1}) and (\ref{wreh1}) are used to determine the normalization constants of the mode-solutions in the FRW and RRW, respectively.

The integrals same as (\ref{ehktyd11}) and (\ref{wreh11}) can be found in \cite{Gradshteyn:1943cpj} (2 of 6.574 and 4 of 6.576), 
however the definition ranges of those and (\ref{ehktyd11}) and (\ref{wreh11}) are different \cite{Gradshteyn:1943cpj}. 
Regarding the remaining (\ref{ehktyd12})-(\ref{ehktyd13}) and (\ref{wreh12})-(\ref{wreh13}), those are not included in \cite{Gradshteyn:1943cpj}.

In what follows, how we have obtained (\ref{ehktyd11}) and (\ref{wreh11}) is shown.
The remaining (\ref{wreh12})-(\ref{ehktyd13}) and (\ref{wreh12})-(\ref{wreh13}) can be obtained in the same way. 
\newline 
 
Using the formula in \cite{Gradshteyn:1943cpj} (2 of 6.574), (\ref{ehktyd11}) can be calculated as follows:
\begin{eqnarray}\label{dthsp1}
\int_0^\infty \! dx \, x^{-1}J_{iu}(x)J_{iv}(x)
=
\frac{\Gamma(\frac{i(u+v)}{2})}
{2
\Gamma(\frac{i(-u+v)}{2}+1)
\,\Gamma(\frac{i( u+v)}{2}+1)
\,\Gamma(\frac{i( u-v)}{2}+1)}
=\frac{-2i}{u^2-v^2}
\frac{\sinh(\pi\frac{u-v}{2})}{\pi}.
\end{eqnarray}

Considering that $u+v=i\varepsilon$, we can write (\ref{dthsp1})  as follows:
\begin{eqnarray}\label{dhsez}
\textrm{(\ref{dthsp1})}
=
\frac{2\sinh(\pi\frac{u-v}{2})}{u-v}\,
\frac{-i}{\pi}
\frac{1}{i \varepsilon}
=
\frac{2\sinh(\pi\frac{u-v}{2})}{u-v}\,
\frac{-i}{\pi}
\lim_{\lambda \to 0}
\frac{1}{i \varepsilon-i\lambda}.
\end{eqnarray}
Then, in general, there is a relation held as the relation of the integrand: 
$\lim_{\lambda \to 0}\frac{1}{x-i\lambda}={\rm p.v.}\frac{1}{x}+\pi i \,\delta(x)$. 
From this relation, since $\delta(x)$ can be written as
$\delta(x)
= \frac{1}{\pi}{\rm Im}\big[\lim_{\lambda \to 0}\frac{1}{x-i\lambda}\big]
= \frac{-i}{\pi}\lim_{\lambda \to 0}\frac{1}{x-i\lambda}
$, (\ref{dhsez}) can be written as follows:
\begin{eqnarray}\label{vsdvbs2}
\textrm{(\ref{dhsez})}
=
\frac{\sinh(\pi u)}{u}\,
\delta(u+v),
\end{eqnarray}
where $i\varepsilon=u+v$, and for the appearance of the $\delta(u+v)$, $v$ has been set as $-u$.
From this result, (\ref{ehktyd11}) is obtained.
\newline

Next, using the formula in \cite{Gradshteyn:1943cpj} (4 of 6.576), (\ref{wreh11}) can be written as follows:
\begin{eqnarray}\label{vsdvbs1}
&& \!\!
\int_0^\infty \! dx \, x^{-(1-\omega)}K_{iu}(x)K_{iv}(x)
\nonumber\\*[1.0mm]
\!\! &=& \!\!
\hspace{4.0mm}
\frac{2^{-2-(1-\omega)}}{\Gamma(1-(1-\omega))}
\Gamma(\frac{1-(1-\omega)+i(u+v)}{2})
\,\Gamma(\frac{1-(1-\omega)+i(-u+v)}{2})
\nonumber\\*[1.0mm]
&& \!\! 
\times \,
\Gamma(\frac{1-(1-\omega)+i(u-v)}{2})
\,\Gamma(\frac{1-(1-\omega)-i(u+v)}{2}), 
\end{eqnarray}
where, in the first line, we have put $x^{-1}$ in  (\ref{wreh11}) as $x^{-(1-\omega)}$ 
($\omega$ is  finally taken to $0$).
We may set all $\omega$ as zero except for that in $\Gamma(\omega)$.  
As a result, (\ref{vsdvbs1}) can be written as follows: 
\begin{eqnarray}\label{vhscu}
\textrm{(\ref{vsdvbs1})}
\!\!\! &=& \!\!\!
\frac{2^{-3}}{\Gamma(\omega)} 
\, \Gamma(\frac{i(u+v)}{2}) 
\, \Gamma(\frac{i(-u+v)}{2})
\, \Gamma(\frac{i(u-v)}{2})
\, \Gamma(\frac{-i(u+v)}{2})
\nonumber\\*[1.0mm]
\!\!\! &=& \!\!\!
2^{-3}\,\omega
\, \frac{\pi}{\frac{u-v}{2}\sinh(\pi\frac{u-v}{2})}
\, \frac{\pi}{\frac{u+v}{2}\sinh(\pi \frac{u+v}{2})}+{\cal O}(\omega^2),
\end{eqnarray}
where $1/\Gamma(\omega)=\omega+{\cal O}(\omega^2)$.

Since $\omega$ is taken to $0$,
(\ref{vhscu}) vanishes  for the case $u \not=v$; namely,
\begin{eqnarray}\label{rnast}
\textrm{(\ref{vhscu})}=0 \quad \textrm{for $u \not=v$}.
\end{eqnarray}
On the other hand, in the case $u =v$, 
we suppose that $u-v=i \varepsilon$ ($\varepsilon$ is finally taken to $0$).
Then, we suppose $\varepsilon$ as $\varepsilon=\omega$ as the same infinitesimal quantity.
As a result, 
(\ref{vhscu}) can be calculated as
\begin{eqnarray}\label{vsdvbs3}
\textrm{(\ref{vhscu})}
= 
\frac{-i}{\pi}
\frac{1}{i\varepsilon}
\frac{\pi^2}{2u \sinh(\pi u)}+{\cal O}(\varepsilon^2),
\end{eqnarray}
where the expansion around $\varepsilon=0$ has been performed making use of the fact that $\varepsilon$ is finally taken to $0$, 
and considering the fact that $\delta(u-v)$ finally appears, we have set as $u=v$ at the stage of (\ref{vsdvbs3}). 
Then, using the expression of the $\delta$-function used in obtaining (\ref{vsdvbs2}) from (\ref{dhsez}),
(\ref{vsdvbs3}) can be written as follows:
\begin{eqnarray}\label{vsdvbs4}
\textrm{(\ref{vsdvbs3})}
=
\frac{-i}{\pi}
\lim_{\lambda \to 0}
\frac{1}{i\varepsilon-i\lambda}\,
\frac{\pi^2}{2u \sinh(\pi u)}+{\cal O}(\varepsilon^2)
=\delta(u-v)\,\frac{\pi^2}{2u \sinh(\pi u)}+{\cal O}(\varepsilon^2).
\end{eqnarray}
From this result and  the result of (\ref{rnast}) for the case of $u\not=v$, (\ref{wreh11}) is obtained. 



\begin{thebibliography}{100} 



\bibitem{Fulling:1972md}  
S.~A.~Fulling,
``Nonuniqueness of canonical field quantization in Riemannian space-time,''
Phys. Rev. D \textbf{7}, 2850-2862 (1973)

\bibitem{Davies:1974th}  
P.~C.~W.~Davies,
``Scalar particle production in Schwarzschild and Rindler metrics,''
J.\ Phys.\ A {\bf 8}, 609 (1975). 

\bibitem{Unruh:1976db} 
W.~G.~Unruh,
``Notes on black hole evaporation,''
Phys.\ Rev.\ D {\bf 14}, 870 (1976).
 

\bibitem{Retzker} 
A.~Retzker, J. I.~Cirac, M.~B.~Plenio and B.~Reznik
``Detection of acceleration radiation in a Bose-Einstein condensate,''
Phys.\ Rev.\ Lett.\  {\bf 101}, 110402 (2008). 
[arXiv:0709.2425 [quant-ph]]


\bibitem{Dvornikov:2015eqa}
M.~Dvornikov,
``Unruh effect for neutrinos interacting with accelerated matter,''
JHEP \textbf{08}, 151 (2015)
[arXiv:1507.01174 [hep-ph]].

\bibitem{Blasone:2018czm}
M.~Blasone, G.~Lambiase, G.~G.~Luciano and L.~Petruzziello,
``Role of neutrino mixing in accelerated proton decay,''
Phys. Rev. D \textbf{97}, no.10, 105008 (2018)
[arXiv:1803.05695 [hep-ph]].

\bibitem{Blasone:2020vtm}
M.~Blasone, G.~Lambiase, G.~G.~Luciano and L.~Petruzziello,
``On the $\beta $-decay of the accelerated proton and neutrino oscillations: a three-flavor description with CP violation,''
Eur. Phys. J. C \textbf{80}, no.2, 130 (2020)
[arXiv:2002.03351 [hep-ph]].


\bibitem{Chen:2021evr}
Y.~Chen, J.~Hu and H.~Yu,
``Entanglement generation for uniformly accelerated atoms assisted by environment-induced interatomic interaction and the loss of the anti-Unruh effect,''
Phys. Rev. D \textbf{105}, no.4, 045013 (2022)
[arXiv:2110.01780 [quant-ph]].

\bibitem{Garay:2016cpf}
L.~J.~Garay, E.~Martin-Martinez and J.~de Ramon,
``Thermalization of particle detectors: The Unruh effect and its reverse,''
Phys. Rev. D \textbf{94}, no.10, 104048 (2016)
[arXiv:1607.05287 [quant-ph]].

\bibitem{Liu:2016ihf}
P.~H.~Liu and F.~L.~Lin,
``Decoherence of Topological Qubit in Linear and Circular Motions: Decoherence Impedance, Anti-Unruh and Information Backflow,''
JHEP \textbf{07}, 084 (2016)
[arXiv:1603.05136 [quant-ph]].

\bibitem{Brenna:2015fga}
W.~G.~Brenna, R.~B.~Mann and E.~Martin-Martinez,
``Anti-Unruh Phenomena,''
Phys. Lett. B \textbf{757}, 307-311 (2016)
[arXiv:1504.02468 [quant-ph]].

\bibitem{Kosior:2018vgx}
A.~Kosior, M.~Lewenstein and A.~Celi,
``Unruh effect for interacting particles with ultracold atoms,''
SciPost Phys. \textbf{5}, no.6, 061 (2018)
[arXiv:1804.11323 [cond-mat.quant-gas]].

\bibitem{Rodriguez-Laguna:2016kri}
J.~Rodriguez-Laguna, L.~Tarruell, M.~Lewenstein and A.~Celi,
``Synthetic Unruh effect in cold atoms,''
Phys. Rev. A \textbf{95}, no.1, 013627 (2017)
[arXiv:1606.09505 [cond-mat.quant-gas]].

\bibitem{Quach:2021vzo}
J.~Q.~Quach, T.~C.~Ralph and W.~J.~Munro,
``Berry Phase from the Entanglement of Future and Past Light Cones: Detecting the Timelike Unruh Effect,''
Phys. Rev. Lett. \textbf{129}, no.16, 160401 (2022)
[arXiv:2112.00898 [gr-qc]].

\bibitem{Martin-Martinez:2010gnz}
E.~Martin-Martinez, I.~Fuentes and R.~B.~Mann,
``Using Berry's phase to detect the Unruh effect at lower accelerations,''
Phys. Rev. Lett. \textbf{107}, 131301 (2011)
[arXiv:1012.2208 [quant-ph]].


\bibitem{Marino:2014rfa}
J.~Marino, A.~Noto and R.~Passante,
``Thermal and Nonthermal Signatures of the Unruh Effect in Casimir-Polder Forces,''
Phys. Rev. Lett. \textbf{113}, no.2, 020403 (2014)
[arXiv:1403.2437 [quant-ph]].

\bibitem{Leonhardt:2017lwm}
U.~Leonhardt, I.~Griniasty, S.~Wildeman, E.~Fort and M.~Fink,
``Classical analog of the Unruh effect,''
Phys. Rev. A \textbf{98}, no.2, 022118 (2018)
[arXiv:1709.02200 [gr-qc]].
  

\bibitem{Lynch:2019hmk}
M.~H.~Lynch, E.~Cohen, Y.~Hadad and I.~Kaminer,
``Experimental observation of acceleration-induced thermality,''
Phys. Rev. D \textbf{104}, no.2, 025015 (2021)
[arXiv:1903.00043 [gr-qc]].

\bibitem{Lochan:2019osm}
K.~Lochan, H.~Ulbricht, A.~Vinante and S.~K.~Goyal,
``Detecting Acceleration-Enhanced Vacuum Fluctuations with Atoms Inside a Cavity,''
Phys. Rev. Lett. \textbf{125}, 241301 (2020)
[arXiv:1909.09396 [gr-qc]].


\bibitem{Ohsaku:2004rv}
T.~Ohsaku,
``Dynamical chiral symmetry breaking and its restoration for an accelerated observer,''
Phys. Lett. B \textbf{599}, 102-110 (2004)
[arXiv:hep-th/0407067 [hep-th]].

\bibitem{Ebert:2006bh}
D.~Ebert and V.~C.~Zhukovsky,
``Restoration of Dynamically Broken Chiral and Color Symmetries for an Accelerated Observer,''
Phys. Lett. B \textbf{645}, 267-274 (2007)
[arXiv:hep-th/0612009 [hep-th]].

\bibitem{Castorina:2007eb}
P.~Castorina, D.~Kharzeev and H.~Satz,
``Thermal Hadronization and Hawking-Unruh Radiation in QCD,''
Eur. Phys. J. C \textbf{52}, 187-201 (2007)
[arXiv:0704.1426 [hep-ph]].

\bibitem{Takeuchi:2015nga}
S.~Takeuchi,
``Bose\textendash{}Einstein condensation in the Rindler space,''
Phys. Lett. B \textbf{750}, 209-217 (2015)
[arXiv:1501.07471 [hep-th]].


\bibitem{Schutzhold:2006gj}
R.~Schutzhold, G.~Schaller and D.~Habs,
``Signatures of the Unruh effect from electrons accelerated by ultra-strong laser fields,''
Phys. Rev. Lett. \textbf{97}, 121302 (2006)
[erratum: Phys. Rev. Lett. \textbf{97}, 139902 (2006)]
[arXiv:quant-ph/0604065 [quant-ph]].

\bibitem{Schutzhold:2009scb}
R.~Schutzhold and C.~Maia,
``Quantum radiation by electrons in lasers and the Unruh effect,''
Eur. Phys. J. D \textbf{55}, 375 (2009)
[arXiv:1004.2399 [hep-th]].

\bibitem{Iso:2010yq}
S.~Iso, Y.~Yamamoto and S.~Zhang,
``Stochastic Analysis of an Accelerated Charged Particle -Transverse Fluctuations-,''
Phys. Rev. D \textbf{84}, 025005 (2011)
[arXiv:1011.4191 [hep-th]].

\bibitem{Oshita:2015xaa}
N.~Oshita, K.~Yamamoto and S.~Zhang,
``Quantum radiation from a particle in an accelerated motion coupled to vacuum fluctuations,''
Phys. Rev. D \textbf{92}, no.4, 045027 (2015)
[arXiv:1508.06338 [hep-th]].

\bibitem{Iso:2016lua}
S.~Iso, N.~Oshita, R.~Tatsukawa, K.~Yamamoto and S.~Zhang,
``Quantum radiation produced by the entanglement of quantum fields,''
Phys. Rev. D \textbf{95}, no.2, 023512 (2017)
[arXiv:1610.08158 [hep-th]].


\bibitem{Parentani:1996gd}
R.~Parentani and S.~Massar,
``The Schwinger mechanism, the Unruh effect and the production of accelerated black holes,''
Phys. Rev. D \textbf{55}, 3603-3613 (1997)
[arXiv:hep-th/9603057 [hep-th]].

\bibitem{Kim:2016dmm}
S.~P.~Kim,
``Schwinger Effect, Hawking Radiation, and Unruh Effect,''
Int. J. Mod. Phys. D \textbf{25}, no.13, 1645005 (2016)
[arXiv:1602.05336 [hep-th]].

\bibitem{Kaushal:2022las}
S.~Kaushal,
``Schwinger effect and a uniformly accelerated observer,''
Eur. Phys. J. C \textbf{82}, no.10, 872 (2022)
[arXiv:2201.03906 [hep-th]].

\bibitem{Prokhorov:2019cik}
G.~Y.~Prokhorov, O.~V.~Teryaev and V.~I.~Zakharov,
``Unruh effect for fermions from the Zubarev density operator,''
Phys. Rev. D \textbf{99}, no.7, 071901 (2019)
[arXiv:1903.09697 [hep-th]].

\bibitem{Prokhorov:2019hif}
G.~Y.~Prokhorov, O.~V.~Teryaev and V.~I.~Zakharov,
``Thermodynamics of accelerated fermion gases and their instability at the Unruh temperature,''
Phys. Rev. D \textbf{100}, no.12, 125009 (2019)
[arXiv:1906.03529 [hep-th]].

\bibitem{Prokhorov:2019yft}
G.~Y.~Prokhorov, O.~V.~Teryaev and V.~I.~Zakharov,
``Unruh effect universality: emergent conical geometry from density operator,''
JHEP \textbf{03}, 137 (2020)
[arXiv:1911.04545 [hep-th]].

\bibitem{Prokhorov:2019sss}
G.~Y.~Prokhorov, O.~V.~Teryaev and V.~I.~Zakharov,
``Calculation of acceleration effects using the Zubarev density operator,''
Particles \textbf{3}, no.1, 1-14 (2020)
[arXiv:1911.04563 [hep-th]].

\bibitem{Parentani:1989gq}
R.~Parentani and R.~Potting,
``The Accelerating Observer and the Hagedorn Temperature,''
Phys. Rev. Lett. \textbf{63}, 945 (1989)


\bibitem{Jacobson:1995ab}
T.~Jacobson,
``Thermodynamics of space-time: The Einstein equation of state,''
Phys. Rev. Lett. \textbf{75}, 1260-1263 (1995)
[arXiv:gr-qc/9504004 [gr-qc]].


\bibitem{Kabat:1994vj}
D.~N.~Kabat and M.~J.~Strassler,
``A Comment on entropy and area,''
Phys. Lett. B \textbf{329}, 46-52 (1994)
[arXiv:hep-th/9401125 [hep-th]].

\bibitem{Dowker:1994fi}
J.~S.~Dowker,
``Remarks on geometric entropy,''
Class. Quant. Grav. \textbf{11}, L55-L60 (1994)
[arXiv:hep-th/9401159 [hep-th]].

\bibitem{Iellici:1996gv}
D.~Iellici and V.~Moretti,
``Thermal partition function of photons and gravitons in a Rindler wedge,''
Phys. Rev. D \textbf{54}, 7459-7469 (1996)
[arXiv:hep-th/9607015 [hep-th]].


\bibitem{Deser:1997ri}
S.~Deser and O.~Levin,
``Accelerated detectors and temperature in (anti)-de Sitter spaces,''
Class. Quant. Grav. \textbf{14}, L163-L168 (1997)
[arXiv:gr-qc/9706018 [gr-qc]].

\bibitem{Parikh:2012kg}
M.~Parikh and P.~Samantray,
``Rindler-AdS/CFT,''
JHEP \textbf{10}, 129 (2018)
[arXiv:1211.7370 [hep-th]].

\bibitem{Fareghbal:2014oba}
R.~Fareghbal and A.~Naseh,
``Rindler/Contracted-CFT Correspondence,''
JHEP \textbf{06}, 134 (2014)
[arXiv:1404.3937 [hep-th]].

\bibitem{Sugishita:2022ldv}
S.~Sugishita and S.~Terashima,
``Rindler bulk reconstruction and subregion duality in AdS/CFT,''
JHEP \textbf{11}, 041 (2022)
[arXiv:2207.06455 [hep-th]].

\bibitem{Rovelli:2013osa}
C.~Rovelli and F.~Vidotto,
``Evidence for Maximal Acceleration and Singularity Resolution in Covariant Loop Quantum Gravity,''
Phys. Rev. Lett. \textbf{111}, 091303 (2013)
[arXiv:1307.3228 [gr-qc]].

\bibitem{Terashima:1999xp}
H.~Terashima,
``Fluctuation dissipation theorem and the Unruh effect of scalar and Dirac fields,''
Phys. Rev. D \textbf{60}, 084001 (1999)
[arXiv:hep-th/9903062 [hep-th]].


\bibitem{Higuchi:2017gcd}
A.~Higuchi, S.~Iso, K.~Ueda and K.~Yamamoto,
``Entanglement of the Vacuum between Left, Right, Future, and Past: The Origin of Entanglement-Induced Quantum Radiation,''
Phys. Rev. D \textbf{96}, no.8, 083531 (2017)
[arXiv:1709.05757 [hep-th]].

\bibitem{Soffel:1980kx}
M.~Soffel, B.~Muller and W.~Greiner,
``DIRAC PARTICLES IN RINDLER SPACE,''
Phys. Rev. D \textbf{22}, 1935-1937 (1980)

\bibitem{Ueda:2021nln}
K.~Ueda, A.~Higuchi, K.~Yamamoto, A.~Rohim and Y.~Nan,
``Entanglement of the Vacuum between Left, Right, Future, and Past: Dirac spinor in Rindler spaces and Kasner spaces,''
Phys. Rev. D \textbf{103}, 125005 (2021)
[arXiv:2104.06625 [gr-qc]].


\bibitem{Higuchi:1992td}
A.~Higuchi, G.~E.~A.~Matsas and D.~Sudarsky,
``Bremsstrahlung and Fulling-Davies-Unruh thermal bath,''
Phys. Rev. D \textbf{46}, 3450-3457 (1992)

\bibitem{Moretti:1996zt}
V.~Moretti,
``Canonical quantization of photons in a Rindler wedge,''
J. Math. Phys. \textbf{38}, 2922-2953 (1997)
[arXiv:gr-qc/9603057 [gr-qc]].

\bibitem{Lenz:2008vw}
F.~Lenz, K.~Ohta and K.~Yazaki,
``Canonical quantization of gauge fields in static space-times with applications to Rindler spaces,''
Phys. Rev. D \textbf{78}, 065026 (2008)
[arXiv:0803.2001 [hep-th]].

\bibitem{Zhitnitsky:2010ji}
A.~R.~Zhitnitsky,
``The Gauge Fields and Ghosts in Rindler Space,''
Phys. Rev. D \textbf{82}, 103520 (2010)
[arXiv:1004.2040 [gr-qc]].

\bibitem{Soldati:2015xma}
R.~Soldati and C.~Specchia,
J. Mod. Phys. \textbf{6}, 1743 (2015)
[arXiv:1504.01880 [hep-th]].

\bibitem{Blommaert:2018rsf}
A.~Blommaert, T.~G.~Mertens, H.~Verschelde and V.~I.~Zakharov,
``Edge State Quantization: Vector Fields in Rindler,''
JHEP \textbf{08}, 196 (2018)
[arXiv:1801.09910 [hep-th]].


\bibitem{Takeuchi:2023nxi}
S.~Takeuchi,
``Canonical quantization of the U(1) gauge field in the right Rindler-wedge in the Rindler coordinates,''
Eur. Phys. J. C \textbf{84}, no.12, 1249 (2024)
[arXiv:2309.09798 [hep-th]].


\bibitem{Gradshteyn:1943cpj}
I.~S.~Gradshteyn, I.~M.~Ryzhik, 
A.~Jeffrey (Editor), and D.~Zwillinger (Editor) 
``Table of Integrals, Series, and Products, Seventh Edition,''
2007, Academic Press.

\end{thebibliography}
\end{document}